\documentclass[a4paper,11pt]{article}
\pdfoutput=1
\usepackage{jheppub}
\usepackage{xcolor,bm,comment,braket,mathtools}
\usepackage[T1]{fontenc}
\usepackage{comment}
\usepackage{graphics}
\usepackage{ulem}

\newcommand{\nn}{\nonumber}
\def\p{\partial}

\newcommand {\beq}{\begin{eqnarray}}
\newcommand {\eeqq}{\end{eqnarray}}

\newcommand {\kahler}{K\"ahler }

\def\hsymbu#1{\smash{\lower2.5ex\hbox{\huge$#1$}}}
\makeatletter
\newcommand{\doublehat}[1]{% 
\begingroup%
  \let\macc@kerna\z@%
  \let\macc@kernb\z@%
  \let\macc@nucleus\@empty%
  \hat{\raisebox{.32ex}{\vphantom{\ensuremath{#1}}}\smash{\hat{#1}}}%
\endgroup%
}
\makeatother

\makeatletter
\@addtoreset{equation}{section}

\renewcommand{\thefootnote}{\fnsymbol{footnote}}
\makeatother

\newcommand{\rme}{\mathrm{e}}
\newcommand{\im}{\mathrm{i}}

\PassOptionsToPackage{unicode}{hyperref}
\PassOptionsToPackage{naturalnames}{hyperref}
\begin{document}
\thispagestyle{empty}
%\title{BPS lumps on a torus with  't Hooft twist}
% \title{2d/4d fractional BPS instantons on a torus with  't Hooft twist}
\title{Fractional instantons in 2d $\mathbb{C}P^{N-1}$ model and \\
4d Yang-Mills theory with  't Hooft twists}

\author[1]{Yui Hayashi,}
\emailAdd{yui.hayashi@yukawa.kyoto-u.ac.jp}
\affiliation[1]{Yukawa Institute for Theoretical Physics, Kyoto University, Kyoto 606-8502, Japan}

\author[2,3]{Tatsuhiro Misumi,}
\emailAdd{misumi@phys.kindai.ac.jp}
\affiliation[2]{Department of Physics, Kindai University, Osaka 577-8502, Japan}
\affiliation[3]{Research and Education Center for Natural Sciences, Keio University, Kanagawa 223-8521, Japan}

\author[4,3,5]{Muneto Nitta,}
\emailAdd{nitta@phys-h.keio.ac.jp}
\affiliation[4]{Department of Physics, 
%$\&$ 
%Research and Education Center for Natural Sciences,
Keio University, 4-1-1 Hiyoshi, Kanagawa 223-8521, Japan}

\affiliation[5]{
International Institute for Sustainability with Knotted Chiral Meta Matter(WPI-SKCM$^2$), Hiroshima University, 1-3-2 Kagamiyama, Higashi-Hiroshima, Hiroshima 739-8511, Japan
}

\author[3,4]{Keisuke Ohashi}
\emailAdd{keisike084@gmail.com}

\author[1]{and\\ Yuya Tanizaki}
\emailAdd{yuya.tanizaki@yukawa.kyoto-u.ac.jp}

\preprint{YITP-25-107}

\abstract{
We derive the explicit formula for fractional BPS lumps (or fractional instantons) in the $\mathbb{C}P^{N-1}$ nonlinear sigma model on a two-dimensional torus under various shift-clock twisted boundary conditions. 
After regularizing the $\mathbb{C}P^{N-1}$ model by an $N$-component Abelian-Higgs model, those twisted boundary conditions introduce nontrivial 't~Hooft fluxes $p/N$ for the $U(1)$ gauge field, and the topological charge becomes fractionalized as $k+p/N\in \mathbb{Z}+p/N$. 
The moduli space is globally determined as the $\mathbb{C}P^{Nk+p-1}$-fiber bundle on a $2$-torus, which is a K\"ahler  manifold of complex dimension $Nk + p$ as predicted by the index theorem. 
We present two different parametrizations of the moduli space: one of them immediately identifies the small-lump singularity appearing in the $\mathbb{C}P^{N-1}$ limit, while the other makes the modular invariance manifest. 

We also discuss the implications of our finding for the $4$d $SU(N)$ Yang-Mills theory on the $4$-torus with 't~Hooft twists. By tuning the aspect ratio of the 4-torus, fractional instantons in the $\mathbb{C}P^{N-1}$ model with a non-Fubini-Study metric are obtained through the dimensional reduction of $4$d Yang-Mills theory, whose moduli space coincides with the one obtained for the standard $\mathbb{C}P^{N-1}$ model as complex manifolds. 
}

\maketitle

\setcounter{page}{1}
\setcounter{footnote}{0}
\renewcommand{\thefootnote}{\arabic{footnote}}

%%%%%%%%%%%%%%%%%%%%%%%%%%%%%%%%%%%%%%%%%%%%%%%
%\newpage
\setcounter{tocdepth}{3}
%\tableofcontents
%%%%%%%%%%%%%%%%%%%%%%%%%%%%%%%%%%%%%%%%%%%%%%%%%%%%%%%%%%%
\section{Introduction}

Instantons in $4$d $SU(N)$ Yang-Mills theory~\cite{Belavin:1975fg} have remarkably rich structure. 
On the $4$-torus, the moduli space of $k$-instantons has the dimension $4Nk$, but the physical meaning of these moduli is mostly wrapped in mystery. 
To see its richness, let us put the $4$d Yang-Mills theory on $\mathbb{R}^3\times S^1$ and consider the situation, where the $3$d gauge field is Abelianized due to the nontrivial holonomy backgrounds~\cite{Lee:1997vp, Lee:1998bb, Lee:1998vu, Kraan:1998kp, Kraan:1998pm, Kraan:1998sn}. 
In this setup, the $4$d $1$-instanton splits into $N$ monopole constituents and each of them carries $4$ zero modes of the $4N$ moduli directions. 

When we just put the Yang-Mills theory on $\mathbb{R}^3\times S^1$, the deconfinement transition occurs for small $S^1$ and the trivial holonomy is preferred by the one-loop effective potential~\cite{Gross:1980br}; in this case, instantons do not enjoy fractionalization. 
It has been discovered that confinement can persist with a double-trace deformation or installation of adjoint fermions on $\mathbb{R}^3 \times S^1$~\cite{Davies:1999uw, Davies:2000nw, Unsal:2007vu, Unsal:2007jx, Shifman:2008ja}, where the monopole constituents play a central role in nonperturbative phenomena. 
This method opens a path to understand the qualitative aspects of quark confinement based on the semiclassical analytic approach of field theory. 
See Refs.~\cite{Gonzalez-Arroyo:2023kqv, Poppitz:2021cxe} for reviews on the uses of fractional instantons.

Instead of adding the center-stabilizing deformation, Yang-Mills theory (and also QCD) on $\mathbb{R}^2 \times T^2$ with a nontrivial 't Hooft flux turns out to have the semiclassical description of confinement for the small $T^2$ regime~\cite{Tanizaki:2022ngt, Tanizaki:2022plm, Hayashi:2023wwi, Hayashi:2024qkm, Hayashi:2024gxv}. 
In this setup, the $4$d instanton again splits into $N$ constituents, but they can be understood as the center vortex that rotates the phases of Wilson loops~\cite{Gonzalez-Arroyo:1998hjb, Montero:1999by, Montero:2000pb, Wandler:2024hsq, GarciaPerez:1989gt, GarciaPerez:1992fj, Itou:2018wkm} (see also Refs.~\cite{Ford:2002pa, Ford:2003vi, Ford:2005sq} for the analytic construction of the $SU(2)$ case via double-periodic instantons). 
The relation between these two semiclassical approaches was unclear until recently, but 
Refs.~\cite{Hayashi:2024yjc, Guvendik:2024umd, Hayashi:2024psa, Hayashi:2025doq} demystify the explicit connection between the monopole constituents on $\mathbb{R}^3\times S^1$ and the center-vortex constituents on $\mathbb{R}^2\times T^2$.

Properties of fractional instantons on the torus with 't~Hooft twists are not fully understood yet even with the help of the Atiyah-Drinfeld-Hitchin-Manin(ADHM)-Nahm technique~\cite{Atiyah:1978ri, Nahm:1979yw}.
In particular, the moduli space of fractional instantons has been a subject of research interest, see Refs.~\cite{Anber:2022qsz, Anber:2023sjn, Anber:2024mco, Anber:2024uwl, Anber:2025yub} for recent studies.
% \blue{Let us point out the work by Ford and Pawlowski~\cite{Ford:2002pa, Ford:2003vi, Ford:2005sq}, which shows, for the case of $SU(2)$, that a doubly-periodic instanton at a special parameter on $\mathbb{R}^2 \times T^2$ can be divided into two fractional instantons. }

These compactified theories with the $SU(N)$ gauge group exhibit deep connections with the two-dimensional ${\mathbb C} P^{N-1}$ models in several aspects as detailed below, where ${\mathbb C} P^{N-1}$ is a complex projective space in the form of a quotient space $\mathbb C P^{N-1}\simeq \frac{SU(N)}{SU(N-1)\times U(1)}$.
Therefore, in this paper, we will examine in detail the properties of fractional instantons in the ${\mathbb C} P^{N-1}$ model, focusing on their moduli space.

The most direct connection between fractional instantons of two theories comes out by considering the $4$d $SU(N)$ Yang-Mills theory on 
\begin{align}
    \mathbb{R}\times \underbrace{(S^1)_B \times (S^1)_C}_{\text{'t Hooft twist}}\times (S^1)_D, 
\end{align}
with small $(S^1)_C\times (S^1)_D$~\cite{Yamazaki:2017ulc, Cox:2021vsa}. 
Then, the $2$d effective theory on $\mathbb{R}\times (S^1)_B$ becomes the $2$d $\mathbb{C}P^{N-1}$ sigma model with the clock-twisted boundary condition along the $B$ direction, while the target $\mathbb{C}P^{N-1}$ variety possesses several singularities instead of the round Fubini-Study metric. 
Fractional instantons of the $\mathbb{Z}_N$-twisted $\mathbb{C}P^{N-1}$ sigma model are first studied in Refs.~\cite{Eto:2004rz, Eto:2006mz}, and the above discussion relates the problem of $4$d fractional instantons to the one in $2$d, which shall be discussed in more details in Section~\ref{sec:4d2d}.
Even before these recent works, the analogy between instantons in the ${\mathbb C} P^{N-1}$ model and those in Yang-Mills theory had been recognized. 
For instance, a well-known similarity is the topological obstruction to single-instanton solutions on periodic tori: just as a 1-instanton solution is absent in $SU(N)$ Yang-Mills theory on a periodic $T^4$  \cite{Schenk:1986xe, Braam:1988qk}, so too is it absent on a periodic $T^2$ in the ${\mathbb C} P^{N-1}$ model \cite{Richard:1981bv, Speight:1997zc}.
In addition, instantons on a periodic $T^2$ in the ${\mathbb C} P^{N-1}$  model have often been studied \cite{Richard:1981bv, Speight:1997zc, Speight:2001he, Aguado:2001xg, Nakamula:2012hi}.
In this paper, we expand upon this line of research to include fractional instantons on $T^2$ with twisted boundary conditions.

The other motivation for studying the $2$d fractional instantons emerges from the study of soliton junctions in the Yang-Mills-Higgs system. 
There, the $2$d ${\mathbb C} P^{N-1}$ model appears on a vortex worldsheet; 
If we consider a 
$U(N) = \frac{SU(N) \times U(1)}{{\mathbb Z}_N}$ gauge theory, 
instead of $SU(N)$, 
coupled with 
$N$ Higgs scalar fields in the fundamental representation, 
%and common $U(1)$ charges,  
it admits a non-Abelian local vortex 
\cite{Hanany:2003hp,Auzzi:2003fs,Eto:2005yh,Eto:2006cx} 
whose low-energy dynamics can be described by the two-dimensional ${\mathbb C} P^{N-1}$ model
(see Refs.~\cite{Eto:2006pg,Shifman:2007ce,Shifman:2009zz} as a review).\footnote{
If the $U(1)$ part is not gauged and is a global symmetry, the model still admits a non-Abelian global vortex 
\cite{Balachandran:2005ev,Nakano:2007dr,Eto:2009kg}
whose low-energy dynamics is still described by the two-dimensional ${\mathbb C} P^{N-1}$ model \cite{Eto:2009bh,Eto:2009tr}.
This is a setup of the color-flavor locked phase of dense QCD, see  Ref.~\cite{Eto:2013hoa} as a review.
}
When absorbed into a vortex, Yang-Mills instantons become 
Bogomol'nyi-Prasad-Sommerfield (BPS) ${\mathbb C}P^{N-1}$ lumps in the vortex worldsheet
\cite{Hanany:2004ea,Shifman:2004dr,Eto:2004rz,Fujimori:2008ee}.
When one considers ${\mathbb R}^3 \times S^1$ with a twisted boundary condition, 
one ${\mathbb C}P^{N-1}$ lump in a vortex worldsheet 
decomposes into $N$ ${\mathbb C}P^{N-1}$ kinks \cite{Eto:2004rz,Eto:2006mz,Eto:2007aw}
%where ${\mathbb C}P^{N-1}$ kinksin the vortex wolrdsheet 
that correspond to magnetic monopoles confined by vortices in the bulk 
\cite{Tong:2003pz,Isozumi:2004vg,Nitta:2010nd}. 
By further considering the Yang-Mills-Higgs system on ${\mathbb R}^2 \times T^2$ with the 't Hooft flux\footnote{In a related but slightly different setup, the Yang-Mills-Higgs system on ${\mathbb R}^2 \times T^2$ with 
${\mathbb Z}_N \times {\mathbb Z}_N$
twisted boundary conditions 
%(of the both C-types) 
admits intersecting vortices 
with instanton charges at their intersections 
\cite{Fujimori:2008ee}. }, we expect this (confined) magnetic monopole to dissolve into a vortex worldsheet wrapping around the torus $T^2$ of ${\mathbb R}^2 \times T^2$, serving as the Higgs-phase analogue of the monopole-vortex continuity (Refs.~\cite{Hayashi:2024yjc, Hayashi:2024psa, Guvendik:2024umd}) that connects two semiclassical approaches.
In this setup, the vortex worldsheet theory will be the ${\mathbb C} P^{N-1}$ model on $T^2$ with the 't Hooft twist.
Hence, the behavior of fractional instantons in the ${\mathbb C} P^{N-1}$ model on the twisted $T^2$ can be interpreted as describing fractional instantons being `confined' inside vortices in the bulk.

% Furthermore, consider Yang-Mills theory on $T^4 =  T^2_{AD}\times T^2_{BC} $ with ‘t Hooft twists applied to the $T^2_{AD}$ and $T^2_{BC}$ sub-tori. In a certain limit, the 2d effective theory is expected to be described by the ${\mathbb C} P^{N-1}$ model (possessing a non-standard metric) on the remaining $T^2$ with 't Hooft flux; see Section \ref{sec:4d2d} for details. 

% Secondly, in QCD(adj.) on $\mathbb{R}^2 \times S^{1}({\rm large}) \times S^{1}({\rm small})$ with the 't Hooft twist, the effective potential of the monopole molecules (bions) coincides, in a certain limit, with that of the fractional instanton molecules in the ${\mathbb Z}_N$-twisted ${\mathbb C} P^{N-1}$ model~\cite{Hayashi:2024psa}.
% These facts imply the deep connection between the moduli space of the center vortex on $\mathbb{R}^2 \times T^2$ and ${\mathbb C} P^{N-1}$ space, and motivate us to study the ${\mathbb C} P^{N-1}$ model on $T^2$ with the 't Hooft twist.

% This provides further motivation for our study to examine the fractional instantons in the ${\mathbb C} P^{N-1}$ model on the twisted $T^2$.

%In this paper, we investigate BPS fractional lumps in the ${\mathbb C} P^{N-1}$ model on $T^2$ with the 't Hooft twist. To this end, we formulate the ${\mathbb C} P^{N-1}$ model as the extended Abelian-$N$ Higgs model, that is a $U(1)$ gauge theory coupled with $N$ Higgs complex scalar fields, where ${\mathbb C} P^{N-1}$ lumps are described as semilocal vortices.

In this paper, we study BPS lump (instanton) solutions in the $\mathbb{C}P^{N-1}$ nonlinear sigma model defined on a two-dimensional torus $T^2\simeq \frac{\mathbb{C}}{L_1\mathbb{Z}+i L_2 \mathbb{Z}}$ with the general clock-shift twisted boundary condition.  
By reformulating the model as an $N$-component Abelian-Higgs theory, the $U(1)$ topological charge becomes fractional as $k+p/N$, where the twist solely determines $p$. 
We then construct general vortex configurations that satisfy twisted boundary conditions. These solutions exhibit fractional topological charges and are characterized by holomorphic data modulo a complexified gauge equivalence (the so-called $V$-transformation in the moduli matrix formalism \cite{Eto:2006pg}). 
We show that the moduli space $\mathcal{M}$ is given by the $\mathbb{C}P^{Nk+p-1}$-fiber bundle over the $2$-torus, 
\begin{equation}
    \mathbb{C}P^{Nk+p-1}\hookrightarrow \mathcal{M} \to \frac{\mathbb{C}}{\frac{N}{Nk+p}\mathbb{Z}+\frac{N}{Nk+p}\tau \mathbb{Z}}, 
\end{equation} 
where $\tau=i\frac{L_2}{L_1}$. 
We note that our construction of the moduli space is for the one of the Abelian-Higgs description, which contains the small-lump singularities in the $\mathbb{C}P^{N-1}$ limit. 
When we parametrize the moduli space by zeros of theta functions, we can systematically identify the location of the small-lump singularity.
We also present an alternative representation of the moduli space, where the invariance under the modular transformation is obvious. 
These two ways of the parametrization can be related by using the product-to-sum formula of the theta functions.

Next, returning to one of our motivations, we address fractional instantons in the $\mathbb{C}P^{N-1}$ model (with a non-Fubini-Study metric) obtained through the dimensional reduction of 4d Yang-Mills theory \cite{Yamazaki:2017ulc}. 
When the aspect ratio of the 4-torus is tuned, a minimal fractional instanton solution for a given twist $p$ (with $Q_{\mathrm{top}} = -p/N$) can be constructed as a 't Hooft-type solution \cite{tHooft:1981nnx, vanBaal:1984ar}. For such a solution, we show that its moduli space coincides with that obtained from the analysis of the standard $\mathbb{C}P^{N-1}$ model.
This coincidence suggests the robustness of our qualitative results, such as the topology of the moduli space.
These results deepen our understanding of the geometric and topological aspects of fractional instantons and vortices.

The organization of the paper is as follows. 
%In Sec.~\ref{sec:CP}, we review the ${\mathbb C}P^{N-1}$ sigma models on ${\mathbb R}^2$. 
%In Sec.~\ref{sec:model}, we introduce the extended Abelian-Higgs model on a torus with twisted boundary conditions and derive the corresponding BPS equations. 
In Sec.~\ref{sec:model}, as a regularized version of the ${\mathbb C}P^{N-1}$ sigma model, we introduce the extended Abelian-Higgs model on a torus with twisted boundary conditions and derive the corresponding BPS equations.
In Sec.~\ref{sec:VS}, we construct explicit BPS solutions using theta functions and analyze the structure of their moduli spaces for various values of the twist parameters. Sec.~\ref{Sec:modulispace} presents an alternative, dual parametrization of the moduli space using Fourier-like theta bases, where the quotient structure and duality symmetries become manifest. 
In Sec.~\ref{sec:4d2d}, we discuss fractional instantons in the $\mathbb{C}P^{N-1}$ model that is obtained by the dimensional reduction from 4d Yang-Mills theory.
In Sec.~\ref{sec:SD}, we summarize our results and discuss their implications for nonperturbative dynamics in compactified gauge theories. In the appendices~\ref{sec:existence}, \ref{sec:proofs_app}, \ref{sec:ThetaSpace}, and \ref{sec:DualityFormulas}, we give detailed proofs of some arguments presented in the main text.

%%%%%%%%%%%%%%%%%%%%%%%%%%%%%%%%%%%%%%%%%%%%%%%%%%%%%%%%%%%%%%%%%%%%%%%%%%%%

%%%%%%%%%%%%%%%%%%%%%%%%%%%%%%%%%%%%%%%%%%%%%%%%%%%%%%%%%%%%%%%%%%%%%%%%%%%%
\section{2d Abelian-Higgs model with 't Hooft twist } \label{sec:model}

Our main subject is the fractional instanton in the $\mathbb CP^{N-1}$ model on $T^2$ with twisted boundary conditions.
For a practical purpose, we instead consider the Abelian-Higgs model (\ref{eq:action_AbelianHiggs}) in Euclidean 2d (or Lorentzian $(1+2)$d) spacetime.
In the strong gauge coupling limit, this Abelian-Higgs model is reduced to the $\mathbb CP^{N-1}$ model.

A brief review on ${\mathbb C}P^{N-1}$ model is provided in Section~\ref{sec:CP}.
The setup is introduced in Section~\ref{sec:model_and_BC}, and then general aspects of the BPS bound are described in Section~\ref{sec:BPS_bound}.
The modular transformation interchanging the two sides of the torus is introduced in Section~\ref{sec:duality}.

\subsection{Brief review on the classical aspects of 2d \texorpdfstring{${\mathbb C}P^{N-1}$}{CP{N-1}} sigma model}
\label{sec:CP}

Here, let us give a quick overview of the classical aspect of the 2d ${\mathbb C}P^{N-1}$ nonlinear sigma model. 
The target space is given by the complex projective space, 
\begin{equation}
    \mathbb{C}P^{N-1}\simeq SU(N)/(S[U(N-1)\times U(1)])\simeq S^{2N-1}/U(1). 
    \label{eq:complex_projective}
\end{equation}
It is convenient to regard $S^{2N-1}\subset \mathbb{C}^{N}$ as the unit sphere of the complex vector space, and we introduce the $N$-component complex scalar field,
\begin{eqnarray}
    \vec{n}:M_2\to S^{2N-1}\subset \mathbb{C}^{N}, \quad x\mapsto \vec{n}(x), 
\end{eqnarray}
where $\vec{n}(x)^\dagger \cdot \vec{n}(x)=1$ and $M_2$ is the $2$d spacetime manifold. 
To take the $U(1)$ quotient in Eq.~\eqref{eq:complex_projective}, we gauge the $U(1)$ transformation, $\vec{n}(x)\mapsto e^{i\alpha}\vec{n}(x)$, so that 
the classical action of the ${\mathbb C}P^{N-1}$ model is given by 
\begin{align}
    \mathcal{S}&=\frac{4\pi}{g^{2}}\int d^{2}x \, ({\cal D}_{\mu}\vec{n})^{\dagger} ({\cal D}_{\mu} \vec{n})\,,
    \label{eq:action_CPN-1}
\end{align}
where $g$ is the coupling constant and ${\cal D}_{\mu}= \partial_{\mu}-iA_{\mu}$ is the covariant derivative.  
As this action is quadratic in $A_\mu$, its path integral is equivalent to solving the equation of motion and we find $A_{\mu}(x) = -i\vec{n}^{\dagger}\cdot {\partial}_{\mu}\vec{n}$; the $U(1)$ gauge field is completely determined by the scalar field $\vec{n}(x)$. 
Field configurations are classified by the topological charge associated with  
$\pi_2({\mathbb C}P^{N-1}) \simeq {\mathbb Z}$, 
\begin{align}
    Q&=\frac{1}{2\pi}\int d^{2}x  i\epsilon_{\mu\nu} ({\cal D}_{\nu}\vec{n})^{\dag} ({\cal D}_{\mu} \vec{n})
    =\frac{1}{2\pi}\int d^{2}x \epsilon_{\mu\nu}\partial_{\mu} A_{\nu}\,.
\label{Qdef}
\end{align}
Each topological sector has the BPS bound, 
\begin{align}
    \mathcal{S}&=\frac{4\pi}{g^2}\int d^2 x \left\{\frac{1}{2}|{\cal D}_\mu \vec{n}\pm i \varepsilon_{\mu\nu} {\cal D}_\nu \vec{n}|^2 \mp i \varepsilon_{\mu\nu}{\cal D}_\mu n^\dagger {\cal D}_\nu n\right\} \ge \frac{8\pi^2}{g^2}|Q|, 
    \label{eq:BPS_CPN-1}
\end{align}
where the equality is saturated if and only if ${\cal D}_1 \vec{n}+ i {\cal D}_2 \vec{n}=0$ for $Q\ge 0$ and ${\cal D}_1 \vec{n}- i {\cal D}_2 \vec{n}=0$ for $Q< 0$. 

The global flavor symmetry of the model is $PSU(N)=SU(N)/{\mathbb Z}_{N}$, where the ${\mathbb Z}_{N}$ center is removed since it coincides with a subgroup of the $U(1)$ gauge symmetry. Because of this overlap of the naive $SU(N)$ global symmetry and the $U(1)$ gauge redundancy, the flavor twisted boundary condition causes the $\frac{1}{N}$ fractional shift of the $U(1)$ topological charge. We shall discuss more details on this point later for the Abelian-Higgs description. 

For $N=2$, we note that $\mathbb{C}P^{1}\simeq S^2$, and we can give the manifestly $U(1)$ gauge-invariant description using the three-component real scalar fields ${\bm m}(x)=(m_1(x),m_2(x),m_3(x))$ with a constraint ${\bm m}(x)^2=1$. 
The relation between the real scalar field ${\bf m}(x)$ and the complex two-component complex field $\vec{n}(x)$ is given by
\begin{eqnarray}
    {\bm m}= \vec{n}^{\dagger} {\bm \sigma} \vec{n}, 
% = \left(\frac{h_1^{*}h_2 + h_2^{*}h_1}{\vec{h}^\dagger \vec{h}} \,,\,
% -i\frac{h_1^{*}h_2 - h_2^{*}h_1}{\vec{h}^\dagger \vec{h}} \,,\,
% \frac{|h^1|^2 - |h^2|^2}{\vec{h}^\dagger \vec{h}} \right) \,,
\label{eq:three-vector}
\end{eqnarray}
with the Pauli matrices ${\bm \sigma}$. 
The configuration $\vec{n} = (1,0)^{T}$ corresponds to the north pole ${\bm m}=(0,0,1)$, while $\vec{n} = (0,1)^{T}$ corresponds to the south pole ${\bm m}=(0,0,-1)$.

% The ${\mathbb C}P^1$ model is equivalent to the $O(3)$ nonlinear sigma model,
% which is described by three real scalar fields ${\bm m}(x)=(m_1(x),m_2(x),m_3(x))$ 
% with a constraint ${\bm m}(x)^2=1$.
% Its action is
% \begin{align}
% S = \frac{\pi}{g^2}
% \int d^2x \, \partial_{\mu} {\bm m} \cdot  \partial_{\mu} {\bm m} \,,
% \end{align}
% where the relation between the real scalar field ${\bf m}(x)$ and the complex two-component complex field $\vec{h}(x)$ is
% \begin{eqnarray}
% {\bm m}= \vec{n}^{\dagger} {\bm \sigma} \vec{n} 
% = \left(\frac{h_1^{*}h_2 + h_2^{*}h_1}{\vec{h}^\dagger \vec{h}} \,,\,
% -i\frac{h_1^{*}h_2 - h_2^{*}h_1}{\vec{h}^\dagger \vec{h}} \,,\,
% \frac{|h^1|^2 - |h^2|^2}{\vec{h}^\dagger \vec{h}} \right) \,,
% \label{eq:three-vector}
% \end{eqnarray}
% with the Pauli matrices ${\bm \sigma}$. 
% The configuration $\vec{n} = (1,0)^{T}$ corresponds to the north pole ${\bm m}=(0,0,1)$, while $\vec{n} = (0,1)^{T}$ corresponds to the south pole ${\bm m}=(0,0,-1)$.

\subsection{Abelian-Higgs model and twisted boundary conditions}
\label{sec:model_and_BC}

2d $\mathbb{C}P^{N-1}$ sigma model is classically scale invariant, and we encounter small-instanton (or small-lump) singularities for the moduli space of the BPS solutions. 
In this paper, let us instead consider the $2$d extended Abelian-Higgs model at the critical coupling to circumvent those singularities in the moduli space, where the classical action is given by
\begin{align}
    \mathcal{S}=\int d^2x \left[\frac1{4e^2} F_{\mu\nu}^2+|{\cal D}_\mu \vec H|^2+\frac{e^2}2 (|\vec H|^2-\xi)^2\right].  \label{eq:action_AbelianHiggs}
\end{align}
Here, $\vec{H}:M_2\to \mathbb{C}^N$ is an $N$-component complex scalar realized as a row vector,  ${\cal D_\mu }=\p_\mu-i A_\mu$ is the $U(1)$ covariant derivative, and $F_{\mu\nu}=i [{\cal D_\mu, {\cal D}_\nu }]=\partial_\mu A_\nu-\partial_\nu A_\mu$ is the $U(1)$ field strength. 
As long as $e^2$ is finite, the $U(1)$ gauge field $A_\mu$ is an independent field from the scalar $\vec{H}$. 
The global flavor symmetry of this model is again given by $PSU(N)$. 

In this action, we set the coefficient of the wine-bottle potential equal to the $U(1)$ gauge coupling $e^2$, which relation is protected for the case of $2$d $\mathcal{N}=(2,2)$ supersymmetry. 
In that context, $\xi \in \mathbb R_{>0}$ denotes the Fayet-Iliopoulos (FI) parameter. 
Let us note, however, that we do not use any properties of $\mathcal{N}=(2,2)$ supersymmetry in this paper; we perform the classical analysis of this action~\eqref{eq:action_AbelianHiggs} to regularize small-instanton singularities of the $\mathbb{C}P^{N-1}$ model~\eqref{eq:action_CPN-1}. 
As we formally take $e^2\to \infty$, the radius of the Higgs field is fixed as $|\vec{H}|^2=\xi$ at the classical level, and we obtain the $2$d $\mathbb{C}P^{N-1}$ model by setting $\vec{H}=\sqrt{\xi}\vec{n}$ with $|\vec{n}|=1$: 
\begin{align}
    \mathcal{S}\xrightarrow{e^2 \rightarrow \infty} \xi \int d^2 x |{\cal D}_\mu \vec n|^2 = \frac{4\pi}{g^2} \int d^2 x |{\cal D}_\mu \vec n|^2,
\end{align}
where we identify $\xi=\frac{4\pi}{g^2}$. 
More generally, one can `regularize' arbitrary K\"ahler flag manifold sigma models to quiver gauge theories in the same manner (see Ref.~\cite{Fujimori:2023wkd}).

% In the large coupling limit $e^2 \to \infty$, this model reduces to the $\mathbb CP^{N-1}$ nonlinear sigma model as our subject of interest.
% Indeed, in the limit $e^2 \to \infty$, the radius of the Higgs field is effectively fixed to $|\vec H|^2 = \xi$.
% Thus, we can write $\vec H = \sqrt{\xi} \vec{n}$, where $\vec n \in \mathbb C^N$ is a $N$-component vector satisfying $\vec{n}^\dagger \vec{n} = 1$.
% On the other hand, since the kinetic term for the gauge field vanishes, the $U(1)$ gauge field plays a role of the auxiliary field.
% We then get
% \begin{align}
% -{\cal L} \xrightarrow{e^2 \rightarrow \infty} \xi |{\cal D}_\mu \vec n|^2 = \frac{4\pi}{g^2} |{\cal D}_\mu \vec n|^2,
% \end{align}
% which is the standard action of the $\mathbb CP^{N-1}$ model.
% Here, the $\mathbb CP^{N-1}$ coupling $g^2$ is defined as $\xi = 4\pi/g^2$. 
% %\red{such that $2\pi \xi=8\pi^2/g^2$}.
% The reason why we consider the Abelian-Higgs model instead of the $\mathbb CP^{N-1}$ is to smooth out the small-lump singularity in the moduli space, which will be explained later.  

Let us consider a torus $T^2 := \mathbb R^2/(L_1\mathbb{Z}\times L_2\mathbb{Z})$ as the $2$d spacetime, and we discuss the relation between the $PSU(N)$ twisted boundary condition and the topological sectors. 
% with `t Hooft twist as the spatial slice of the model, \red{where the global symmetry is often broken to }. 
The torus coordinate is written as $(x_1,x_2)\in \mathbb{R}^2$ with the equivalence relation,  
\begin{align}
    (x_1,x_2) \sim   \left(x_1+n_1 L_1,\, \, x_2+ n_2 L_2\right) ,  
\end{align}
for $n_1,n_2\in \mathbb{Z}$. We take the fundamental domain of the torus as $\mathcal{F}(T^2)=[0,L_1)\times [0,L_2)$, 
and we impose the following flavor-twisted boundary condition: 
\begin{align}
    &\left\{\begin{array}{l}
        A_\mu(x_1+L_1,x_2)=A_\mu(x_1,x_2)+ \p_\mu \lambda_s(x_1,x_2), \\
        \vec H(x_1+L_1,x_2)=e^{i \lambda_s(x_1,x_2)}\vec H(x_1,x_2) S^{p_s},
    \end{array}\right.\label{eq:HbcS}\\
    &\left\{\begin{array}{l}
        A_\mu(x_1,x_2+L_2)=A_\mu(x_1,x_2)+ \p_\mu \lambda_c(x_1,x_2),\\
        \vec H(x_1,x_2+L_2)=e^{i \lambda_c(x_1,x_2)}\vec H(x_1,x_2) C^{p_c},
    \end{array}
    \right.\label{eq:HbcC}
\end{align}
where $0\le p_s,p_c<N$, $e^{i\lambda_{s,c}}$ are $U(1)$ transition functions, 
% \footnote{
% The $U(1)$-gauge transition functions $\lambda_s, \lambda_c$ are assumed to be continuous functions without any brunch cut.  
% Therefore, their differences are single-valued whereas the transition functions are originally phases of $U(1)$-valued functions, hence defined modulo $2\pi \mathbb Z$. }, 
and $S, C$ are $N\times N$ shift and clock matrices defined by 
\begin{align}
(S)^a{}_{b}\coloneqq \delta^a_{b-1}+\delta^a_{N}\delta^1_{b},\quad (C)^a{}_{b}\coloneqq \omega^{a-1} \delta^a_{b},\quad {\rm with~} \omega=e^{\frac{2\pi i}{N}}. \label{eq:shift_and_clock_matrices}
\end{align}
They satisfy $S^{\rm T}=S^{-1}, C^{\rm T}=C$ and  $S^N=C^N={\mathbf 1}_N$.  For example, with $N=3$, the matrices $S,C$ are given as
 \begin{align}
S=\begin{pmatrix}
0&1&0\\0&0&1\\1&0&0
\end{pmatrix}, \quad  
C=\begin{pmatrix}
1&0&0\\0&\omega&0\\0&0&\omega^2
\end{pmatrix}.
\end{align}
This boundary condition explicitly breaks the $PSU(N)$ symmetry to a smaller subgroup unless $p_s=p_c=0$. Especially when $\gcd(N,p_s)=\gcd(N,p_c)=1$, it is completely broken. 

Since the matrices $S,C$ are noncommutative as $SC=\omega \,CS$, the above flavor twists cause also a nontrivial twist on the $U(1)$ gauge sector as follows.
Considering the relationship between  $\vec H(x_1+L_1,x_2+L_2)$ and $\vec H(x_1,x_2)$ in two different ways using the two conditions \eqref{eq:HbcS} and \eqref{eq:HbcC},
it can be seen that the following constraint  must hold for the pair $(\lambda_s,\lambda_c)$:
\begin{align}
\exists k \in \mathbb Z:\quad
[\lambda_s(x_1,x_2+L_2)-\lambda_s(x_1,x_2)]-[\lambda_c(x_1+L_1,x_2)-\lambda_c(x_1,x_2)]= 2\pi \left(k +\frac{ p}N\right), \label{eq:lambdaCC}
\end{align}
with an integer $ p \in \{0,1,\ldots, N-1\} $ defined by
\begin{align}
    p \equiv p_s p_c \quad {\rm mod~} N.
\end{align}
This constraint forces the magnetic flux to be fractional as
\begin{align}
\frac1{2\pi} \int_{{\cal F}(T^2)} d^2x F_{12}=\frac1{2\pi} \Big\{
[\lambda_s(0,L_2)-\lambda_s(0,0)]-[\lambda_c(L_1,0)-\lambda_c(0,0)] \Big \}= k +\frac{ p}N, \label{eq:fractionalized_topcharge}
\end{align}
where ${\cal F}(T^2)$ is the fundamental domain of the torus $T^2$.
The integer $k$ characterizes a topological sector of the model we consider. 
This is nothing but the $\mathbb{Z}_N$ 't Hooft-twisted boundary condition for the $U(1)$ gauge field. 
In this sense, we call this flavor-twisted boundary condition (\ref{eq:HbcS}) as the 't~Hooft twist for the Abelian-Higgs model in this paper. 

\subsection{BPS configurations and the moduli space}
\label{sec:BPS_bound}

The $2$d classical action~\eqref{eq:action_AbelianHiggs} can be rewritten as 
\begin{align}
    \mathcal{S}&=\int d^2 x\left\{|{\cal D}_1 \vec{H}\pm \im {\cal D}_2\vec{H}|^2 + \frac{e^2}{2}\left|\frac{1}{e^2}F_{12}\pm (|\vec{H}|^2-\xi)\right|^2 \pm \xi F_{12}\right\}, 
\end{align}
where the $\pm$ sign should be understood coherently, and the BPS bound is given by
\begin{align}
    \mathcal{S}  \ge \xi \left|\int_{{\cal F}(T^2)} d^2x F_{12}\right|
    =
    \left\{\begin{array}{ccc}
    \displaystyle   2\pi \xi \left(k+\frac{ p}N\right)& {\rm for~} k\ge0, &\vspace{2mm} \\ 
    \displaystyle    2\pi \xi \left(k_{\rm a}+\frac{ p_{\rm a}}N\right) & {\rm for~} k<0 &(k_{\rm a}\ge 0),
    \end{array} \right.
\end{align}
with $k_{\rm a}\coloneqq-1-k$ and $p_{\rm a}\coloneqq N-p$. 
In this paper, we concentrate on the topological sectors with the nonnegative integer $k\in \mathbb Z_{\ge 0}$ without loss of generality. 
Then, the BPS bound is saturated if and only if the BPS equation is satisfied, 
\begin{align}
    &({\cal D}_1+i {\cal D}_2)\vec H =0 ,\label{eq:BPSH}\\
    &\frac1{e^2}F_{12}+|\vec H|^2-\xi=0. \label{eq:BPSF}
\end{align}

Note that in generic cases ($p\not = N/2$), the minimal BPS bound can be satisfied in a topological sector with either $k=0$ or $k=-1$.
For $0\leq 2p<N$, the BPS states with $k=0$ saturate the minimal bound, whereas the anti-BPS states with $k=-1$ saturate the minimal bound for $N<2p<2N$.
When $2p=N$, both BPS states ($k=0$) and anti-BPS states ($k=-1$) saturate the minimal bound and become degenerate.

By integrating the both sides of Eq.~\eqref{eq:BPSF}, we find the so-called Bradlow bound~\cite{Bradlow:1990ir},
\begin{align}
L_1 L_2  \xi \ge   \frac{1}{e^2}\int_{{\cal F}(T^2)} d^2x  F_{12}=\frac{2\pi }{e^2} \left(k +\frac{ p}N\right), 
\label{eq:BradlowBound}
\end{align}
where the equality holds if and only if $\vec{H}=0$ everywhere. 
In the following, we always assume the Bradlow bound is satisfied as a strict inequality so that $\vec{H}\not=0$ somewhere. We note that this is always true in the limit $e^2 L_1L_2\to \infty$, which includes the case of the $\mathbb{C}P^{N-1}$ model. 
% the FI parameter $\xi$ satisfies this condition for a given $k$ and $p$ of our interest\footnote{If one is interested only in the $\mathbb CP^{N-1}$ model, which is $e^2 \rightarrow \infty$ limit, this bound is trivial in this limit. The  Bradlow bound becomes nontrivial when the $\mathbb CP^{N-1}$ model is `regularized' to the Abelian-Higgs model.}.

With introducing a complex coordinate (and the modulus of the torus)
\begin{align}
    z:= \frac{x_1+i x_2}{L_1}, \quad \tau :=\frac{iL_2}{L_1},
\end{align}
Eq.~\,\eqref{eq:BPSH} can be solved by the following ansatz, 
\begin{align}
A_{\bar z}(x_1,x_2)&=  i\,\p_{\bar z }\psi(x_1,x_2), \label{eq:psi_h_repr_A} \\
\vec H(x_1,x_2)&=\sqrt{\xi}e^{-\psi (x_1,x_2)} \vec h(z). \label{eq:psi_h_repr_H}
\end{align}
Here, we assume that $\psi:\mathbb{R}^2 \to \mathbb C$ is a non-singular function and that  $\vec h(z)$ is an $N$-component row vector of entire functions with respect to $z$. 
This representation has the following redundancy of multiplying an entire function $v(z)$ as 
\begin{align}
(e^\psi, \vec h(z))\, \quad \mapsto \quad  \, e^{v(z)}(e^\psi, \vec h(z)), \label{eq:Vtrf}
\end{align} 
and we call this the $V$-transformation.
The pseudo-periodic boundary conditions  in terms of $\psi$ and $\vec h(z)$ are defined up to the above transformation as 
\begin{align}
    &\left\{\begin{array}{l}
        \psi(x_1+L_1,x_2)=\psi(x_1,x_2)+v_s(z)-i\lambda_s(x_1,x_2),\\
        \vec h(z+1)= e^{v_s(z)} \vec h(z) S^{p_s},
    \end{array}
    \right.\label{eq:hzbcS}\\
    &\left\{\begin{array}{l}
        \psi(x_1,x_2+L_2)=\psi(x_1,x_2)+v_c(z)-i\lambda_c(x_1,x_2),\\
        \vec h(z+\tau)= e^{v_c(z)} \vec h(z) C^{p_c}, 
    \end{array}\right.\label{eq:hzbcC}
\end{align}
with introducing a pair of entire transition functions $v_s(z),v_c(z)$. 
From the single-valuedness of $e^{-\psi}$ and \eqref{eq:lambdaCC}, the transition function should satisfy a constraint
\begin{align}
    [v_s(z+\tau)-v_s(z)]-[v_c(z+1)-v_c(z)]= 2\pi i\, \left(k +\frac{p}N\right),
    \label{eq:vCC}
\end{align}
with the same $k$ appeared  in  \eqref{eq:lambdaCC}.\footnote{
If $k$ is replaced with a different integer $k' (\not =k)$,  we can show that $e^{-\psi}$ must have zeros or poles, which is prohibited by the assumption. }
Under the transformation \eqref{eq:Vtrf} the pair $(v_s,v_c)$ also transforms as
\begin{align}
    (v_s(z),v_c(z))\quad \mapsto\quad  \left( v_s(z)+v(z+1)-v(z), \,v_c(z)+v(z+\tau)-v(z) \right),\label{eq:vsvcVtrf}
\end{align}
under which the above constraint \eqref{eq:vCC} is invariant.

Substituting the ansatz~\eqref{eq:psi_h_repr_A} and \eqref{eq:psi_h_repr_H} into Eq.~\,\eqref{eq:BPSF}, we obtain
\begin{align}
(\p_1^2+\p_2^2) {\rm Re}\, \psi =e^2 \xi (1-e^{-2{\rm Re}\,\psi} |\vec h(z)|^2)\,, \label{eq:master}
\end{align}
which is called the Taubes equation 
for $N=1$ \cite{Taubes:1979tm}.
This equation with the boundary conditions described above  is assumed to be uniquely solved with respect to ${\rm Re}\,\psi(x_1,x_2)$ with a given $\vec h(z)$
if and only if the Bradlow bound \eqref{eq:BradlowBound} is strictly satisfied. See Appendix \ref{sec:existence} for the detail. 
We can always remove the imaginary part of $\psi$ using the $U(1)$ gauge redundancy, and thus the physical information of $\psi$ is uniquely determined with given $\vec h(z)$. 
Thus, finding the self-dual solutions turns out to be equivalent to find an $N$-component vector of entire functions $\vec{h}(z)$ that satisfies the boundary condition~\eqref{eq:hzbcS} and \eqref{eq:hzbcC}: Let ${\cal M}^{N;(p_s,p_c)}_{k}$ denote the moduli space of the BPS solutions with the $(p_s,p_c)$ twisted boundary condition and the topological charge $k+\frac{p}{N}$, and then it is given by
\begin{align}
    &\quad {\cal M}^{N;(p_s,p_c)}_{k} \notag \\
    &\simeq 
    \left\{
    (\vec{h}(z),v_s(z),v_c(z)) \left| 
    \begin{array}{c}
        \exists z\in \mathbb{C} \,\, \text{such that} \,\, |\vec h(z)|\not= 0, \\
        \vec{h}(z+1)= e^{v_s(z)} \vec h(z) S^{p_s}, \quad 
        \vec h(z+\tau)= e^{v_c(z)} \vec h(z) C^{p_c},  \\
        (v_s(z+\tau)-v_s(z))-(v_c(z+1)-v_c(z))= 2\pi i \left(k +\frac{p}N\right)
    \end{array}
    \right. \right\}/\stackrel{V} \sim,
    \label{eq:general_formula_moduli}
\end{align}
where we introduce the $V$-equivalence relation, 
\begin{align}
    (\vec h(z), v_s(z), v_c(z)) \, \stackrel{V} \sim \,  
    (e^{v(z)} \vec h(z), v_s(z)+v(z+1)-v(z), v_c(z)+v(z+\tau)-v(z) ),\label{eq:Vequiv}
\end{align}
to remove the redundancy by the $V$-transformation~\eqref{eq:Vtrf} and \eqref{eq:vsvcVtrf}. 
We note that $\mathcal{M}^{N;(p_s,p_c)}_k$ becomes a complex manifold, whose dimension is determined by the index theorem, 
% Therefore moduli parameters $\{\phi^\alpha \in\mathbb C\}$ in $\vec h(z)=\vec h(z,\phi^\alpha)$ 
% characterizing V-equivalent class  are complex coordinates of the moduli space of the solutions, ${\cal M}^{N;(p_s,p_c)}_{k}$
% which is a complex manifold.
% The number of zero modes around the BPS states can be predicted by the index theorem. 
% See Appendix.A in \cite{Eto:2009bg} for the model with $\mathbb R^2$ spatial slice. 
% Applying the result in $\mathbb R^2$ to our model in $T^2$, we obtain 
\begin{align}
M\,:=\,{\rm dim}_{\mathbb C} {\cal M}^{N;(p_s,p_c)}_{k}= \frac{N}{2\pi} \int_{{\cal F}(T^2)}d^2x  F_{12}=Nk+p. \label{eq:index_theorem_defM}
\end{align}
We will see in the following sections that this is indeed true.

In the large coupling limit $e^2\to \infty$,  the solution of \eqref{eq:master} is explicitly given by
\begin{align}
{\rm Re}\,\psi =\frac12 \log |\vec h(z)|^2 + O(e^{-2}\partial^2), 
\end{align}
where the leading order term makes sense only if 
\begin{align}
\forall z\in \mathbb{C}  \quad  |\vec h(z)|^2\not=0.\label{eq:nonsingular}
\end{align}
Again, by removing the imaginary part of $\psi$ via $U(1)$ gauge transformations, we obtain $\vec{H}=\sqrt{\xi}\vec{h}(z)/|\vec{h}(z)|=:\sqrt{\xi}\vec{n}(z,\bar{z})$, and we can check that the $U(1)$ gauge field becomes $A_\mu=-i \vec{n}^\dagger \partial_\mu \vec{n}$ as we have seen in Sec.~\ref{sec:CP}. 

In this paper, we basically consider the case of $2$d Euclidean spacetime, but let us comment on the case of the $(2+1)$d Lorentzian case, $\mathbb{R}_{\mathrm{time}} \times T^2$. 
In this case, we can regard the BPS solutions as the finite-energy BPS states, 
and its low-energy dynamics is described by 
the moduli (or Manton) approximation
\cite{Manton:1981mp,Eto:2006uw},
which promotes the moduli parameters $\{\phi^\alpha \}$ to $t$-dependent functions $\{\phi^\alpha(t)\}$. 
Then, the problem is reduced to the quantum mechanics of a particle living on the moduli space ${\cal M}^{N;(p_s,p_c)}_{k}$.
Especially in the limit of $e^2\to \infty$,  the effective Lagrangian is explicitly given by
\begin{align}
    {\cal L}_{\rm eff}= \frac{\p^2 K_{\rm vtx}}{\p \phi^\alpha \p \bar \phi^\beta } \dot \phi^\alpha \dot{\bar \phi}^\beta,\quad
    K_{\rm vtx} =\xi \int_{{\cal F}(T^2)}d^2x \log |\vec h(z)|^2, \label{eq:kahlermetric}
\end{align}
which define a K\"ahler metric for ${\cal M}^{N;(p_s,p_c)}_{k}$.
It is obvious that if the coordinates are chosen appropriately, the above K\"ahler potential gives a positive definite metric. 
Furthermore, this metric is infinitely differentiable at every point that satisfies the condition \eqref{eq:nonsingular}.
 Conversely, this metric is considered to have the so-called small-lump singularity on submanifolds where the condition \eqref{eq:nonsingular} is broken and thus descriptions in the $\mathbb CP^{N-1}$ nonlinear sigma model are ill-defined.

\subsection{Modular transformation on \texorpdfstring{$T^2$}{T2}}\label{sec:duality}
Here let us discuss the duality of interchanging the $x_1$- and $x_2$-axes with keeping its complex structure. 
To do so we must also interchange the matrices $S$ and $C$ except in the case of $(p_s,p_c)=(0,0)$.
Let us consider the following  transformation of $\vec h(z)$
\begin{align}
\vec h^{\rm D}(z):=\vec h(z) U_{\rm dF}\,,
\end{align}
with a unitary matrix $U_{\rm dF}$ given by
\begin{align}
(U_{\rm dF})^a{}_b :=\frac1{\sqrt{N}} e^{-\frac{2\pi i}N (a-1)(b-1)},
\end{align}
which causes a discrete Fourier transform (DFT) and satisfies
\begin{align}
   U_{\rm dF}U_{\rm dF}^\dagger ={\bf 1}_N,\quad U_{\rm dF}^{\rm T}=U_{\rm dF},\quad  (U_{\rm dF})^4={\bf 1}_N. 
\end{align}
Since this matrix $U_{\rm dF}$ transforms $S,C$ as follows 
\begin{align}
SU_{\rm dF}=U_{\rm dF}C^{-1},\quad CU_{\rm dF}=U_{\rm dF}S \,,
\end{align}
the pseudo-periodic boundary conditions are rewritten as
\begin{align}
\vec h^{\rm D}(z+\tau)&=e^{v_c(z)}\vec h^{\rm D}(z) S^{p_c},\\
\vec h^{\rm D}(z-1) &=e^{-v_s(z-1)}\vec h^{\rm D}(z) C^{p_s}. 
\end{align}
%\begin{align}
%[-v_c((z-\tau)+1)+v_c(z-\tau)]-[v_s(z-\tau)-v_s(z)]= 2\pi i\, \left(k +\frac{p}N\right),\label{eq:vCC2}
%\end{align}
Therefore, we find that there must be an isomorphic mapping  on vortex solutions in our models  under  the following replacement,
\begin{align}
\mathfrak{D}:& \left\{(z,\tau),  (p_s,p_c), \,\big(v_s(z),v_c(z)\big)\right\}\nonumber \\ \quad \mapsto \quad  
&\left\{(z/\tau, -1/\tau),  (p_c,p_s), \big(v_c(z),-v_s(z-1)\big)\right\}.
\end{align}
%Here we consider a conformal transformation $z\to z/\tau$ and thus 
Note that this modular transformation $\mathfrak{D}$ is an element of $\mathbb Z_4$.
This mapping indicates that the following two moduli spaces are isomorphic:
\begin{align}
    {\cal M}^{N:(p_s,p_c)}_k  \simeq  {\cal M}^{N:(p_c,p_s)}_k.
\end{align}

%\begin{align}
%\vec h^{\rm D}(z+\tau)&=e^{v_c(z)}\vec h^{\rm D}(z) S^{-p_c}\\
%\vec h^{\rm D}(z-L_1) &=e^{-v_s(z-L_1)}\vec h^{\rm D}(z) C^{-p_s} 
%\end{align}

%begin{align}
%\left\{(L_1,\tau),  (p_s,p_c), \big(v_s(z),v_c(z)\big)\right\}  \quad \leftrightarrow \quad  
%left\{(\tau, -L_1),  (-p_c,-p_s), \big(v_c(z),-v_s(z-L_1)\big)\right\}
%\end{align}

\section{Explicit construction of fractional BPS lumps}
\label{sec:VS}

In this section, we will determine the explicit form of the BPS lumps that solve (\ref{eq:BPSH}) and (\ref{eq:BPSF}). 
In Sec.~\ref{sec:BPS_bound}, we find the ansatz (\ref{eq:psi_h_repr_A}) and (\ref{eq:psi_h_repr_H}) that automatically solves one of the BPS equations (\ref{eq:BPSH}), and the other BPS equation (\ref{eq:BPSF}) determines $\psi(x_1,x_2)$ once the $N$-component entire functions $\{ h_a(z) \}$ are given (if the Bradlow bound is satisfied). 
Therefore, the remaining task is to find $\vec{h}(z)=\{ h_a(z) \}$ that satisfies the given boundary conditions with choosing a representative of the $V$-equivalence class.
% Hence, we will explicitly construct general solutions to the boundary conditions \eqref{eq:hzbcS} and \eqref{eq:hzbcC} 
% by choosing the representative of the V-equivalence class \eqref{eq:Vequiv}.

It is convenient to use the Jacobi theta function for the construction of the solution. As there are several standard notational systems for the theta function, let us write down the explicit formula for our choice of the definition: 
\begin{align}
\theta_1(z|\tau)
% & \coloneqq \theta_1(z,q)   \quad   {\rm with~}q\coloneqq  e^{\pi i \tau}\\
&\coloneqq 2 q^{\frac14} \sin (\pi z) \prod_{n=1}^\infty \left\{(1-q^{2n})(1-2q^{2n}\cos (2\pi z) +q^{4n})\right\} \notag\\
&= \sum_{n\in\mathbb Z}q^{\left(n+\frac12\right)^2}e^{2\pi i \left(n+\frac12 \right)\left(z-\frac12\right)}, \label{eq:thetadef}
\end{align}
where $q=e^{\pi i\tau}$. It is the most general entire function (up to constant multiplications) satisfying the pseudo-periodic boundary condition,
\begin{align}
\theta_1(z+1| \tau)=-\theta_1(z|\tau),\quad \theta_1(z+\tau|\tau)=-q^{-1}e^{-2\pi i z }\theta_1(z|\tau). 
\end{align}
The set of zeros are given by $\{z\in \mathbb{C}\,|\, \theta_1(z|\tau)=0\}=\mathbb{Z}+\tau \mathbb{Z}$. 
% Especially, we use the following properties of $\theta_1(z|\tau)$: a set of zeros 
% \begin{align}
% \left\{z\in \mathbb C \big|\theta_1(z|\tau)=0\right\}=\left\{ n+m \tau\big| n,m\in \mathbb Z\right\}
% \end{align}

%%%%%%%%%%%%%%%%%%%%%%%%%%
\subsection{Description of (fractional) BPS lumps by their zeros}
\label{sec:BPSlumps_formula_by_zeros}
% \subsection{Set of zeros $\hat\sigma_a$ and Extended Torus $\hat T^2$}
\def\setp{{\,\, .\!+\,\,}}

The shift-twisted boundary condition \eqref{eq:hzbcS} implies that $h_{a-p_s}, h_{a-2p_s},\ldots, h_{a-(\frac{N}{\gcd(N,p_s)}-1)p_s}$ are completely determined by $h_a(z)$, and thus the problem to find (fractional) BPS lumps is reduced to determine entire functions $h_{\bar{a}}(z)$ for $\bar{a}=1,\ldots, \gcd(N,p_s)$. 
Repeatedly applying \eqref{eq:hzbcS} by $\frac{N}{\gcd(N,p_s)}$ times, we obtain the boundary condition that is closed for the single flavor index,
\begin{equation}
    h_{a}\left(z+\frac{N}{\gcd(N,p_s)}\right) = \rme^{v_s(z)+\cdots + v_s(z+\frac{N}{\gcd(N,p_s)}-1)} h_{a}(z). 
\end{equation}
Thus, it is convenient to extend the torus $T^2=\mathbb{C}/(\mathbb{Z}+\tau \mathbb{Z})$ to the one, 
% Therefore it is convenient to introduce an extended torus $\hat T^2$ as
\begin{align}
    \hat T^2:= \mathbb C /\hat \Gamma_s\qquad {\rm with~}\quad\hat \Gamma_s := \frac{N}{\gcd(N,p_s)}\, \mathbb Z+ \tau \,\mathbb Z, 
\end{align}
and we write its fundamental domain as ${\cal F}(\hat T^2):= \{\frac{N}{\gcd(N,p_s)}\, s+\tau\, u \big|s,u\in \mathbb [0,1)\}$. 

Let us choose a flavor index $a$ such that $h_{a}(z)$ is not identically zero. As we have assumed the Bradlow bound is strictly satisfied, $|\vec{h}|$ cannot be identically zero, and thus such $\bar{a}$ exists. 
We write the set of zeros of $h_a(z)$ in the fundamental domain of $\hat T^2$ as
\begin{align}
    \hat \sigma_a &:= \set{ z \in {\cal F}(\hat T^2)\big| h_a(z)=0}  \notag \\
    &=\set{ z_{(a,n)} \big|  n=1,2,\cdots,\hat M}, 
    %\simeq \Sigma_a/\hat \Gamma_s^{\hat M},
\end{align}
where the zeros are counted with their multiplicity. 
The number of zeros $\hat{M}$ of $h_{\bar{a}}(z)$ inside $\mathcal{F}(\hat{T}^2)$ is determined by the topological charge (see Appendix \ref{sec:NumberZero}), 
\begin{equation}
    \hat{M}=\frac{N}{\gcd(N,p_s)}\left(k+\frac{p}{N}\right)=\frac{M}{\gcd(N,p_s)},  \label{eq:numberofzeros}
\end{equation}
which does not depend on the flavor index as long as $h_a(z)$ is not identically zero. 
The set of zeros of $h_a(z)$ on the entire complex plane is given by the lattice,
\begin{equation}
    \Sigma_a=\hat{\sigma}_a+\hat{\Gamma}_s. 
\end{equation}
One of entire functions whose zeros are exactly $\Sigma_a$ can be given by 
\begin{align}
    \prod_{n=1}^{\hat M}\theta_1\left(\frac{\gcd(N,p_s)}{N} (z-z_{(a,n)}) \Big|\frac{\gcd(N,p_s)}{N}\tau\right), 
\end{align}
which should give $h_a(z)$ up to some multiplications of exponential of entire functions. 

Like Abel's theorem on elliptic functions, 
we can show that there is the following constraint on the sum of  the zeros of $h_a(z)$ satisfying the boundary conditions \eqref{eq:hzbcS} and \eqref{eq:hzbcC}: For $\bar a=1,2,\cdots,\gcd(N,p_s)$, there exists $\langle z \rangle\in \mathbb C$ such that
\begin{align}
    \sum_{n=1}^{\hat M} z_{(\bar a, n)} \equiv \hat M \langle z \rangle +\frac{\bar a-1}{\gcd(N,p_s)} p_c \quad {\rm mod~} \hat \Gamma_s. 
    \label{eq:Abel}
\end{align}
Here, $\langle z \rangle$ is independent of the index $\bar a$ and depends only on the pair of holomorphic transition functions, $(v_s,v_c)$.
See Appendix~\ref{sec:Abel} for the proof.

Below, we shall see explicit solutions in the cases with $p_s=0$, $p_s =1$, and general $p_s$ separately. 

%%%%%%%%%%%%%%%%%%
\subsubsection{BPS lumps for \texorpdfstring{$p_s=0$}{ps=0}}\label{sec:ps0}
Let us consider the case with $p_s=0$, where we do not need to extend the torus $\hat T^2 = T^2$.  
In this case, according to \eqref{eq:numberofzeros} and \eqref{eq:Abel}, each $h_a(z)$ has $k (=\hat M)$ zeros, 
\begin{align}
 \hat \sigma_a= \{z_{(a,n)} \in {\cal F}(T^2)| n=1,2,\cdots,k\} ,
\end{align}
and their sum for each $a$ is given by
\begin{align}
    \sum_{n=1}^k z_{(a,n)}= k \langle z \rangle +\frac{a-1}Np_c+m_a+l_a\,\tau\quad  {\rm with~} \exists\,(m_a,l_a)  \in \mathbb Z^2,
    \label{eq:sumzero1}
\end{align}
where the zeros and $k \langle z \rangle$ are assumed to take values in ${\cal F}(T^2)$.
With  this set of zeros, we can write down the representative of the $V$-equivalence for each entry $h_a(z)$ as 
\begin{align}
    h_a(z)\quad \stackrel{V}{\sim}\quad h_a^{\rm rep}(z)=c_a \,e^{-2\pi i\,l_a (z-\langle z\rangle)}\prod_{n=1}^{k} \theta_1(z-z_{(a,n)}| \tau), 
    \label{eq:sol1}
\end{align}
where $\vec{c}\in \mathbb{C}^N\setminus\{\vec{0}\}$. 
Here, the factor $e^{-2\pi i\,l_a z}$ in $h_a^{\rm rep}(z)$ is introduced so that $(v_s,v_c)$ is independent of the flavor index $a$ as required.
Indeed, we can check that $\vec h^{\rm rep}(z)$ given above satisfies the boundary conditions \eqref{eq:hzbcS} and \eqref{eq:hzbcC} with the holomorphic transition functions, 
\begin{align}
(v_s(z),v_c(z))= 2\pi i k \times    \Big(\frac12 ,\, -(z- \langle z \rangle)-\frac{1+\tau}2 \Big).
\end{align}
We can also show that any solution to the boundary conditions \eqref{eq:hzbcS}, \eqref{eq:hzbcC} is $V$-equivalent to the above form, and thus Eq.~\eqref{eq:sol1} gives the most general solution\footnote{This can be attributed to the property that holomorphic functions on the extended torus are uniquely determined by their zeros; thus, if two such functions have identical zeros, they are necessarily the same up to constant.}. See Appendix \ref{sec:Uniquness} for the proof.
In the above expression of $\vec h(z)$, the redundancy by the $V$-equivalence is mostly fixed,
and the only remaining one is given by
\begin{align}
    (c_1,c_2,\cdots,c_N)\stackrel{V}{\sim} \lambda (c_1,c_2,\cdots,c_N)  \quad {\rm with~} \lambda \in \mathbb C^* (\coloneqq \mathbb C\backslash \{0\}),
\end{align}
that is, $\{c_a\}$ parametrize $\mathbb CP^{N-1}$.
In this case, therefore, the complex dimension of the moduli space can be calculated as follows,
\begin{align}  
    {\rm dim}_{\mathbb C} {\cal M}^{N;(0,p_c)}_{k}
    &=\sum_{a=1}^N( |\hat \sigma_a|-\# \{\langle z\rangle\})+{\rm dim}_{\mathbb C} \mathbb C P^{N-1} + \# \{ \langle z\rangle \} \nn \\
&=N(k-1)+(N-1)+1=Nk =M.
\end{align}

Next, let us consider the topology of ${\cal M}^{N;(0,p_c)}_k$. 
While the above expression contains discrete parameters $\{(m_a,l_a)|a=1,2,\cdots,k\}$, we can see that the moduli space ${\cal M}^{N;(0,p_c)}_k$ is a connected space.
To understand this point,  it is convenient to allow redundant expressions and extend the domain of each $z_{(a,n)}$ and $k\langle z \rangle$ from ${\cal F}(T^2)$ 
to $\mathbb C$. Then, for example  by replacing $z_{(a,1)}$ to $z_{(a,1)}+m_a+l_a\tau$ for all $a$, we can always set $m_a=l_a=0$ so that %in Eq.~\eqref{eq:sumzero1} as follows,
\begin{align}
%\exists\,\hat \sigma_a'\simeq \Sigma_a/\hat \Gamma_s^k,\quad z_{(a,n)}\in \hat \sigma_a':\quad     
    \sum_{n=1}^k z_{(a,n)}= k \langle z \rangle +\frac{a-1}Np_c. \label{eq:sumzero2}
\end{align}
Consequently, $h_a^{\rm rep}(z)$ is more simply expressed as follows while redefining  $c_a$, 
\begin{align}
h_a(z)\quad \stackrel{V}{\sim}\quad h_a^{\widehat{\rm rep}}(z)=c_a\prod_{n=1}^{k} \theta_1(z-z_{(a,n)}| \tau)\,. \label{eq:sol2}
\end{align}
In this representation,  it is clear that the moduli space ${\cal M}_k^{N:(0,p_c)}$ is connected  and has no boundaries.
As a trade-off for revealing these properties, this representation has redundancy.
We can confirm that there are the following equivalence relations,
\begin{align}
\left(\{z_{(a,n)}\},\{c_a\},k \langle z\rangle\right) \quad \sim \quad\left(\{z_{(a,n)}+\omega_{(a,n)}\},\{c_a\,e^{\Delta_a}\},k\langle z\rangle+\omega\right)\,,
\end{align}
where $\omega_{(a,n)}=m_{(a,n)}+l_{(a,n)}\,\tau $ with $m_{(a,n)},l_{(a,n)} \in\mathbb Z$ and $\omega  \in \hat \Gamma_s$ have constraints 
\begin{align}
 \sum_{n=1}^k \omega_{(1,n)}=\sum_{n=1}^k \omega_{(2,n)}=\cdots =\sum_{n=1}^k \omega_{(N,n)}=\omega,
\end{align}
and $\Delta_a\in \mathbb C$ is given by
\begin{align}
    \Delta_a= 2\pi i\sum_{n=1}^k l_{(a,n)} \left( z_{(a,n)}+\frac12 l_{(a,n)}\tau \right)\,.
\end{align}
Furthermore, there are also equivalence relations for permutations of the zeros $\{z_{(a,n)}\}$ for each $a$.
Note that each of the zeros parameterizes $T^2$,
but in the moduli space they do not appear in direct product with $\mathbb CP^{N-1}$ described by $\{c_a\}$. 
The above equivalence relations mean that 
the parameter $c_a\in \mathbb CP^{N-1}$ exhibits monodromy as the zeros complete a full cycle.
This monodromy makes it difficult to understand the topology of the moduli space.
In Sec.\ref{Sec:modulispace}, we will propose an alternative expression of the solution with a different parametrization,
where the moduli space is expressed as a quotient space.

\subsubsection{BPS lumps for \texorpdfstring{$p_s=1$}{ps=1}} %with ${\rm gcd}(p,N)=1$}

Next, let us consider the case $p_s=1$, where the extended torus becomes $\hat{T}^2=\mathbb{C}/(N\mathbb{Z}+\tau \mathbb{Z})$. 
In this case, all the flavor indices are related by the shift-twisted boundary condition~\eqref{eq:hzbcS}, and we only need to determine $h_1(z)$ with the transition function $(v_s(z),v_c(z))$. 
Let us write its zeros in $\mathcal{F}(\hat{T}^2)$ as $z_{n}:=z_{(1,n)}$ with $n=1,\ldots, \hat{M}=M$, and then we may set 
\begin{align}
    h_1^{\mathrm{rep}}(z)=\prod_{n=1}^{M} \theta_1\left(\frac{z-z_{n}}{N}\Big| \frac{\tau}{N}\right). 
\end{align}
We can determine $v_c(z)$ by $h_1(z+\tau)=e^{-2\pi \im \left(k+\frac{p}{N}\right)\left(z-\langle z\rangle +\frac{N+\tau}{2}\right)}h_1(z)$ with $\langle z \rangle=\frac1{M} \sum_{n=1}^{M}z_n$. 
As this $v_c(z)$ satisfies $-[v_c(z+1)-v_c(z)]=2\pi i\left(k+\frac{p}{N}\right)$, we may set $v_s(z)$ as constant for the constraint \eqref{eq:vCC}. 
For constant $v_s$, we can determine $\rme^{v_s(z)+\cdots v_s(z+N-1)}=\rme^{Nv_s}$ by $h_1(z+N)=(-1)^M h_1(z)$, but there is an ambiguity when taking the $N$-th root to fix $v_s$. 
Thus, we have to introduce a discrete parameter $\alpha=1,\ldots, N$ for the holomorphic transition function, 
\begin{align}
(v_s(z),v_c(z))=2\pi i \left(k+\frac{p}N\right) \times \Big(\frac12 ,\, -(z-\langle z \rangle)-\frac{N+\tau}2\Big)+\left( \frac{2\pi i}N \alpha,\,0  \right).
\end{align}
If we would like to write down $h_a(z)$ for all the flavor indices, they can be obtained by using \eqref{eq:hzbcS} as follows,
\begin{align}
    h_a(z)\quad \stackrel{V}{\sim} \quad h_a^{\rm rep}(z)&= e^{\frac{2\pi i}N \alpha (a-1)}\prod_{n=1}^{M} e^{\frac{\pi i}{N}(a-1)} \theta_1\left(\frac{z-z_n-a+1}N\Big| \frac{\tau}N\right). 
    %\varphi_a^N(z-z_n),\quad {\rm with~}\alpha \in \mathbb Z,\\
    % \varphi_a^N(z)&\coloneqq e^{\frac{\pi i}{N}(a-1)} \theta_1\left(\frac{z-a+1}N\Big| \frac{\tau}N\right),\quad (\varphi_{a+N}^N=\varphi_a^N), 
\end{align}
% where we introduced the integer $\alpha$ to construct the most general one. 
This is the most general form of the solution, and see Appendix \ref{sec:Uniquness} for the proof. 
The factor $e^{\pi i\frac{a-1}N}$ comes from $v_s(z)$, which ensures $h_{a+N}^{\rm rep}(z)=h_a^{\rm rep}(z)$. 
We note that the $V$-transformation is completely fixed in the above expression.

% In this case, $\hat N=N$, $p=p_c$ and $\hat M=M$ since $\bar N={\rm gcd}(N,p_s=1)=1$. 
% According to the relation  \eqref{eq:sigmarelation} with $\bar r(a)=1$ and $\hat s(a)=a$, 
% the zeros  $z_{(a,n)} \in  \hat \sigma_a$ are given as  
% \begin{align}  z_{(a,n)}\equiv z_n+ a-1 \quad {\rm mod~} N,
% \end{align}
% with  $z_n\in {\cal F}(\hat T^2), (n=1,\cdots,M)$.

The discrete parameter $\alpha$ can be absorbed into the continuous moduli parameter by extending, e.g., $z_1\in \mathcal{F}(T^2)$ to $z_1\in \mathbb{C}$; 
\begin{align}
    \vec h^{\rm rep}(z)\Big|_{z_1\to z_1+\tau} \quad\stackrel{V}{\sim}\quad \vec h^{\rm rep}(z)C^{-1} =\vec h^{\rm rep}(z)\Big|_{\alpha\to \alpha+1}.
\end{align}
Therefore, we can always rewrite $h_a^{\rm rep}(z)$ into the following representation 
by extending the domains of the zeros,
\begin{align}
     h_a(z)\quad \stackrel{V}{\sim}\quad   h_a^{\widehat {\rm rep}}(z)&= \prod_{n=1}^{M} e^{\frac{\pi i}{N}(a-1)} \theta_1\left(\frac{z-z_n-a+1}N\Big| \frac{\tau}N\right), 
     \label{eq:repr_ps1_extendedmoduli}
\end{align}
where the equivalence relation on the moduli parameter $\{z_n \in \mathbb C |n=1,2,\cdots,M\}$ becomes
\begin{align}
    \{z_n\} \quad \sim \quad \{ z_n +N m_n+l_n\, \tau \},\quad {\rm with~} m_n, \, l_n \in \mathbb Z,\quad \sum_{n=1}^M l_n \in N \mathbb Z. \label{eq:equi_rel_zn_ps1}
\end{align}
Unlike the case with $p_s=0$, there is no monodromy in this model. 
Instead, however, the modular duality discussed in Sec.~\ref{sec:duality} is not manifest in this representation.
In Sec.~\ref{Sec:modulispace}, we propose the alternative representation so that the modular duality becomes more apparent.

%%%%%%%%%%%%%%%%%%%%%%%%%%%%%%%%%%%%%%%
\subsubsection{General \texorpdfstring{$p_s$}{ps}}
\label{sec:BPSlump_generalps}

Let us consider the case of general $p_s$, which can be constructed by combining the cases with $p_s=0$ and $p_s=1$. 
Due to the shift-twisted boundary condition, we need to specify $h_{\bar{a}}(z)$ for $\bar{a}=1,\ldots, \gcd(N,p_s)$ as the other flavor indices are determined by them. 
The zeros $z_{(\bar{a},n)}$ of $h_{\bar{a}}$ are taken from $\mathbb{C}$ instead of $\mathcal{F}(\hat{T}^2)$ so that the following equality holds on the nose,
\begin{align}
     \sum_{n=1}^{\hat M} z_{(\bar a,n)}= \hat M \langle z \rangle +\frac{p_c(\bar a-1)}{\gcd(N,p_s)},  \label{eq:modified_abel_constraint_extended}
\end{align}
for some $\langle z\rangle \in \mathbb{C}$. Then, we can take the representative of $h_{\bar{a}}$ as
\begin{align}
    h^{\mathrm{rep}}_{\bar{a}}(z)=c_{\bar{a}}\prod_{n=1}^{\hat{M}} \theta_1 \left(\frac{\gcd(N,p_s)}{N}(z-z_{(\bar{a},n)})\Big| \frac{\gcd(N,p_s)}{N}\tau\right), 
    \label{eq:BPSlump_generalps_abar}
\end{align}
where $(c_{\bar{a}})_{\bar{a}=1,\ldots,\gcd(N,p_s)}\in \mathbb{C}^{\gcd(N,p_s)}\setminus\{\vec{0}\}$. 
We can choose the holomorphic transition function $(v_s,v_c)$ as
\begin{align}
    (v_s(z),v_c(z))=2\pi i \left(k+\frac{p}N\right) \times \Big(\frac12 ,\, -(z-\langle z \rangle)-\frac{N/\gcd(N,p_s)+\tau}2\Big). 
    \label{eq:holomorphic_transition_function_generalpspc}
\end{align}
Strictly speaking, $v_s(z)$ has an ambiguity by $\frac{2\pi \gcd(N,p_s)}{N}\alpha$ with $\alpha=1,\ldots, \frac{N}{\gcd(N,p_s)}$, but we omit this discrete parameter because it turns out to be absorbed via extension of the continuous moduli $z_{(\bar{a},n)}$. 
As discussed in the previous subsections, there are equivalence relations on the moduli parameters
$( \{z_{(\bar a,n)} \}, \{c_{\bar a}\}, \langle z \rangle )$ including monodromy,
but we will not discuss them here because it is better to use the alternative expressions in the next section to understand the topology.
Here, we just note that the $\mathbb{C}^*$ multiplication on $\vec{c}$ is the remnant of the $V$-transformation, and thus the dimension of the moduli space is calculated as 
\begin{align}
    {\rm dim}_{\mathbb C} {\cal M}^{N:(p_s,p_c)}_k 
    &= \sum_{\bar{a}=1}^{\gcd(N,p_s)} (|\hat{\sigma}_{\bar{a}}|-\#\{\langle z\rangle\}) + \dim_{\mathbb{C}} \mathbb{C}P^{\gcd(N,p_s)-1} + \#\{\langle z\rangle\} \notag\\
    &=\gcd(N,p_s)(\hat{M}-1)+(\gcd(N,p_s)-1)+1 \notag\\
    &= Nk+p =M,
    %=\bar N(\hat N k+\hat p-1)+(\bar N-1)+1=N k+p=M,
\end{align}
which is consistent with the result of the index theorem. 

To write down the formula for $h_{a}(z)$ for all the flavor indices, we just need to apply \eqref{eq:hzbcS} for the above $h_{\bar{a}}(z)$ with the given $(v_s(z),v_c(z))$. 
To treat the flavor label concisely, it is convenient to consider a bijective map as a set from $\mathbb Z_N$ to $\mathbb Z_{N/\gcd(N,p_s)}\times \mathbb Z_{\gcd(N,p_s)}$ 
\begin{align}
    a\in \mathbb Z_N \quad \mapsto \quad    (\hat s(a), \bar r(a))\in \mathbb Z_{N/\gcd(N,p_s)}\times \mathbb Z_{\gcd(N,p_s)} \,,
    % \mathbb Z_{\hat N}\times \mathbb Z_{\bar N} 
    \label{eq:a2sr}
\end{align}
where maps $\bar r(a), \hat s(a)$ are defined by
\begin{align}
    \bar r(a)& :=\bar r_0(a-1)+1,  \quad ( 1\le \bar r(a) \le \gcd(N,p_s)),\\
    \hat s(a) &:=q_s\, \hat q(a-1) +1,
\end{align}
with the remainder $\bar r_0(b)$ and the quotient $\hat q(b)$ of $b\in \mathbb Z$ divided by $\gcd(N,p_s)$ and the B\'ezout's coefficients $q_s, r_s \in \mathbb Z_{\ge 0}$ uniquely given such that they satisfy  $q_s p_s-r_s N=\gcd(N,p_s) \, (1\le q_s <\frac{N}{\gcd(N,p_s)})$.
Note that there is an identity map given by $\bar r, \hat s$ as
\begin{align}
 a = 1+\bar r_0(a-1)+\gcd(N,p_s) \hat q(a-1) \equiv \bar r(a) + p_s\, (\hat s(a)-1) \quad {\rm mod~} N,   
\end{align}
which guarantees the map \eqref{eq:a2sr} is bijective. 
% There are the following relations between the zeros of $ \vec h(z)$,
% \begin{align}
%     z_{(a,n)}\equiv z_{(\bar r(a),n)}+\hat s(a)-1\quad {\rm mod~} \hat N, 
% \end{align}
% and by taking $\mathbb C$ as the domain of definition of $z_{(a,n)}$, we can assume that $\{z_{(\bar a,n)}\}$ satisfy
% the following relations without loss of generality
% \begin{align}
% \exists\,\langle z\rangle\in \mathbb C,\, \bar a=1,2,\cdots, \bar N:  \quad \sum_{n=1}^{\hat M} z_{(\bar a,n)}= \hat M \langle z \rangle +\frac{p_c(\bar a-1)}{\bar N}.
% \end{align}
Using this notation, the general solution of $\vec h(z)$ is $V$-equivalent to the following form,
\begin{align}
    h_a^{\widehat {rep}}(z)
  &= c_{\bar r(a)}\, e^{\pi i (k+\frac{p}{N})(\hat{s}(a)-1)} \prod_{n=1}^{\hat M} \theta_1\left(\frac{\gcd(N,p_s)}{N}(z-z_{(\bar{r}(a),n)}-\hat{s}(a)+1)\Big| \frac{\gcd(N,p_s)}{N}\tau\right). 
  %\varphi_{\hat s(a)}^{\hat N}(z-z_{(\bar r(a),n)})\quad {\rm with~}c_{\bar r(a)} \in \mathbb C.  
  \label{eq:generalPsol}
\end{align}
% Here the factor $e^{\frac{2\pi i} {\hat N} \alpha (\hat s(a)-1)}$ appeared in Appendix \ref{sec:Uniquness} 
% has been absorbed by redefining zeros and $\langle z \rangle$.  
By construction, $\vec h^{\widehat{rep}}(z)$ given above satisfies the boundary conditions \eqref{eq:hzbcS} and \eqref{eq:hzbcC}. 
To check this explicitly, it is convenient to use the following equations: 
\begin{align}
 \bar r(a+p_s)=\bar r(a),\quad   \hat s( a+p_s)& \equiv \hat s(a)+1 \quad {\rm mod~} \frac{N}{\gcd(N,p_s)},\\
  \bar r(a+N)=\bar r(a),\quad   \hat s( a+N)& \equiv \hat s(a)\qquad \quad {\rm mod~} \frac{N}{\gcd(N,p_s)},
\end{align}
and
\begin{align}
    M (\hat s(a)-1)+p_c(\bar r(a)-1) \equiv p_c (a-1)\quad {\rm mod~} N.
\end{align}

%%%%%%%%%%%%%%%%%%%%%%%%%%%%%%%%%%%%%%%%%%%%%%%%%%%%%%%
%%%%%%%%%%%%%%%%%%%%%%%%%%%%%%%%%%%%%%%%%%%%%%%%%%%%%
\subsection{Illustration of fractional BPS lumps}
% \subsection{Fractional vortices}
\label{sec:fractionalV}

In this section, let us visually explore the behavior of the fractional BPS lumps in simple setups. 
For this purpose, we will show heatmaps of the action density $\rho$ in the large coupling limit $e^2\to \infty$, which is given by
\begin{align}
   \rho(z,\bar{z}) := \frac{\xi}{2} \p_i^2\log |\vec h(z)|^2. 
   % \quad  {\rm for~} e^2\to \infty.  
   \label{eq:energy_density}
\end{align}
In Sec.~\ref{sec:QMlimit},  we consider the quantum mechanical limit by taking $L_1\gg L_2$ or $L_2\ll L_1$, which correspond to $-i\tau\to +0$ or $+\infty$, respectively, and we discuss the minimal fractional BPS lump with $p_s=p_c=1$ and $k=0$. 
In Sec.~\ref{sec:singularity}, we consider how the small-lump singularity at $e^2\to \infty$ appears in our expression of fractional BPS lumps.

%%%%%%%%%%%%%%%%%%%%%%%%%%%%%%%%%%
\subsubsection{Fractional instantons with \texorpdfstring{$p_s=p_c=1$ and $k=0$}{ps=pc=1 and k=0} on \texorpdfstring{$\mathbb{R}\times S^1$}{RxS1}}
\label{sec:QMlimit}
% \subsubsection{Domain walls in the limit of $\tau\to 0,i \infty $ }

\begin{figure}[th]
\centering
\includegraphics[height=3.5cm]{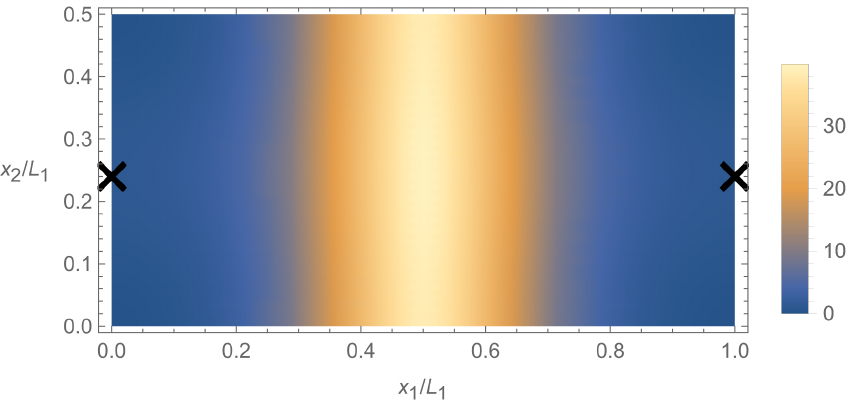}
\includegraphics[height=3.5cm]{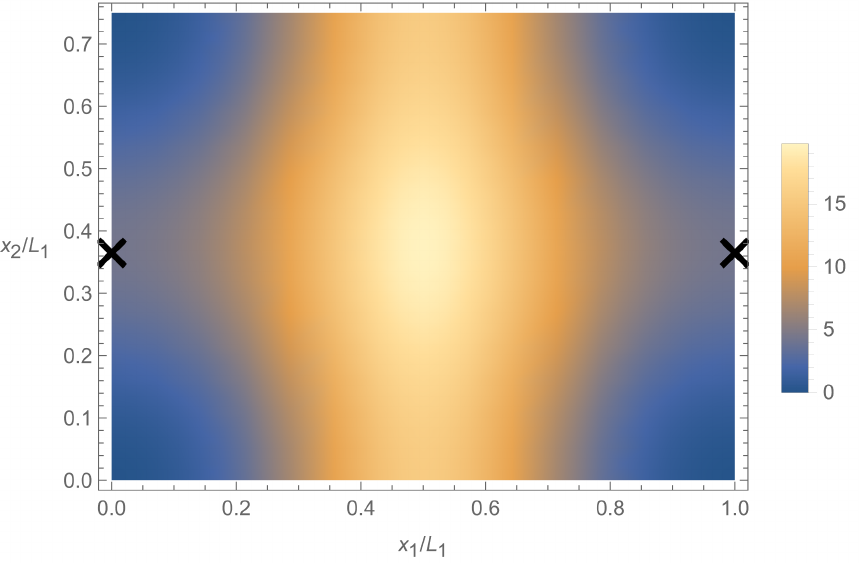}\\
\includegraphics[height=5cm]{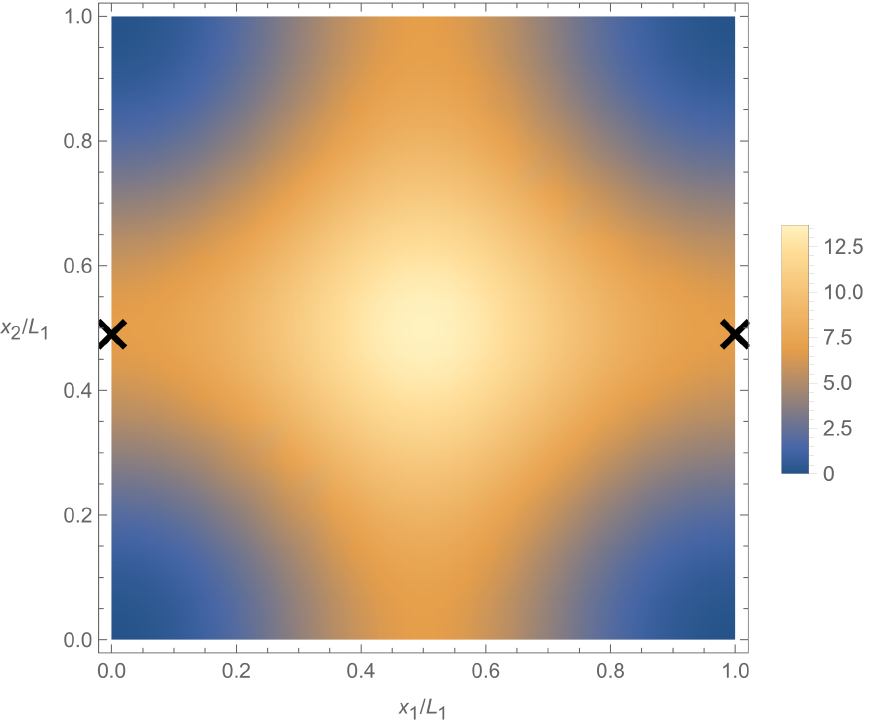}
\includegraphics[height=5cm]{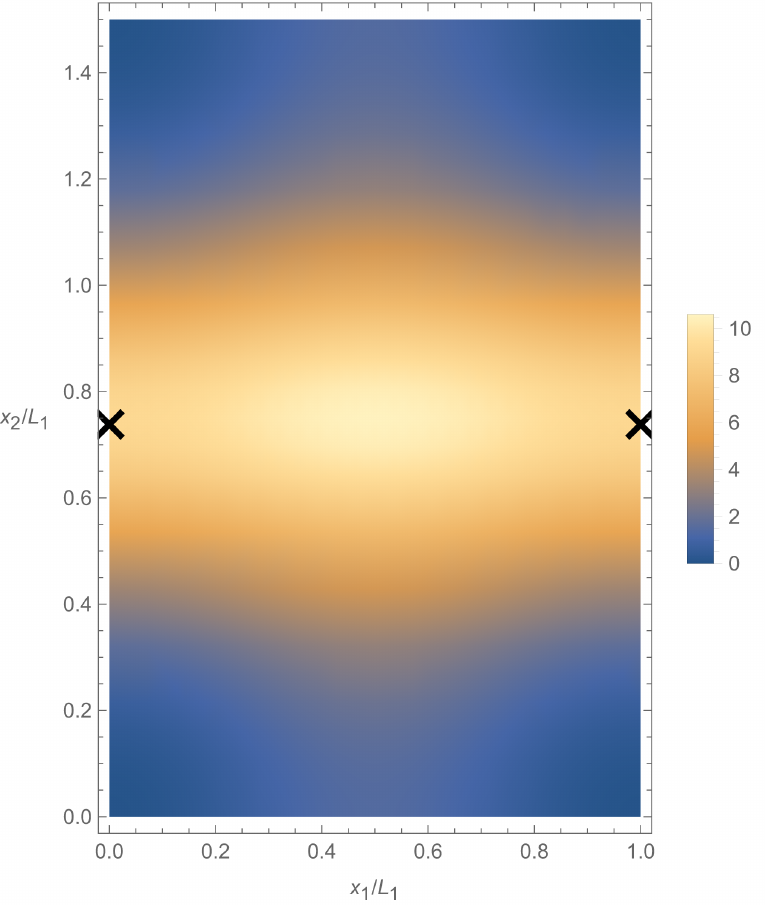}
\includegraphics[height=5cm]{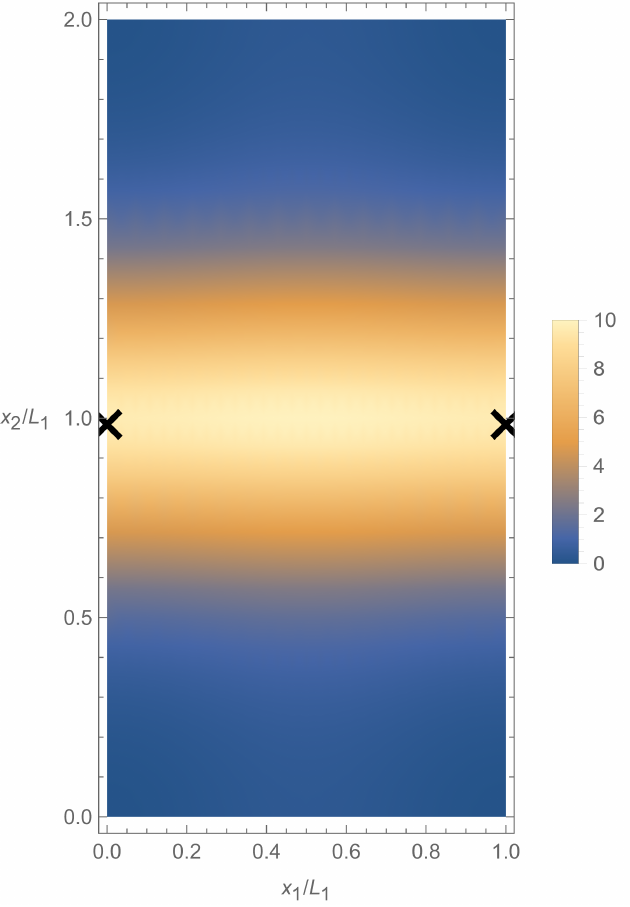}
\caption{Heatmaps of the dimensionless action density $\rho ~(\times L_1^2)$ with $\xi=2$ for a single fractional vortex in the cases of $-i\tau=1/2,3/4,1,3/2,2$.
The zeros are marked with crosses.
The left one at $(x_1,x_2) = (0,L_2/2)$ is zero of $h_2$, while the right one at $(x_1,x_2) = (L_1,L_2/2)$ is zero of $h_1$.
\label{fig:taudependence}}
\end{figure}

In this subsection, we consider the minimal fractional BPS lumps with $p_s=p_c=1$ and $k=0$. 
Let us set $z_1=\frac{\tau+N}{2}$ for concreteness, and then \eqref{eq:repr_ps1_extendedmoduli} becomes 
\begin{align}
    h^{\widehat{\mathrm{rep}}}_a(z)=e^{\frac{\pi i}{N}(a-1)}\theta_1\left(\frac{z-z_1-a+1}{N}\Big|\frac{\tau}{N}\right)
    \stackrel{V}\sim\,  \theta_3\left(\frac{z-a+1}{N}\Big|\frac{\tau}{N}\right),  
\end{align}
where $\theta_3(z|\tau)=\sum_{n\in\mathbb{Z}}q^{n^2}e^{2\pi i n z}$. 
Figure~\ref{fig:taudependence} presents heatmaps of the action density $\rho$ with $N=2$ for several choices of $-i\tau=\frac{L_2}{L_1}$. 
% The parameters are defined as $N=2$, $(p_s,p_c)=(1,1)$ and  $k=0$, which characterizes a system possessing a single fractional vortex, $M=1$.
% In these plots, we put the zero at $z_1 = \frac{\tau + 2}{2}$ in the representation $h_a^{\widehat {\rm rep}}(z)$ given by Equation (\ref{eq:repr_ps1_extendedmoduli}).
In these figures, the locations of the zeros are marked with crosses: $h_1(L_1,L_2/2) = 0$ and $h_2(0,L_2/2) = 0$.
One can observe the formation of a domain-wall-like configuration when $|\tau|$ becomes very large or small (see $-\im \tau =1/2,2$ cases of Figure~\ref{fig:taudependence}). 
We shall examine this behavior more closely.

Let us regard the limit $\tau=i\frac{L_2}{L_1}\to +0\times i$ as $L_1\to \infty$ with fixed $L_2$. 
We can then think of this setup as the quantum mechanical reduction of the $2$d $\mathbb{C}P^{N-1}$ sigma model with the $C$-twisted boundary condition for $x_2\sim x_2+L_2$, and the uncompactified coordinate $x_1$ becomes the imaginary-time direction. 
Up to the $U(1)$ gauge transformation, the classical vacua with the $C$-twisted boundary condition is obtained by the $x_2$-independent configurations with 
\begin{align}
    \vec H C=\vec H \omega^\ell, 
\end{align}
for some $\ell\sim \ell+N$, and we find the $N$ distinct classical vacua for the $\mathbb{C}P^{N-1}$ field, 
\begin{align}
    \vec{n}=\frac{\vec{H}}{|\vec{H}|}=(1,0,\ldots, 0),\, (0,1,\ldots, 0),\ldots,\, (0,0,\ldots, 1). 
\end{align}
The minimal fractional BPS lumps with $Q=1/N$ are the fractional instanton solutions that cyclically permute these $N$ classical vacua, and the $S$-twisted boundary condition along $x_1$ controls the boundary condition at infinities to single out the fractional instanton under the limit $L_1\to \infty$. 
Since the $C$-twisted boundary condition opens the gap for the Kaluza-Klein (KK) spectrum by $\frac{2\pi}{NL_2}$ without having the zero mode, the action density of the instanton configuration should have a sharp peak controlled by this KK mass gap, which explains the domain-wall-like configuration  at $-\im \tau =1/2$ of Figure~\ref{fig:taudependence}. 

Let us explicitly see that $\vec{h}^{\widehat{\mathrm{rep}}}(z)/|\vec{h}^{\widehat{\mathrm{rep}}}(z)|$ with $p_s=p_c=1$ and $k=0$ gives the fractional instanton discussed above in the limit $L_1\to \infty$. 
We should recall, however, that our definition of the complex coordinate is given by $z=\frac{x_1+i x_2}{L_1}$, which naively goes to zero in the limit $L_1\to \infty$, and we need to rewrite the expression using Jacobi's imaginary transformation so that the natural variable becomes $\hat{z}=\frac{z}{-i\tau}=\frac{x_1+i x_2}{L_2}$: Let us again set $z_1=\frac{\tau+N}{2}$, then Jacobi's imaginary transformation gives
\begin{align}
    h^{\widehat{\mathrm{rep}}}_a(z) 
    &\stackrel{V}\sim\, \frac{1}{(-i \tau/N)^{1/2}}e^{-i \frac{\pi}{N \tau}(z-a+1)^2}\theta_3\left(\frac{z-a+1}{\tau}\Big|-\frac{N}{\tau}\right) \notag\\
    &\stackrel{V}\sim\, e^{\frac{2\pi i }{N}\frac{(a-1)}{\tau}(z-\frac{a-1}{2})} \theta_3\left(\frac{z-a+1}{\tau}\Big|-\frac{N}{\tau}\right),  
\end{align}
up to the $V$-equivalence. 
Let us take the $L_1\to \infty$ limit with fixed $x_1-\frac{L_1}{2}$, then $\theta_3\to 1$ and the only $a=1$ and $2$ components have non-vanishing limit. As a result, we find 
\begin{align}
 \vec h(z)
\quad &\stackrel{V}\sim\quad
 \left(1,\,\,
e^{\frac{2\pi }{N L_2}\left(x_1 - \frac{L_1}{2}\right)+\frac{2\pi i }{N}\frac{x_2}{L_2}},\,\,
0,\,\,\cdots,\,\,0
 \right)\,,  
\end{align}
which behaves (up to the $U(1)$ gauge redundancy) as
\begin{align}
 \frac{\vec{h}(z)}{|\vec{h}(z)|} \,\,\to\,\, 
 \begin{cases}
\left( 1,0, \ldots,0\right) \quad x_{1}-\frac{L_1}{2}\to -\infty, \\
\left(0,1, \ldots,0\right) \quad x_{1}-\frac{L_1}{2}\to +\infty. 
 \end{cases}\,
\end{align}
This is nothing but the fractional instanton solution with the minimal topological charge $Q=1/N$ in the ${\mathbb C}P^{N-1}$ sigma model on ${\mathbb R}\times S^1$ with the ${\mathbb Z}_N$ twisted boundary condition~\cite{Eto:2004rz, Eto:2006mz, Eto:2007aw}.
Let us note that the composite configuration (complexified solution) of fractional instantons and anti-instantons, called bions, play a pivotal role in the resurgence analysis of ${\mathbb C}P^{N-1}$ models as discussed in the literature \cite{Dunne:2012ae,Dunne:2012zk,Misumi:2014jua,Misumi:2014bsa,Misumi:2014rsa,Misumi:2015dua,Misumi:2016fno,Fujimori:2016ljw,Fujimori:2017oab,Fujimori:2017osz,Fujimori:2018kqp,Fujimori:2019skd,Misumi:2019upg,Fujimori:2020zka}.\footnote{The $C$-twisted boundary condition in the context of $4$d QCD was first introduced in Refs.~\cite{Kouno:2012zz, Sakai:2012ika, Kouno:2013zr, Kouno:2013mma, Kouno:2015sja} for the Polyakov-loop effective model, which observed that the Polyakov loop becomes an exact order parameter for the confinement-deconfinement transition when $N_c=N_f=N$. This was also confirmed for $\mathbb{Z}_N$-twisted QCD by the numerical lattice computation~\cite{Iritani:2015ara}. 
Ref.~\cite{Cherman:2017tey} clarifies the existence of the color-flavor center symmetry in the $C$-twisted boundary condition of $N_f=N_c$-flavor QCD on $\mathbb{R}^3\times S^1$, which acts as the center rotation on the Polyakov loop and the shift operation on the quark flavors, and Refs.~\cite{Shimizu:2017asf, Tanizaki:2017qhf, Tanizaki:2017mtm, Dunne:2018hog} uncover the preservation of various 't~Hooft anomalies thanks to the presence of the color-flavor center symmetry. 
}

In the opposite limit, $\tau \to i \infty$, we regard $L_1$ as the fixed size and take the limit $L_2\to \infty$. 
This gives the quantum mechanical reduction of $\mathbb{C}P^{N-1}$ model with the $S$-twisted boundary condition for $x_1\sim x_1+L_1$. 
Here, the classical vacua are fixed points of $S$-twist, $\vec{H}S=\vec{H}\omega^\ell$ %with some $\ell\sim \ell+N$
, which are given by 
\begin{align}
    \vec{n}=\frac{\vec{H}}{|\vec{H}|}=\frac{1}{\sqrt{N}}(1, e^{\frac{2\pi i \ell}{N}}, \ldots, e^{\frac{2\pi i (N-1)\ell}{N}}), 
\end{align}
with $\ell=0,1,\ldots, N-1$. 
The fractional instantons cyclically permute these classical vacua, and the explicit formula in the $\tau=i\infty$ limit with $z_1=\frac{\tau+N}{2}$ and fixed $x_2-\frac{L_2}{2}$ is given by
\begin{align}
    h_a(z) &\stackrel{V}\sim\,  \theta_3\left(\frac{z-a+1}{N}\Big|\frac{\tau}{N}\right) \notag\\ 
    &\to 1+e^{\frac{2\pi}{NL_1}(x_2-\frac{L_2}{2})+\frac{2\pi i}{N}(-\frac{x_1}{L_1}+(a-1))}, 
\end{align}
which behaves (up to the $U(1)$ gauge redundancy) as
\begin{align}
 \frac{\vec{h}(z)}{|\vec{h}(z)|} \,\,\to\,\, 
 \begin{cases}
\left( 1,\, 1,\, 1,\cdots,1\right) /\sqrt{N}
\quad\, & x_{2}-\frac{L_2}{2}\to -\infty,  \\
\left( 1,\,  e^{\frac{2\pi i}{N}},\, e^{\frac{4\pi i}{N}},\cdots,e^{\frac{2\pi i(N-1)}{N}}\right)/\sqrt{N}
\quad & x_{2}-\frac{L_2}{2}\to +\infty.
 \end{cases}\,
\end{align}
Unlike the $C$-twisted case, the zeros ($h_a=0$ for some $a$) are located at the midpoints between these vacua instead of the locations at the classical vacua.
As shown in Figure~\ref{fig:taudependence} for $-\im \tau =2$,  the zeros are indeed located on a peak of the action density.

Note that these two limits are interchanged by taking the modular duality with discrete Fourier transformation (Section \ref{sec:duality}).
Here, $C$-twist and $S$-twist are interchanged each other in this duality, so the above observation is consistent with the duality.

\subsubsection{Local vortex and small-lump singularity}
\label{sec:singularity}

In this paper, we study the self-dual configurations of the $N$-component Abelian-Higgs model instead of directly analyzing those of the $\mathbb{C}P^{N-1}$ model. 
We can obtain the self-dual configuration of the $\mathbb{C}P^{N-1}$ model by taking the formal limit $e^2\to \infty$, which requires \eqref{eq:nonsingular} as a necessary condition. 

Here, let us consider the ``bad'' situation for the $\mathbb{C}P^{N-1}$ limit, 
where the solution of $\psi$ becomes ill-defined in the large coupling limit $e^2\to \infty$: 
\begin{align}
    \exists z_0 \in  T^2 \quad \text{such that}\quad  \vec{h}(z_0)=\vec{0}. 
\end{align}
In this case, $h'_a(z)=h_a(z)/\theta_1(z-z_0|\tau)$ becomes an entire function. 
To see it explicitly, we note that the $S$-twisted boundary condition~\eqref{eq:hzbcS} shows $\vec{h}(z_0+1)=e^{v_s(z_0)}\vec{h}(z_0)S^{p_s}=\vec{0}$. 
Thus, $\{z_0, z_0+1,\ldots, z_0+\frac{N}{\gcd(N,p_s)}-1\}$ are all common zeros of $h_a(z)$ in $\mathcal{F}(\hat{T}^2)$, and $h_a(z)$ should contain the following product, 
\begin{align}
    &\quad \prod_{l=1}^{N/\gcd(N,p_s)} \theta_1\left(\frac{\gcd(N,p_s)}{N}(z-z_0-l+1)\Big| \frac{\gcd(N,p_s)}{N}\tau \right) \notag\\
    &=(-1)^{N/\gcd(N,p_s)} \prod_{n=1}^\infty \frac{(1-q^{2n})^{N/\gcd(N,p_s)}}{(1-q^{2nN/\gcd(N,p_s)})}\times \theta_1(z-z_0|\tau),
\end{align}
which formula can be confirmed by direct computation using the infinite-product representation \eqref{eq:thetadef}. 
As a result, the common factor $\theta_1(z-z_0|\tau)$ can be extracted from $\vec h(z)$ as
\begin{align}
    \vec h(z)  = \theta_1(z-z_0|\tau)\times  \vec h'(z) . 
    %\quad \stackrel{{\rm singular~V}}{\sim }\quad \vec h'(z).
\end{align}
We note that $\vec{h}(z)$ and $\vec{h}'(z)$ are not $V$-equivalent as $\theta_1(z-z_0|\tau)$ has a zero. 
The total action of $\vec{h}(z)$ is given by $\mathcal{S}=2\pi \xi (k+\frac{p}{N})$; the $\theta_1(z-z_0|\tau)$ factor contributes by  $2\pi \xi$ and the rest, $2\pi \xi(k-1+\frac{p}{N})$, is carried by $\vec{h}'(z)$. 
This can be confirmed explicitly by the formal computation of the action density~\eqref{eq:energy_density} in the $e^2\to \infty$ limit,
\begin{align}
    \rho(z,\bar{z})=\frac{2\pi\xi}{L_1^2} \delta^2(z-z_0)+ \frac{\xi}{2}\partial_i^2 \ln |\vec{h}'(z)|^2. 
\end{align}
For finite $e$, the gauge field has a nonzero interaction range, and the delta function above is therefore replaced by a regularized vortex-like profile with a size of order $1/(e\sqrt{\xi})$. 
In this sense, the common $\theta_1(z-z_0|\tau)$ factor deserves to be called the local vortex for finite $e$, or the small-lump singularity in the $e^2\to \infty$ limit.

\begin{figure}[t]
\centering
\begin{minipage}{.40\textwidth}
\includegraphics[width= 0.95\textwidth]{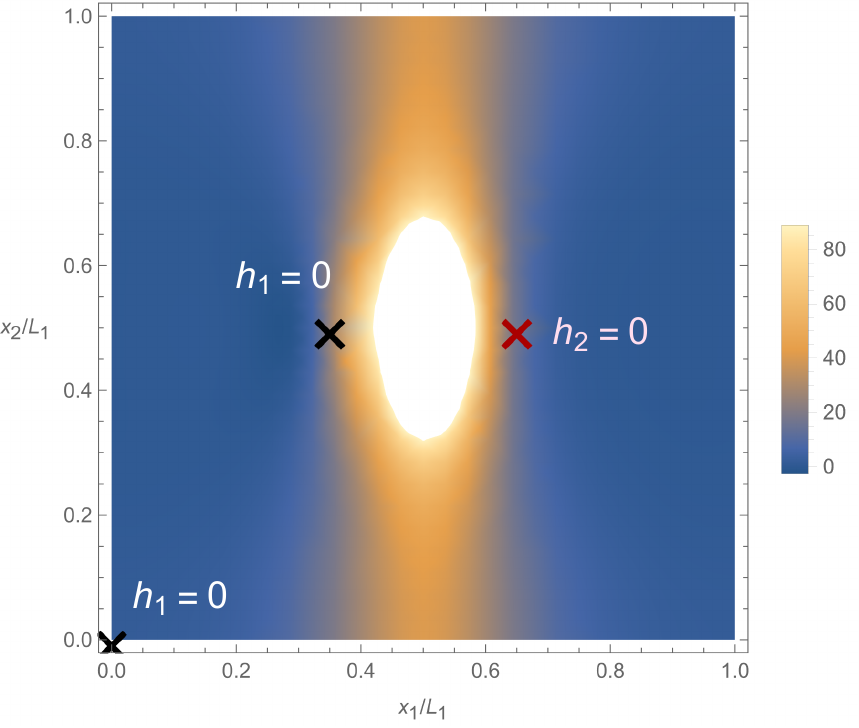}
\end{minipage}\quad
\begin{minipage}{.40\textwidth}
\includegraphics[width= 0.95\textwidth]{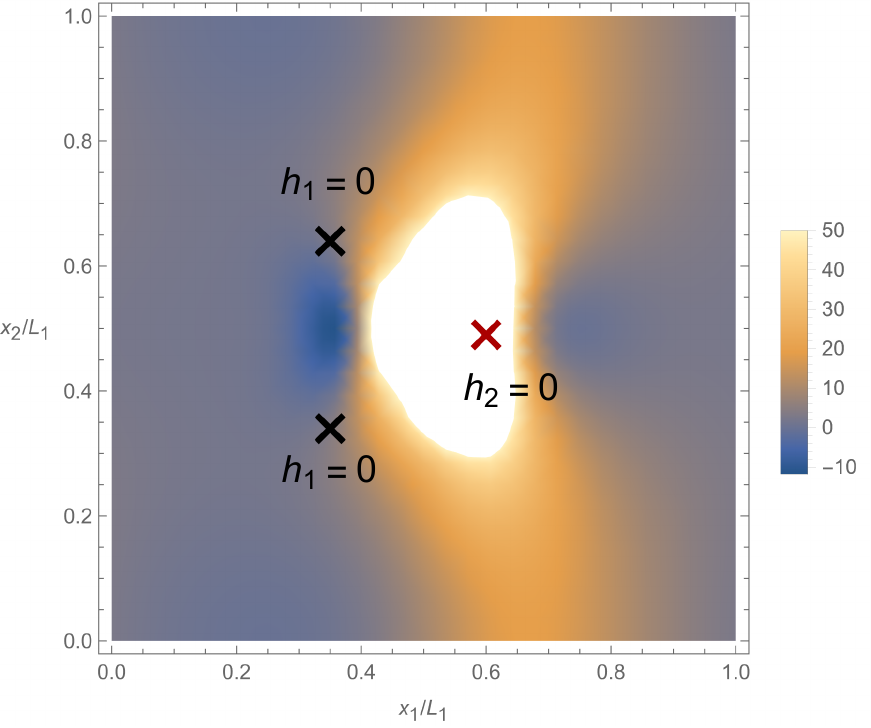}
\end{minipage}
\caption{
The configuration becomes singular if different-flavor zeros approach each other.
When $N=2$, the zeros are merged $h_1=h_2=0$ means $\vec{H} = \vec{0}$, which is not a point of $\mathbb{C}P^1$. 
}
\label{fig:diff_type_collision}
\end{figure}

\begin{figure}[t]
\centering
\begin{minipage}{.40\textwidth}
\includegraphics[width= 0.95\textwidth]{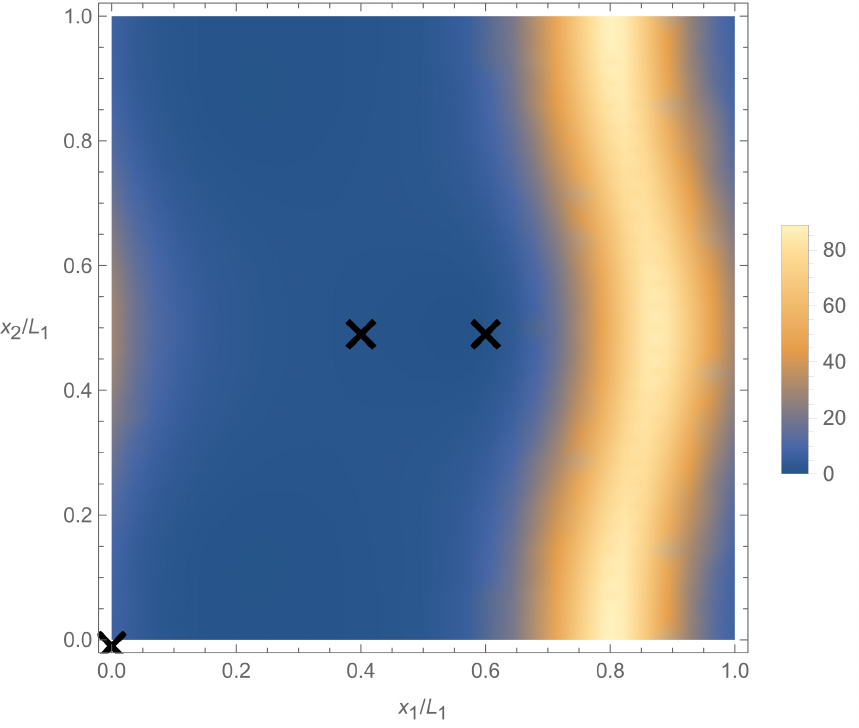}
\end{minipage}\quad
\begin{minipage}{.40\textwidth}
\includegraphics[width= 0.95\textwidth]{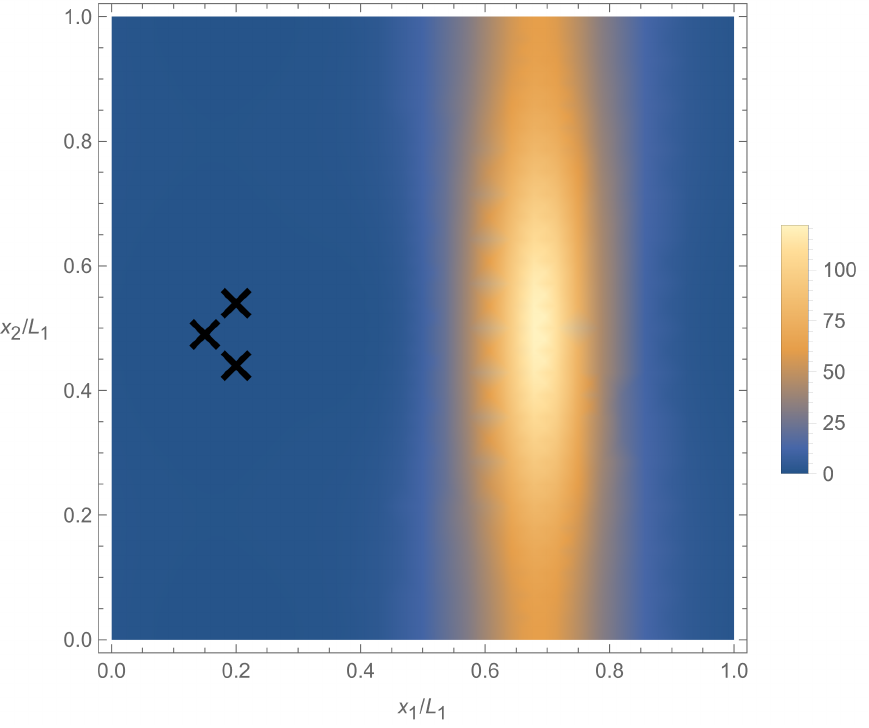}
\end{minipage}
\caption{
If the zeros of $h_1$ collide, they merge into a higher zero of $h_1$.
No singular behavior appears even if these three zeros are located at the same point. (The zeros are separated only for the sake of presentation.)
}
\label{fig:same_type_collision}
\end{figure}

Here, we would like to emphasize that the local vortex appears only when all the flavors have a common zero $z_0$ so that $|\vec{h}(z_0)|=0$, and that the local vortex is not formed even if some flavors have a zero with higher multiplicity. 
To visualize collisions of zeros of the same/different flavors, let us consider the simplest case with $N=2$, $p_s=p_c=1$, and $k=1$.
In this case, there are $M=3$ zeros in total. 
In Figure \ref{fig:diff_type_collision}, we show the case when the zeros of different flavors become closer, while Figure \ref{fig:same_type_collision} describe the case when the zeros of a single flavor become closer. 

\begin{enumerate}
    \item Collision of different-flavor zeros

    Here, we look at the collision of zeros of $h_1$ and $h_2$. 
    If there exists a point with $h_1 = h_2 =0$, i.e., $\vec{H}=\vec{0}$, the energy density in the limit $e^2 \rightarrow \infty$ (\ref{eq:energy_density}) diverges.
    The point $\vec{H}=\vec{0}$ is not included in $\mathbb{C}P^{N-1}$, and such a singularity is called the small-lump singularity. 
    Figure \ref{fig:diff_type_collision} depicts the heatmap for the energy density of a few configurations where zeros of $h_1$ and $h_2$ come close with each other, and we can observe the formation of the local vortex.

    \item Collision of same-flavor zeros

    When zeros of the same flavor collide, nothing singular happens.
    Figure \ref{fig:same_type_collision} plots a few configurations where zeros of $h_1$ are close.

    Notice that a domain wall is formed in this figure.
    For example, suppose all three zeros of $h_1$ are located at the same point, say $z = z_0$. A domain wall is created between the zero of $h_1$ at $z = z_0$ and the zero of $h_2$ at $z = z_0 + 1$ \footnote{We note that, for the case of the minimal fractional lump, the energy density shows a cross shape instead of the domain wall at $L_1=L_2$.
    That is, there is another ``wall'' along the horizontal direction (See Figure \ref{fig:taudependence}).
    However, if the order of the zero is higher, such a ``horizontal wall'' is suppressed. The action density becomes more concentrated in an intermediate region. As a result, a domain-wall-like energy density is visible instead of the cross-shaped one. }.
    
\end{enumerate}

For a sufficiently large size of $T^2$, 
a set of $N$ different flavors of zeros close together can be regarded as a single semilocal vortex, of which size is roughly given by 
%the maximal distance 
 root mean square of their distances
in $T^2$ (not $\hat T^2$). 
 This can be easily observed by considering the case of $\mathbb R^2$ space with  $z=x_1+ix_2$. Since there is $SU(N)$ symmetry in the model,\footnote{ Introduction of a local $SU(N)$ breaking term~\cite{Eto:2009bz,Eto:2016mqc} might 
give the peak of the action density at each of the zeros.} the following set of zeros  
\begin{align}
   \vec h(z)\quad \stackrel{V}\sim\quad  (z-z_1,z-z_2,\dots,z-z_N)\,,
   %=\sqrt{N}(z-\langle z\rangle, c, 0,\dots,0)U,\quad  UU^\dagger={\bf 1},
\end{align}
gives an axial symmetric action density, 
\begin{align}
    \rho=\frac{2\xi|c|^2}{(|z-\langle z\rangle|^2+|c|^2)^2},\quad \langle z\rangle =\frac1N \sum_a z_a \,,
\end{align}
with a single peak and a size $|c|$ given by,
\begin{align}
    |c|^2= \frac1N \sum_{a=1}^N |z_a-\langle z\rangle|^2=\frac1{N^2} \sum_{b>a}|z_b-z_a|^2.
    \end{align}
In this sense, $k$ is the number of semilocal/local vortices, whereas $M(=Nk +p)$ is the number of zeros.
If $k=0$, i.e., for a minimal fractional instanton, since $M=p <N$, 
no local vortex appears, and thus no singularity appears in the moduli space.

%%%%%%%%%%%%%%%%%%%%%%%%%%%%%%%%%%%%%%%%%%%%%%%%%%%%%%%
%%%%%%%%%%%%%%%%%%%%%%%%%%%%%%%%%%%%%%%%%%%%%%%%%%%%%%%
\section{Global structure of the moduli space \texorpdfstring{$\mathcal{M}^{N;(p_s,p_c)}_k$}{}}
\label{Sec:modulispace}

\def\matC{\mathcal A}

In this section, we determine the moduli space $\mathcal{M}^{N;(p_s,p_c)}_k$ as a \kahler manifold. 
To fully determine the global structure, we need to give an alternative expression of BPS lumps by using the ``product-to-sum'' formula of the theta functions, which shall be explained later. 
 
From the expression~\eqref{eq:general_formula_moduli} of the moduli space, we can readily show that $\mathcal{M}^{N;(p_s,p_c)}_k$ is a fiber bundle with the base space $T^2\simeq \mathbb{C}/(\frac{N}{M}\mathbb{Z}+\frac{N}{M}\tau \mathbb{Z})\ni \langle z\rangle $ and the fiber $\mathbb{C}P^{M-1}$: 
\begin{align}
    \mathbb{C}P^{M-1}\,\hookrightarrow\, \mathcal{M}^{N;(p_s,p_c)}_k \to \frac{\mathbb{C}}{\frac{N}{M}\mathbb{Z}+\frac{N}{M}\tau \mathbb{Z}}. 
    \label{eq:modli_CPfiber}
\end{align}
To see this fiber-bundle structure, let us note that the holomorphic transition function $(v_s(z),v_c(z))$ can be taken as \eqref{eq:holomorphic_transition_function_generalpspc}, which depends only on $\langle z\rangle$ among the moduli parameters. 
Thus, we can have the map from $\mathcal{M}^{N;(p_s,p_c)}_k$ to the base manifold by extracting $\langle z\rangle$ out of the transition function $v_c(z)=-2\pi i \frac{M}{N}(z-\langle z\rangle)+\mathrm{const.}$ with $M=Nk+p$. 
Since $e^{v_c(z)}$ is physically meaningful, we have $\langle z\rangle \sim \langle z\rangle +\frac{N}{M}$. 
We can also perform the $V$-transformation \eqref{eq:vsvcVtrf} by $v(z)=2\pi i z$ relating $\langle z\rangle \sim \langle z\rangle+\frac{N}{M}\tau$ without affecting $e^{v_s(z)}$, and we get $\mathbb{C}/(\frac{N}{M}\mathbb{Z}+\frac{N}{M}\tau \mathbb{Z})$ as the base manifold\footnote{Alternatively, this base space can be understood as follows.
The average of the zeros $\langle z\rangle$ is determined by, e.g., $   \langle z \rangle =   \frac{1}{\hat M } \sum_{n=1}^{\hat M} z_{(\bar a=1,n)}$.
Here, from its definition (\ref{eq:BPSlump_generalps_abar}), each zero $z_{(\bar a=1,n)}$ should be understood with (at least) the following identification: $z_{(\bar a,n)} \sim  z_{(\bar a,n)} + \frac{N}{\operatorname{gcd}(N,p_s)} (n + \tau m)$ with $n,m \in \mathbb{Z}$.
Thus, the average $\langle z\rangle$ is subject to the identification $\langle z\rangle \sim  \langle z\rangle + \frac{N}{M} (n + \tau m)$ and take a value in $\mathbb{C}/(\frac{N}{M}\mathbb{Z}+\frac{N}{M}\tau \mathbb{Z})$. %\green{(note: the precise equivalence relation should be something like (\ref{eq:equi_rel_zn_ps1}), but this becomes cumbersome in general $p_s$; here only `necessary' identification is enough)}
}. 
Once we fix $\langle z\rangle$ to identify the fiber, \eqref{eq:general_formula_moduli} says 
\begin{align}
    &\quad \left.\mathcal{M}^{N;(p_s,p_c)}_k\right|_{\langle z\rangle: \text{fixed}} \notag\\ 
    &=\frac{\left\{\vec{h}(z) \,\Big|\, \vec{h}(z+1)= e^{v_s(z)} \vec h(z) S^{p_s}, \quad 
        \vec h(z+\tau)= e^{v_c(z)} \vec h(z) C^{p_c} \right\}\setminus\{\vec{0}\}}{\mathbb{C}^*}, 
\end{align}
where the $\mathbb{C}^*$ quotient comes from the remnant of the $V$-equivalence relation after fixing $v_s(z)$ and $v_c(z)$. 
Since the conditions of $\vec{h}$ are linear, the numerator becomes the complex vector space with removing the origin. This is nothing but the complex projective space, which has to be $\mathbb{C}P^{M-1}$ by the dimensional reasoning, and we find \eqref{eq:modli_CPfiber}. 

% Under a lump solution \red{on non-compat base spaces like $\mathbb R^2$ or $\mathbb R \times S^1$}, the position of each lump can be defined as a point that maps to an antipodal point relative to a point of the vacuum in the target space $\mathbb CP^{N-1}$,
% where zeros and a pole at $z=\infty$ of a certain entry of $\vec h(z)$ correspond to the positions and the vacuum, respectively. However, in a compact spatial slice such as the torus $T^2$, 
% the definition of the vacuum itself is ambiguous, and therefore the position of each lump must also be ambiguous.
% %The center of the zeros is generally unrelated to the center of mass in this case.

In Sec.~\ref{sec:BPSlumps_formula_by_zeros}, we have obtained an explicit formula of the BPS lumps by specifying the zeros of $h_a(z)$, and the moduli parameter is identified as $\{c_{\bar{a}}, z_{(\bar{a},n)}\}_{\bar{a}=1,\ldots,\gcd(N,p_s);n=1,\ldots, \hat{M}}$. 
With this expression, however, it is very unclear how these moduli parameters form the $\mathbb{C}P^{M-1}$ manifold after fixing $\langle z\rangle$. 
Moreover, the description by zeros is not suitable to understand the modular duality as the modular transformation does not preserve zeros. 
This is why we look for an alternative expression in the following.

Let us make a remark that BPS lumps on the periodic torus have been investigated in several studies, and see, e.g., \cite{Richard:1981bv, Speight:1997zc, Speight:2001he, Aguado:2001xg, Nakamula:2012hi}.
In particular, Ref.~\cite{Aguado:2001xg} determines the moduli space by constructing all solutions of BPS instantons on the periodic torus, and our following discussion include their result for the case of the periodic boundary condition.

% we discussed that a special arrangement of $N$ zeros generates a sharp energy peak, but in general, 
% zeros of $h_a(z)$ and energy peaks are often unrelated.
% Furthermore, generally, the duality transformation does not maintain the information of the zeros. 
% It is not necessary to stick to the set of zeros as parameters for describing the moduli space.

\subsection{Product-to-sum formula of the theta functions}
\label{sec:PtS_formula_alternative_expression}

Let us give a nice analogy that explains the basic idea for the alternative representation. 
The degree-$k$ polynomials form the $\mathbb{C}$-vector space with the basis $\{1,z,\cdots, z^k\}$. Even if we specify the degree-$k$ polynomials by its $k$ zeros, $\{z_1,\ldots, z_k\}$, it can be expanded as 
\begin{align}
a_0\prod_{n=1}^k(z-z_n)= \sum_{n=0}^k (-1)^n a_n z^{k-n}. \label{eq:concept_repara_monic}
\end{align}
Here, $\{z_n\}/S_k$ and $\{a_n/a_0\}$ are isomorphic to each other via the elementary symmetric polynomials when $a_0\not=0$. 
Although coordinate systems such as $\{z_n\}/S_k$ may exhibit conical singularities, if we choose a coordinate system based on the moduli parameters that appear linearly in $\vec h(z)$, we can ensure that the metric defined by Eq.~\eqref{eq:kahlermetric} does not have any coordinate singularities on the moduli space ${\cal M}_k^{N;(p_s,p_c)}$.
In this section we will rather discuss that it is better to consider such a linearized coordinate system for the moduli space of the solutions 
to understand the topology of the moduli space in terms of linear algebra.

For $K\in \mathbb Z_{>0}$, let us define a vector space of entire functions, 
\begin{align}
    {\cal F}_{K}(\tau)[z]:=\{f(z):\text{entire}\, |\, f(z+1)=(-1)^Kf(z),\, f(z+\tau)=(-q^{-1}e^{-2\pi i z})^K f(z)\}. 
    \label{eq:FKelement}
\end{align}
% ${\cal F}_{K}(\tau)[z]$ with 
%  where  an element $f(z) \in {\cal F}_{K}(\tau)[z]$ is an entire function of $z$ and 
% satisfies the following pseudo-periodic boundary conditions:  
% \begin{align}
% f(z+1)=(-1)^Kf(z),\quad f(z+\tau)=(-q^{-1}e^{-2\pi i z})^K f(z), \label{eq:FKelement}
% \end{align}
As we have seen in Sec.~\ref{sec:BPSlumps_formula_by_zeros}, 
\begin{align}
  \prod_{n=1}^K \theta_1(z-z_n|\tau) \,  \in\, {\cal F}_{K}(\tau)[z-\langle z \rangle],
\end{align}
with $\langle z \rangle =\frac1{K} \sum_{n=1}^Kz_n$. 
The Fourier expansion of $f(z)\in {\cal F}_{K}(\tau)[z]$ is given by  
\begin{align}
    f(z)=\sum_{m\in \mathbb Z} c_m \, e^{ 2\pi i \left(m-\frac{K}2\right)  z},
\end{align}
 with a certain series $\{c_m\in \mathbb C| m\in\mathbb Z \}$ satisfying $c_{m+K}=(-1)^K q^{2m}  c_m$, and we find $ {\cal F}_{K}(\tau)[z] \simeq \mathbb C^K$. 
Let us choose the following basis of ${\cal F}_{K}(\tau)[z]$ denoted by $\{\vartheta_1^K, \vartheta_2^K,\cdots \vartheta_K^K\}$:
\begin{align}
    \vartheta^{K}_n(z|\tau)
    &:=\sum_{m\in \mathbb Z}q^{K\left(m+\frac12 +\frac{n-1}K\right)^2}e^{2\pi i K\left(m+\frac12 +\frac{n-1}K\right)\left(z-\frac12\right)} \notag\\
    &=q^{\frac{(n-1)^2}K}e^{2\pi i (n-1)(z-\frac12)}\theta_1\left(Kz+(n-1)\tau-\frac{K-1}2\Big| K\tau\right),
    \label{eq:basis}
\end{align}
which satisfies $\vartheta_{n+K}^K=\vartheta_n^K$. %One can check that these satisfy the conditions in \eqref{eq:FKelement}.

Therefore, a product of the theta functions can always be expanded by the above basis,
and in fact we can show that 
there is a product-to-sum formula for the theta functions as
\begin{align}
\prod_{n=1}^K\theta_1(z-z_n|\tau)=\sum_{n=1}^K a_n \, \vartheta^{K}_n(z-\langle z\rangle |\tau)\quad{\rm with~} 
a_n=\Theta_{n-1}(\vec w) \,,  \label{eq:product-to-sum-maintext}
\end{align}
where a vector $\vec w\in \mathbb C^K$ is defined as $(\vec w)_a=\langle z\rangle-z_a$ and 
$\Theta_{n-1}(\vec w)$ are functions giving a map from $S[(T^2)^K]/S_K$\footnote{$S[(T^2)^K]$ stands for the traceless part of $(T^2)^K$; for $\vec w \in S[(T^2)^K]$, the average vanishes $\sum_{n=1}^K(\vec w)_a= 0$.} to $\mathbb CP^{K-1}$ \cite{Yamazaki:2017ulc, GriffithsHarris1994}.
See Appendix \ref{sec:ThetaSpace} for the detail.

We can show this theta-function map from $S[(T^2)^K]/S_K$ to $\mathbb CP^{K-1}$ is bijective.
Indeed, with an arbitrary given set $\{a_n\}$, Eq.~\eqref{eq:FKelement} tells us that 
a ratio of the right-hand side to the left-hand side is always an elliptic function.
Therefore, by setting a set of zeros $\{z_n\}$ to a set of $K$ zeros of the right-hand side, the ratio must become constant.
That is, there always exists a map from $\{\{a_n\}/\mathbb C^*,\langle z\rangle\}$ to $\{z_n\}/S_K$.
Hence, this map from $S[(T^2)^K]/S_K$ to $\mathbb CP^{K-1}$ is bijective, whereas an explicit formula is not available, analogous to the lack of a general solution for quintic equations.

We note that  a shifting $z\to z+\Delta$ with certain special values of $\Delta$
gives  non-trivial automorphism  
on the basis $\{\vartheta_1^K,\vartheta_2^K,\cdots,\vartheta_K^K\}$:
\begin{align}
\vartheta^{K}_n\left(z+\frac1K\Big|\tau\right) &=-e^{\frac{2\pi i}K  (n-1)}\vartheta^{K}_n(z|\tau),
\label{eq:theta_auto1}\\
\vartheta^{K}_n\left(z+\frac{\tau}K\Big|\tau\right) &=-q^{-\frac1K}e^{-2\pi i z}\vartheta^{K}_{n+1}(z|\tau). 
\label{eq:theta_auto2}
\end{align}
The first identity directly guarantees linear independence of each element of the basis and so does the second one.
It is easy to check that the conditions \eqref{eq:FKelement} are reproduced by repeatedly applying the above equations $K$ times.

%%%%%%%%%%%%%%%%%%%%%%%%%%%%%%%%%%%%%%%%%
\subsection{Alternative expression of fractional BPS lumps and the moduli spaces}
\label{sec:modulispace_quotient}

% Using the product-to-sum formula for $K=k$, the solution of $\vec h(z)$ given in Eq.~\eqref{eq:sol2} with $p_c=0$ can be rewritten to
% \begin{align}
%     h_a(z)\quad \stackrel{V}\sim \quad \sum_{n=1}^k \matC^n{}_{a}\, \vartheta_n^k (z-\langle z\rangle | \tau). 
% \end{align}
% with a certain coefficients $\{\matC^n{}_a\} \in \mathbb CP^{Nk-1}$.  

Let us give an alternative expression of the fractional BPS lumps~ \eqref{eq:BPSlump_generalps_abar} to obtain the global structure of the moduli space. 
As we have discussed in Sec.~\ref{sec:BPSlump_generalps}, we only need to specify $h_{\bar{a}}(z)$ for $\bar{a}=1,\ldots, \gcd(N,p_s)$ as other flavors are completely determined by the $S$-twist boundary condition. 
Applying the product-to-sum formula with $K=\hat{M}=\frac{M}{\gcd(N,p_s)}$ for \eqref{eq:BPSlump_generalps_abar}, there exists $\mathbb{C}$-valued coefficients $\{\matC^n{}_{\bar a}\}$ such that 
\begin{align}
    h_{\bar{a}}(z) \quad \stackrel{V}\sim \quad \sum_{n=1}^{\hat M} \matC^n{}_{\bar a}\,  \vartheta_n^{\hat M}\left(\frac{\gcd(N,p_s)}{N}\left(z-\langle z\rangle-\frac{p_c(\bar{a}-1)}{M}\right)\Big| \frac{\gcd(N,p_s)}{N}\tau \right), 
\end{align}
for $\bar{a}=1,\ldots, \gcd(N,p_s)$. The remnant of the $V$-equivalence relation gives 
\begin{align}
    \mathbb{C}^*:\quad \{\langle z \rangle, \matC^n{}_{\bar a}\}\quad \sim\quad  \{ \langle z \rangle, \lambda \,\matC^n{}_{\bar a}\},
    %\quad{\rm with ~}\lambda\in \mathbb C^*.
\end{align}
with $\lambda \not =0$, and the coefficient $\{\matC^n{}_{\bar a}\}$ correctly parametrizes the fiber $\mathbb{C}P^{M-1}$ of Eq.~\eqref{eq:modli_CPfiber}. 
This formula indicates that the set of moduli parameters $\{c_a,\,z_{(\bar{a},n)}\}$ in \eqref{eq:BPSlump_generalps_abar} is mapped to the other parametrization $\{\langle z\rangle, \matC^n{}_{\bar a}\}$.
Conversely, the definition of  ${\cal F}_K[z]$ ensures that the above formula with arbitrary $\{\langle z\rangle, \matC^n{}_{\bar a}\}$ gives the general solution of $\vec h(z)$.
% That is, the set of equivalent classes $[\{c_a,\,z_{(a,n)}\}]$ 
% is isomorphic to the set of $[\{ \matC^n{}_a, \langle z \rangle\}]$ with equivalent relations discussed in
% Sec.\ref{sec:ps0} and the next, which gives the moduli space ${\cal M}^{N:(0,0)}_k$. 

The automorphisms~\eqref{eq:theta_auto1} and \eqref{eq:theta_auto2} give the following equivalence relation on $\{\langle z\rangle, \matC^n{}_{\bar a}\}$:
\begin{eqnarray}
    \mathbb Z_s:&\quad   \{ \langle z \rangle,\, \matC^n{}_{\bar a}\} \quad \sim &\quad \left\{\langle z \rangle +\frac{N}M, \,e^{2\pi i\frac{n-1}{\hat M}} \matC^n{}_{\bar a}\right\}, \label{eq:Zs} \\
\mathbb Z_c:&\quad    \{ \langle z \rangle,\, \matC^n{}_{\bar a}\}\quad \sim &\quad \left\{\langle z \rangle +\frac{N}M\tau , \,\matC^{n-\frac{N}{\gcd(N,p_s)}}{}_{\bar a}\, e^{2\pi i \frac{\bar a-1}{M}p_c}\right\}.\label{eq:Zc}
\end{eqnarray}
These define a tiny torus $\mathbb C/(\frac{N}{M}\mathbb Z+\frac{N}{M}\tau \mathbb Z)$ with monodromy, as described at the beginning of this section.
We can confirm that no other equivalence relation  exists for $\{ \langle z \rangle,\, \matC^n{}_{\bar a}\}$, which shall be proven in Appendix~\ref{sec:EqRel}.  
The above relations lead to
\begin{eqnarray}
   \frac{M}{\gcd(N,p_s)} \mathbb Z_s:&\quad   \{ \langle z \rangle,\, \matC^n{}_{\bar a}\} \quad \sim &\quad \left\{\langle z \rangle +\frac{N}{\gcd(N,p_s)}, \, \matC^n{}_{\bar a}\right\},  \\
    \frac{M}{\gcd(N,p_c)} \mathbb Z_c:&\quad    \{ \langle z \rangle,\, \matC^n{}_{\bar a}\}\quad \sim &\quad \left\{\langle z \rangle +\frac{N}{\gcd(N,p_c)}\tau , \,\matC^n{}_{\bar a} \right\}.
\end{eqnarray}
Therefore  $\langle z \rangle$ parametrizes the doubly-extended torus: \footnote{
The group actions $\mathbb Z_{s,c}$ are equivalent to $\mathbb Z$ as a group, but they should not be confused with the lattice defining the torus. They differ by a constant factor:
\begin{align}
    \mathbb Z_s\times \mathbb Z_c \simeq \frac{N}{M}\mathbb Z+\frac{N}{M}\tau \mathbb Z\, (\hbox{as a lattice on }\mathbb C).
\end{align}
}
\begin{align}
    \doublehat{T}^2 := \frac{\mathbb C }{\frac{N}{\gcd(N,p_s)}\mathbb{Z}+\frac{N}{\gcd(N,p_c)}\tau \mathbb{Z}},
\end{align}
and the total moduli space is expressed as
\begin{align}
{\cal M}^{N;(p_s,p_c)}_k\simeq   \frac{\mathbb C P^{M-1}\times \mathbb C}{\mathbb Z_s\times \mathbb Z_c} %\,\red{\left(=\mathbb C P^{M-1}\rtimes \frac{\mathbb C}{\frac{N}{M}\mathbb Z+\frac{N}{M}\tau \mathbb Z}\right)} \,
=\frac{\mathbb C P^{M-1}\times \doublehat{T}^2}{\mathbb Z_{M/\gcd(N,p_s)}^s\times \mathbb Z_{M/\gcd(N,p_c)}^c}, \label{eq:moduli_space_ps_pc}
\end{align}
where
\begin{align}
     \mathbb Z_{M/\gcd(N,p_s)}^s:=\mathbb Z_s/ \left(\frac{M}{\gcd(N,p_s)}\mathbb Z_s\right),\quad
    \mathbb Z_{M/\gcd(N,p_c)}^c:=\mathbb Z_c/ \left(\frac{M}{\gcd(N,p_c)}\mathbb Z_c\right).
\end{align}
This result (\ref{eq:moduli_space_ps_pc}) indeed indicates the $\mathbb C P^{M-1}$-fiber structure over the tiny torus, as given in (\ref{eq:modli_CPfiber}).

Note that this moduli space is derived from the Abelian-Higgs model, where the small-lump singularity of the original $\mathbb{C}P^{N-1}$ model is smoothed out 
thanks to the appearance of a local vortex. 
Therefore, the presented moduli space should be interpreted as the closure of the true moduli space for the $\mathbb{C}P^{N-1}$ model, since the small-lump singularities appear as the boundary points of the original moduli space.
To obtain the accurate moduli space of the $\mathbb{C}P^{N-1}$ model, we need to eliminate the small-lump singularities.

Let us temporarily consider the $p_s=p_c=0$ case for simplicity.
A (trivial) necessary and sufficient condition for the regular lump solution can be stated as follows.
We define the map $\vartheta:\mathbb{C} \rightarrow \mathbb{C}^k$ by $z \mapsto (\vartheta_1^k(z|\tau),\cdots,\vartheta_k^k(z|\tau))$.
We also regard the coefficients as a matrix $\mathcal{A}:  \mathbb{C}^k \rightarrow  \mathbb{C}^N$.
The necessary and sufficient condition is $\operatorname{Im}(\vartheta) \cap \operatorname{Ker} (\mathcal{A}) = \emptyset$.
This is nothing but the moduli space on the periodic $T^2$ obtained in Ref.~\cite{Aguado:2001xg}.
That condition should be able to be described only by the matrix, $\mathcal A$.
A necessary condition for the lump moduli space is ${\rm rank}\matC>1$, otherwise all flavors are proportional to the same function, and thus all vortices are of local type.
In the case of $1<k\le N$, a sufficient condition is ${\rm rank} \matC=k$ since
$\{\vartheta^k_a(z)\}$ have no common zero. However, the exact necessary and sufficient condition in terms of the matrix $\matC$ only is still unclear.
For the $N=1$ or $k=1$ case, there are no corresponding lump solutions in the $\mathbb CP^{N-1}$ model on the torus. For the $N\not =1$ and $k=2$ case, the necessary and sufficient condition is just 
${\rm rank} \matC=2$.
We leave the explicit removal of these singular points as a task for future investigation.

\subsection{Modular transformation on the moduli space for some simple cases}\label{sec:DFT}

Under a conformal transformation $z\mapsto z'=z/\tau$ with $\tau\mapsto -1/\tau$, 
there must be an isomorphic mapping on the solutions and the moduli space discussed above with exchanging $(p_s,p_c)$ and $(p_c,p_s)$. In the following, we give this explicit modular transformation for the case of $(p_s,p_c)=(0,0)$ and that of $(p_s,p_c)=(1,p), (p,1)$. 

\subsubsection{Explicit modular transformation for \texorpdfstring{$(p_s,p_c)=(0,0)$}{(ps,pc)=(0,0)}}

First, let us consider the explicit form of the modular duality on the periodic torus $(p_s,p_c)=(0,0)$.
The first observation is the isomorphism between the spaces of pseudo-periodic holomorphic functions
\begin{align}
{\cal F}_K(\tau)[z]&\simeq {\cal F}_K\left(-\frac1{\tau}\right) \left[ \frac{z}\tau \right],\notag \\
f(z)&\mapsto f(z)e^{\frac{\pi i}\tau Kz^2}\,,
\end{align}
which can be shown by checking the pseudo-periodic boundary conditions.
%\begin{align}
%\forall f(z) \in {\cal F}_K(\tau)[z]    \quad \Leftrightarrow  \quad  f(z)e^{\frac{\pi i}\tau Kz^2}\in  {\cal F}_K\left(-\frac1{\tau}\right) \left[ \frac{z}\tau \right],
%\end{align}

This isomorphism indicates that there must also be a linear isomorphism from the basis of $ {\cal F}_K(\tau)[z] $ to that of ${\cal F}_K\left(-\frac1{\tau}\right) \left[ \frac{z}\tau \right]$.
Note that in Eqs.~\eqref{eq:Zs} and  \eqref{eq:Zc}, 
there appears 't Hooft twist not on the flavor space $\mathbb C^N$ but on the internal space $\mathbb C^k$ as a result of the map $S[(T^2)^k]/S_k$ to $\mathbb CP^{k-1}$ (Appendix \ref{sec:ThetaSpace}). 
Therefore, we can use the same analogy discussed in Sec.~\ref{sec:duality} and that is,
such a linear mapping between those two bases must also be a DFT.
In fact, we find the following identity:
\begin{align}
    \frac1{\sqrt{K}}\sum_{l=1}^Ke^{-2\pi i \frac{(n-1)(l-1)}K}\vartheta^{K}_l(z|\tau)
    =\sqrt{\frac{i}\tau} \left(ie^{-\pi i \frac{z^2}\tau}\right)^K \vartheta_n^K\left(\frac{z}\tau \Big|- \frac1\tau\right),
\end{align}
which is an extended version of 
Jacobi's imaginary transformation
\begin{align}
\theta_1(z|\tau)=i \sqrt{\frac{i}\tau} e^{-\frac{\pi i}\tau z^2} \theta_1\left(\frac{z}\tau\Big|-\frac1{\tau}\right).
\end{align}
See Appendix \ref{sec:DualityFormulas} for the proof.
As a result, we find that under the duality transformation the moduli parameters transform as
\begin{align}
    \mathfrak{D}: \{\matC^n{}_a, \langle z \rangle\} \quad \mapsto \quad 
    \left\{\matC_D^n{}_a, \frac{\langle z \rangle}\tau\right\}
\end{align}
where $\matC_D^n{}_a$ is obtained by DFT  as
\begin{align}
    \matC_D^n{}_a= \frac1{\sqrt{k}}\sum_{l=1}^ke^{2\pi i \frac{(n-1)(l-1)}k}\matC^l{}_a.
\end{align}

\subsubsection{Explicit modular transformation between \texorpdfstring{$(p_s,p_c)=(1,p), (p,1)$}{(ps,pc)=(1,p),(p,1)} }

Let us consider the modular duality between $(p_s,p_c)=(1,p)$ and $(p_s,p_c)=(p,1)$ with gcd$(N,p)=1$.  
For this purpose, we need to write down the formula for $h_a(z)$ with all the flavor indices $a=1,\ldots, N$: 
\begin{align}
    h_a(z) \quad \stackrel{V}\sim \quad \sum_{n=1}^{\hat M} \tilde{\matC}^n{}_{\bar r(a)}\,  \vartheta^{k,N;(p_s,p_c)}_{n,a}(z-\langle z\rangle|\tau),
\end{align}
where $\vartheta^{k,N;(p_s,p_c)}_{n,a}$ is defined by
\begin{align}
\vartheta^{k,N;(p_s,p_c)}_{n,a}(z|\tau)
&:= e^{\pi i \frac{u(a)}N}\vartheta_n^{\hat M}\left(\frac{\gcd(N,p_s)}{N}\left(z-\frac{u(a)}{M}\right)\Big| \frac{\gcd(N,p_s)}{N}\tau \right)\,,
   % &:= \theta_{n,\hat s(a)}^{\hat M,\hat N}\left(z-\frac{(\bar r(a)-1) p_c}{ M} \Big|\tau\right)e^{\frac{\pi i}N p_c(\bar r(a)-1)}\\
   %  &=q^{\frac{(n-1)^2}{\hat M \hat N}}e^{2\pi i (n-1)\left(\frac{z}{\hat N}-\frac{u(a)}{\hat N M}-\frac12\right)
   % +\frac{\pi i}N  u(a)  } \nonumber\\
   %  &\qquad \times \theta_1\left( \frac{M z- u(a)+\bar N(n-1)\tau}N-\frac{\hat M-1}2\Big| \frac{M}N \tau\right)
\end{align}
with an integer $u(a)$ defined by
\begin{align}
    u(a):= M(\hat s(a)-1)+p_c(\bar r(a)-1).
\end{align}
Let us write $\vartheta_{n,a}(z):=\vartheta^{k,N;(p_s,p_c)}_{n,a}(z|\tau)$ briefly.
Using the identities for $\vartheta_n^K$ and $u(a)$, we can derive the following identities for $\vartheta_{n,a}(z)$.
This function has  the periodicity on the indices
\begin{align}
   \vartheta_{n+\hat M,a}(z) =  \vartheta_{n,a}(z),\quad\vartheta_{n,a+N}(z)=\vartheta_{n,a}(z),
\end{align}
and satisfies the pseudo-periodic boundary conditions
\begin{align}
% {\cal T}_s: \quad 
\vartheta_{n,a}(z+1)&= e^{\pi i\frac{M}N }\,\vartheta_{n,a-p_s}(z), \label{eq:theta-ppdc1}\\ 
% {\cal T}_c:\quad   
\vartheta_{n,a}(z+\tau)&=\left(-q^{-\frac1{\hat N}}e^{-2\pi i\frac{z}{ \hat N}}\right)^{ \hat M }e^{\frac{2\pi i}N p_c(a-1)}\,\vartheta_{n,a}(z).\label{eq:theta-ppdc2}
\end{align}

Since ${\rm gcd}(N,p_s)={\rm gcd}(N,p_c)=1$, $\hat M=M(=Nk+p)$ and $\bar{r}(a)=1$ for any $a$. 
Thus, we  may drop the label $\bar{r}(a)$ for $\tilde{\matC}^n{}_{\bar r(a)}$, and solutions on the original and dual coordinate can be written as
\begin{align}
    h_a(z) \quad \stackrel{V} \sim& \quad \sum_{n=1}^M \matC_{n}\, \vartheta_{n,a}^{k,N;(1,p)}\left(z-\langle z\rangle\Big|\tau\right),\\
    h^D_a(z)\quad \stackrel{V} \sim& \quad \sum_{n=1}^M \matC_{n}^D\, \vartheta_{n,a}^{k,N;(p,1)}\left(z_D-\langle z\rangle_D\Big|\tau_D\right),
\end{align}
with $z_D:=z/\tau, \tau_D:=-1/\tau$.

With reference to the discussion in Sec.~\ref{sec:DFT}, we can derive the following formula for the transition between the two bases for those two cases:
\begin{align}
 &\,  \frac{1}{\sqrt{NM}}\sum_{n=1}^M e^{\frac{2\pi i}{M} \nu(n-1)(m-1)}\sum_{a=1}^{N} e^{-\frac{2\pi i}N (a-1)(b-1)}\,\,\vartheta_{n,a}^{k,N;(1,p)}\left(z+\frac{N-1}2\Big|\tau\right)\nonumber \\
 =&\sqrt{\frac{i}\tau} \left(-i\,
 e^{-\frac{\pi i}{\tau}z^2}\right)^{\frac{M}N} \,\vartheta^{k,N;(p,1)}_{m,b}\left(\frac{z}\tau+\frac{N+1}2\Big| -\frac1\tau\right),
\end{align}
where the integer $\nu$ is defined by
\begin{align}
    \nu:= k q_{\rm b}+r_{\rm b}\quad \in \mathbb Z_{>0}\,,
\end{align}
with  a pair of integers $q_{\rm b}, r_{\rm b}$ satisfying $q_{\rm b} p-r_{\rm b} N=1, (1\le q_{\rm b}<N)$. See Appendix \ref{sec:DualityFormulas} for the proof.
Note that 
the operation multiplying $\nu$ on $\mathbb Z_M$ gives a bijective map as, for $m,m'\in \mathbb Z_M$
\begin{align}
    m' \equiv -\nu (m-1)+1 \quad {\rm mod~} M\quad \Leftrightarrow \quad   m \equiv N (m'-1)+1 \quad {\rm mod~} M
\end{align}
since $\nu$ is the inverse of $-N$ on $\mathbb Z_M$ as
\begin{align}
    q_{\rm b} M- \nu N= q_{\rm b} p-r_{\rm b} N=1.
\end{align}

Using the above formula, 
we find the transition functions between $\{\matC_m, \langle z\rangle \}$ and 
$\{\matC_m^D, \langle z\rangle_D \}$ as
\begin{align}
    \matC_{m}^D &= \frac{1}{\sqrt{M}}\sum_{n=1}^M e^{-\frac{2\pi i}M \nu (n-1)(m-1)} \matC_n,\\
    \langle z\rangle_D&= \frac1\tau \left({\langle z\rangle}+\frac{N-1}2\right)-\frac{N+1}2.
\end{align}

\section{From 4d Yang-Mills theory to 2d \texorpdfstring{$\mathbb{C}P^{N-1}$}{CP{N-1}} model}
\label{sec:4d2d}

Before concluding this paper, let us turn to a different aspect. In this section, we give another viewpoint on fractional instantons in the 2d $\mathbb{C}P^{N-1}$ model from $4$d $SU(N)$ Yang-Mills theory.

In Section~\ref{sec:4d2d_setup}, %\ref{sec:4d2d_reduction}, and \ref{sec:4d2d_holonomy_to_CPN}, 
we review the connection between $4$d $SU(N)$ Yang-Mills theory and $2$d $\mathbb{CP}^{N-1}$ model, based on the dimensional reduction by Yamazaki and Yonekura \cite{Yamazaki:2017ulc}.
This dimensional reduction will also clarify why the twist by the clock and shift matrices is referred to as 't Hooft twist for $\mathbb{C}P^{N-1}$.
It should be noted that the metric of the resulting $\mathbb{C}P^{N-1}$ model is not the Fubini-Study metric; however, the following discussion suggests the potential significance of our results for fractional instantons in $4$d Yang-Mills theory.

In Section \ref{sec:torons_modulispace}, we will further discuss the moduli space of fractional instantons from this perspective.
It is known that, by tuning the aspect ratio of the torus $T^4$, fractional instantons in 4d Yang-Mills theory on $T^4$ that saturate the BPS bound can be constructed using Abelianized fields (known as 't Hooft toron \cite{tHooft:1981nnx, vanBaal:1984ar}). Using this fact, BPS solutions can also be constructed in the $\mathbb{C}P^{N-1}$ model obtained via dimensional reduction. We will then examine the moduli space of these 't Hooft-like solutions.
As the main focus of this paper is the standard $\mathbb{C}P^{N-1}$ model, we will confine our investigation here to simple cases $(k=0,p_s=p,p_c=1)$ and demonstrate that the topologies of the moduli spaces are identical to those for the standard $\mathbb{C}P^{N-1}$ model, obtained in the previous section. 
This observation also indicates that our qualitative results, for instance, the topology of the moduli space, are robust to changes in the metric.

\subsection{Setup: \texorpdfstring{$SU(N)$}{SU(N)} Yang-Mills theory on \texorpdfstring{$T^4$}{T4} with an intersecting 't Hooft flux}
\label{sec:4d2d_setup}

We begin with the $4$d $SU(N)$ Yang-Mills theory on $T^4 = S^1_{A} \times S^1_{B} \times S^1_{C} \times S^1_{D} $ with an intersecting 't Hooft flux for $S^1_{A} \times S^1_{D}$ and $S^1_{B} \times S^1_{C}$.
The setup is detailed as follows:
\begin{itemize}
    \item We assume $T^2_{CD}=S^1_{C} \times S^1_{D}$ is a small torus:
    \begin{align}
        L_C,L_D \ll L_A,L_B,
    \end{align}
    where $L_*$ is the length of the circle $S^1_*$ (with $* = A,B,C,D$, respectively).

    \item We insert two 't Hooft fluxes on $S^1_{A} \times S^1_{D}$ and $S^1_{B} \times S^1_{C}$.

    This setup corresponds to inserting a $\mathbb{Z}_N^{[1]}$ symmetry background $B$ such that 
    \begin{align}
        \int_{S^1_{A} \times S^1_{D}} B = \frac{2\pi}{N}, ~~\int_{S^1_{B} \times S^1_{C}} B = \frac{2\pi}{N}.
    \end{align}
    These two fluxes intersect, and the quantization of the topological charge is modified as~\cite{vanBaal:1982ag}
\begin{align}
    Q_{\mathrm{top}} =\frac{1}{8\pi^2} \int_{T^4}\operatorname{tr} \left( (\Tilde{f} - B) \wedge (\Tilde{f} - B)\right) \in - \frac{1}{N} + \mathbb{Z},
\end{align}
where we have promoted the original $SU(N)$ field strength $f$ to the $U(N)$ field strength $\Tilde{f}$ such that $\operatorname{tr} \Tilde{f} = NB$ due to the existence of the $\mathbb{Z}_N^{[1]}$ symmetry background.

\end{itemize}

In this setup, the solution satisfying the BPS bound is the fractional instanton: $S= \frac{8\pi^2}{Ng^2},~Q_{\mathrm{top}} = - \frac{1}{N}$.
In other words, these twists force the theory to have a fractional topological charge, and we can study various properties of fractional instantons in this setup.

\subsubsection{Dimensional reduction from 4d to 2d}
\label{sec:4d2d_reduction}

Since the torus $T^2_{CD}=S^1_{C} \times S^1_{D}$ is chosen to be small, the 4d Yang-Mills theory on $T^4$ will be effectively described by a 2d theory on $T^2_{AB}=S^1_{A} \times S^1_{B}$. 
We first review a simple dimensional reduction and then take into account the effect of the 't Hooft twist.

We ignore all higher Kalzua-Klein modes on the small torus $T^2_{CD}$.
The gauge field components in the $S^1_C$ and $S^1_D$ directions are represented solely by the holonomies $(P_C,P_D)$, which are adjoint scalars from the $2$d perspective.
Thus, the small compactification of $T^2_{CD}$ leads to the reduction:
\begin{align}
    4\mathrm{d}&~SU(N)~ \mathrm{gauge~on}~T^4 \notag \\
    &\Rightarrow (2\mathrm{d}~SU(N)~ \mathrm{gauge~} + \mathrm{two~adjoint~scalars}~P_C,P_D)~\mathrm{on}~~T^2_{AB}
\end{align}
From the flatness condition $F_{CD}=0$ on $T^2_{CD}$, the holonomies $P_C, P_D$ must commute with each other, 
\begin{align}
    P_C P_D = P_D P_C. 
\end{align}
The commuting holonomies can be parameterized by their eigenvalues (or equivalently, can be represented in the Polyakov gauge):
\begin{align}
    P_C = \operatorname{diag}(\rme^{\im \phi^{(C)}_1},\cdots,\rme^{\im \phi^{(C)}_N}),~~     P_D = \operatorname{diag}(\rme^{\im \phi^{(D)}_1},\cdots,\rme^{\im \phi^{(D)}_N}),
\end{align}
with the constraint $\sum_{i=1}^N \phi^{(C)}_i = \sum_{i=1}^N \phi^{(D)}_i = 0$.
Here, the holonomies $(\Vec{\phi}^{(C)}, \Vec{\phi}^{(D)})$ are identified up to the Weyl permutation and $2\pi$ periodicity: 
\begin{align}
     &(\phi^{(C)}_i, \phi^{(D)}_i) \sim (\phi^{(C)}_{\sigma(i)}, \phi^{(D)}_{\sigma(i)}),~~(\sigma \in S_N), \notag \\
     &\phi^{(C)}_i \sim \phi^{(C)}_i + 2\pi,~~~\phi^{(D)}_i \sim \phi^{(D)}_i + 2\pi,
\end{align}
where $S_N$ is the symmetric group.
The holonomy space, or the space of flat connections on $T^2_{CD}$, can be represented as
\begin{align}
    \mathcal{M}_{\mathrm{holonomy}} = \frac{U(1)^{N-1} \times U(1)^{N-1}}{S_N}\,.  \label{eq:holonomy_space}
\end{align}

At almost all points of the holonomy space, the adjoint higgsing occurs and the low-energy gauge group is given by the Cartan subgroup: $SU(N) \rightarrow U(1)^{N-1}$.
Then, the 2d $U(1)^{N-1}$ gauge field decouples, so the nontrivial part of the classical equation of motion only involves the holonomy variables $(\Vec{\phi}^{(C)}, \Vec{\phi}^{(D)})$. 
Thus, we have the 2d effective theory of the holonomies by the dimensional reduction.
\begin{align}
    4\mathrm{d}&~SU(N)~ \mathrm{gauge~on}~T^4 \notag \\
    &\Rightarrow (2\mathrm{d}~U(1)^{N-1}~ \mathrm{gauge~} + \mathcal{M}_{\mathrm{holonomy}}~\textrm{-valued~non-linear~sigma~model})~\mathrm{on}~~T^2_{AB}, \notag \\
    &\approx 2\mathrm{d}~\mathcal{M}_{\mathrm{holonomy}}~\textrm{-valued~non-linear~sigma~model}~\mathrm{on}~~T^2_{AB}
\end{align}

Next, let us take the 't Hooft twist into account.
In our setup, the 't Hooft twists are imposed on the tori $S^1_{A} \times S^1_{D}$ and $S^1_{B} \times S^1_{C}$.
The 't Hooft twist is regarded as the twist by the 1-form symmetry $\left( \mathbb{Z}_N^{[1]} \right)_{\mathrm{4d}}$.
Upon the $T^2_{CD}$ compactification, the 1-form symmetry $\left( \mathbb{Z}_N^{[1]} \right)_{\mathrm{4d}}$ is decomposed as follows in the 2d effective theory:
\begin{align}
    \left( \mathbb{Z}_N^{[1]} \right)_{\mathrm{4d}} \rightarrow \left( \mathbb{Z}_N^{[1]} \right)_{\mathrm{2d}} \times \left( \mathbb{Z}_N^{[0]} \right)_{\mathrm{2d}} \times \left( \mathbb{Z}_N^{[0]} \right)_{\mathrm{2d}},
\end{align}
where $\left( \mathbb{Z}_N^{[1]} \right)_{\mathrm{2d}}$ acts on the Wilson loop on the large torus $T^2_{AB}$, and $\left( \mathbb{Z}_N^{[0]} \right)_{\mathrm{2d}} \times \left( \mathbb{Z}_N^{[0]} \right)_{\mathrm{2d}}$ act on the holonomies $(P_C,P_D)$, respectively.
In terms of the holonomy, the 't Hooft twist simply becomes the twist by the $\left( \mathbb{Z}_N^{[0]} \right)_{\mathrm{2d}}$.
Therefore, the 't Hooft twist on $S^1_{A} \times S^1_{D}$ becomes the twisted boundary condition on $P_D$ for $S^1_A$: $P_D (x_A+L_A,x_B) = \rme^{2 \pi \im /N} P_D (x_A,x_B)$.
Similarly, the 't Hooft twist on $S^1_{B} \times S^1_{C}$ is translated as $P_C (x_A,x_B+L_B) = \rme^{2 \pi \im /N} P_C (x_A,x_B)$.
With the parametrization $(\Vec{\phi}^{(C)}, \Vec{\phi}^{(D)})$, these twisted boundary conditions are summarized as follows:
\begin{align}
   &\begin{cases}
    (P_C(x_A+L_A,x_B), P_D(x_A+L_A,x_B)) = (P_C(x_A,x_B), \rme^{2 \pi \im /N} P_D(x_A,x_B) )  \\
    (P_C(x_A,x_B+L_B), P_D(x_A,x_B+L_B)) = (\rme^{2 \pi \im /N} P_C(x_A,x_B) , P_D(x_A,x_B) ),
    \end{cases}
    \notag \\
    & \Longleftrightarrow
    \begin{cases}
    (\Vec{\phi}^{(C)}(x_A+L_A,x_B), \Vec{\phi}^{(D)}(x_A+L_A,x_B)) = (\Vec{\phi}^{(C)}(x_A,x_B), \Vec{\phi}^{(D)}(x_A,x_B) - 2 \pi \Vec{\mu}_1)  \\
    (\Vec{\phi}^{(C)}(x_A,x_B+L_B), \Vec{\phi}^{(D)}(x_A,x_B+L_B)) = (\Vec{\phi}^{(C)}(x_A,x_B) - 2 \pi \Vec{\mu}_1, \Vec{\phi}^{(D)}(x_A,x_B) ),
    \end{cases}
    \label{eq:twisted_BC_4dto2d}
\end{align}
where $\Vec{\mu}_1$ is the fundamental weight:
\begin{align}
    \Vec{\mu}_1 = (1,0,\cdots,0) - \frac{1}{N} (1,1,\cdots,1).
\end{align}
Due to the $2\pi$ periodicity, the first term does not contribute; however, we included it to render the tracelessness manifest.

After the dimensional reduction, we have the following classical action for the 2d effective theory\footnote{The inclusion of one-loop quantum corrections generates a holonomy potential that destabilizes the center symmetry for pure Yang-Mills theory~\cite{Gross:1980br, Luscher:1982ma, vanBaal:1986ag}. 
% To keep the center symmetry at the quantum level, some specific prescription must be employed.
% For this purpose, Ref.~\cite{Yamazaki:2017ulc} uses the $\mathbb{R} \times S^1_B \times (S^1_C \times S^1_D)$ compactification with the 't Hooft flux to $S^1_B \times S^1_C$. 
However, we will work solely with the classical action because our current focus is to analyze the classical Yang-Mills equation on $T^4$.}:
\begin{align}
    S_{\mathrm{holonomy}} = \int \frac{L_C}{g^2 L_D} |d \Vec{\phi}^{(D)}|^2 + \frac{L_D}{g^2 L_C} |d \Vec{\phi}^{(C)}|^2.
\end{align}
For later convenience, let us define the complex variable $\Vec{\phi}$ to represent the holonomy $(\Vec{\phi}^{(C)}, \Vec{\phi}^{(D)})$
    \begin{align}
        \Vec{\phi} := \frac{1}{2\pi} \left( -\Vec{\phi}^{(D)} + \tau \Vec{\phi}^{(C)} \right), \label{eq:def_complex_phi}
    \end{align}
where $\tau = \im L_D/L_C$. Then, the $2 \pi$ periodicity is represented by $\phi_i \sim \phi_i + 1,~\phi_i \sim \phi_i + \tau$.
With this parametrization of $\mathcal{M}_{\mathrm{holonomy}}$, the 2d action can be rewritten as,
\begin{align}
    S_{\mathrm{holonomy}} = \frac{4\pi^2 L_C}{g^2 L_D} \int |d \Vec{\phi}|^2 \,.\label{eq:2daction_with_vecphi}
\end{align}

Incidentally, we can derive the BPS bound in this totally Abelianized regime as follows,
\begin{align}
    S_{\mathrm{holonomy}} &= \frac{L_C}{g^2 L_D} \int \left|\partial_\mu \Vec{\phi}^{(D)} \pm \frac{L_D}{L_C} \epsilon_{\mu \nu} \partial_\nu \Vec{\phi}^{(C)} \right|^2 \mp \frac{8 \pi^2}{g^2} \int \frac{ d \Vec{\phi}^{(C)} \wedge d\Vec{\phi}^{(D)}}{4\pi^2} \notag \\
    &\geq \frac{8 \pi^2}{g^2} |Q_{\mathrm{top}}|.
\end{align}
The bound is saturated if and only if
\begin{align}
    \partial_\mu \Vec{\phi}^{(D)} \pm \frac{L_D}{L_C} \epsilon_{\mu \nu} \partial_\nu \Vec{\phi}^{(C)} = 0.
\end{align}
This BPS equation can be simply expressed in terms of a complex variable:
\begin{align}
    \left( \partial_A \pm \im \partial_B \right) \Vec{\phi}= 0.
\end{align}
This equation tells that $\Vec{\phi}$ needs to be holomorphic $(-)$ or anti-holomorphic $(+)$ with respect to $x_A + \im x_B$ to saturate the BPS bound.

\subsubsection{An isomorphism from holonomy space to \texorpdfstring{$\mathbb{C}P^{N-1}$}{CP{N-1}}}
\label{sec:4d2d_holonomy_to_CPN}

In Ref.~\cite{Yamazaki:2017ulc}, an explicit map from the holonomy space to $\mathbb{C}P^{N-1}$ has been constructed. 
This map is defined in Appendix \ref{sec:YYmap_prodtosum}, which was utilized to express the product-to-sum formula. 
Although we do not write its explicit formula here as we do not use it in later sections, let us comment on some important features:
\begin{itemize}
    \item Fubini-Study metric has the $PSU(N)$ global symmetry, but the dimensional reduction of $4$d Yang-Mills theory only has a smaller symmetry, $\mathbb{Z}_N\times \mathbb{Z}_N\subset PSU(N)$. 
    Thus, it is natural that the target space metric is distorted so that the $PSU(N)$ symmetry is explicitly broken to $\mathbb{Z}_N\times \mathbb{Z}_N$, which are realized by the clock-shift matrices. 
    \item Moreover, the target space metric has a singularity due to the appearance of the extra massless mode at fixed points of some Weyl permutation. For the case of $\mathbb{C}P^1$, there are $4$ singular points with massless $W$ boson. 
\end{itemize}

In terms of the complex variable $\Vec{\phi}$ (\ref{eq:def_complex_phi}), we can apply the map in Appendix \ref{sec:YYmap_prodtosum}. 
Then, the isomorphism between $\mathcal{M}_{\mathrm{holonomy}}$ and $\mathbb{C}P^{N-1}$ is given by (\ref{eq:YY-map-K}) and (\ref{eq:def_large_Theta}) at $K=N$:
\begin{align}
    \mathcal{M}_{\mathrm{holonomy}} &\longrightarrow \mathbb{C}P^{N-1} \notag \\
    \Vec{\phi}  & \mapsto [h_1 = \Theta_1(\Vec{\phi}), \cdots, h_N = \Theta_N(\Vec{\phi})]. \label{eq:Yamazaki-Yonekura-map}
\end{align}
Hence, we can regard the theory of the two holonomies as the $\mathbb{C}P^{N-1}$ model.
Moreover, with the homogeneous coordinates $[h_1,\cdots,h_N] = [\Vec{h}]$, the twisted boundary conditions (\ref{eq:twisted_BC_4dto2d}) are translated as\footnote{From these relations, we could regard $x_A$ as $x_2$ and $x_B$ as $x_1$. However, we still use $(x_A,x_B,x_C,x_D)$ in this section.}:
\begin{align}
    &[\Vec{h}] (x_A+L_A,x_B) = [C^{-1} \Vec{h}] (x_A,x_B) , \notag \\
    &[\Vec{h}] (x_A,x_B + L_B) = [S^{-1} \Vec{h}] (x_A,x_B). \label{eq:CPN-1-twisted-BC-from4d}
\end{align}
Indeed, from the definition of (\ref{eq:def_large_Theta}), we have
\begin{align}
    [\Vec{h}] (x_A+L_A,x_B) &= [\Theta_1(\Vec{\phi} + \Vec{\mu}_1), \cdots, \Theta_N(\Vec{\phi}+ \Vec{\mu}_1)] \notag \\
    &= [\rme^{- \frac{2 \pi \im}{N}} \Theta_1(\Vec{\phi}), \rme^{- \frac{4 \pi \im }{N}} \Theta_2(\Vec{\phi}) \cdots, \Theta_N(\Vec{\phi})] \notag \\
    &= [\rme^{- \frac{2 \pi \im}{N}} C \Vec{h}] \sim [C^{-1} \Vec{h}],
\end{align}
with the clock matrix $C$.
Also, noting that
\begin{align}
    \Theta_k(\Vec{\phi} - \tau \Vec{\mu}_1) = \rme^{- \pi \im \tau \Vec{\mu}_1^2 + 2 \pi \im \Vec{\mu}_1 
 \cdot \Vec{\phi}} \Theta_{k-1}(\Vec{\phi}),
\end{align}
we get
\begin{align}
    [\Vec{h}] (x_A,x_B + L_B) &= [\Theta_1(\Vec{\phi} - \tau \Vec{\mu}_1), \cdots, \Theta_N(\Vec{\phi} - \tau \Vec{\mu}_1)] \notag \\
    &\sim [S^{-1} \Vec{h}].
\end{align}

As a summary of this review part, we have considered the 4d Yang-Mills theory on $T^4$ with 't Hooft twists so that the topological charge is in $-1/N + \mathbb{Z}$.
After the small compactification of $T^2_{CD}$, we have a 2d effective theory of holonomies on $T^2_{AB}$.
The space of the holonomies can be mapped to $\mathbb{C}P^{N-1}$, so the 2d effective theory can be expressed as the $\mathbb{C}P^{N-1}$ model (with an unusual metric).
Then, the 't Hooft twists are translated as the twisted boundary condition by the shift and clock matrices for the $\mathbb{C}P^{N-1}$ variable (\ref{eq:CPN-1-twisted-BC-from4d}).
This is also justifies to call the $S$ and $C$ twist as ``the 't Hooft twist for $\mathbb{C}P^{N-1}$ model''.

\subsection{'t Hooft torons, \texorpdfstring{$\mathbb{C}P^{N-1}$}{CP{N-1}} fractional instantons, and moduli space}
\label{sec:torons_modulispace}

Since the topological charge is in $-1/N + \mathbb{Z}$ in our setup, the minimal-action solution is the fractional instanton.
Although the metric in the $\mathbb{C}P^{N-1}$ variable is highly complicated, simple solutions can be written in terms of the holonomies $\Vec{\phi}$.
When the aspect ratio of the torus $\frac{L_AL_D}{L_BL_C}$ is set to a special value, the minimal action solution satisfying the BPS bound can be explicitly constructed.

The 4d origin of these solutions is 't Hooft's toron: 't Hooft constructed self-dual solutions of (Abelianized) constant field strength on the twisted torus \cite{tHooft:1981nnx, vanBaal:1984ar} with tuned aspect ratio.
For generic sizes of the torus, the self-dual solutions should be more non-Abelian; so the BPS solutions are not necessarily expressed only in terms of the low-energy Abelian degrees of freedom.
If the aspect ratio of the 4-torus is tuned at a special value, the self-dual fractional instanton becomes the (Abelian) constant field strength, which is included within the low-energy degrees of freedom.

Note that we can arbitrarily tune the aspect ratio of the torus $\frac{L_AL_D}{L_BL_C}$  while keeping the hierarchy $L_C,L_D \ll L_A,L_B$.
Assuming that tuning the aspect ratio does not change the global structure of the moduli space, we can observe the moduli space of the fractional instantons in the $\mathbb{C}P^{N-1}$ model.

Below, we consider the minimal solutions at $Q_{\mathrm{top}}=-1/N$ and generalizations at $Q_{\mathrm{top}}=-p/N$ with $0<p<N$. 
The following solutions are essentially the 2d reduced forms of 't Hooft's torons.

\subsubsection{Minimal solution at \texorpdfstring{$Q_{\mathrm{top}}=-1/N$}{Qtop=-1/N} and \texorpdfstring{$L_AL_D/L_BL_C=1/(N-1)$}{LA LD/LB LC=1/(N-1)}}

The minimal solution can be constructed as follows.
\begin{align}
    \vec{\phi}(x_A,x_B) = \frac{x_A}{N(N-1)L_A} 
    \begin{pmatrix}
N-1  \\
-1   \\
-1   \\
-1   \\
\vdots \\
-1
\end{pmatrix}
+
\frac{1}{N-1}
    \begin{pmatrix}
0  \\
0   \\
1   \\
2   \\
\vdots \\
N-2
\end{pmatrix}
- \tau \frac{x_B}{NL_B}     \begin{pmatrix}
N-1  \\
-1   \\
-1   \\
-1   \\
\vdots \\
-1
\end{pmatrix}\,. \label{eq:minimal_solution_holonomy}
\end{align}
This configuration trivially satisfies the equation of motion from the action (\ref{eq:2daction_with_vecphi}), so we will check the twisted boundary conditions.
The main trick is that the holonomies are identified up to the Weyl permutations as the holonomy space is $\mathcal{M}_{\mathrm{holonomy}} = \frac{U(1)^{N-1} \times U(1)^{N-1}}{S_N}$.

We define a particular Weyl permutation $P_W^{(N-1)}$, which permutes the latter $N-1$ components:
\begin{align}
P_W^{(N-1)}
    \begin{pmatrix}
\phi_1  \\
\phi_2   \\
\phi_3   \\
\vdots \\
\phi_{N-1}   \\
\phi_N
\end{pmatrix}
=
\begin{pmatrix}
\phi_1  \\
\phi_3   \\
\phi_4   \\
\vdots \\
\phi_N  \\
\phi_2
\end{pmatrix}\,.
\end{align}
Through some calculations, we have
\begin{align}
P_W^{(N-1)}  \vec{\phi}(x_A+L_A,x_B) &= \vec{\phi}(x_A,x_B) - \vec{\mu}_1 \notag \\
\vec{\phi}(x_A,x_B+L_B) &= \vec{\phi}(x_A,x_B) - \tau \vec{\mu}_1 \,,\label{eq:BC_minimal_sol_holonomy}  
\end{align}
modulo $\mathbb{Z} -\tau \mathbb{Z}$.
Since the holonomy space is the quotient by $S_N$, the given configuration (\ref{eq:minimal_solution_holonomy}) indeed satisfies the twisted boundary condition. 

This solution has the following action:
\begin{align}
    S_{\mathrm{holonomy}} = \frac{4 \pi^2}{N g^2} \left( \frac{L_B L_C}{ (N-1) L_A L_D}  + \frac{(N-1)L_A L_D}{ L_B L_C }  \right) \geq \frac{8 \pi^2}{N g^2},
\end{align}
and the BPS bound is saturated if $(N-1) L_A L_D = L_BL_C$.
Note that at this ratio, the solution \eqref{eq:minimal_solution_holonomy} only depends on $(x_A-\im x_B)$ and is thus an anti-holomorphic solution.
This BPS configuration corresponds to 't Hooft's constant-field-strength solution on the torus with the tuned aspect ratio.

Furthermore, once the boundary condition \eqref{eq:BC_minimal_sol_holonomy} (for the lift to $\mathbb{C}^{N-1}$) is specified, the expression \eqref{eq:minimal_solution_holonomy} becomes the general solution, modulo a constant. 
In other words, the only remaining degrees of freedom to keep the BPS bound correspond to the addition of anti-holomorphic functions, and any such function without a topological charge must be constant.
Hence, the moduli of this solution can be parametrized only by one complex number $c \in \mathbb{C}$,
\begin{align}
    \vec{\phi}(x_A,x_B) \rightarrow \vec{\phi}(x_A,x_B) + c \vec{\mu}_1,
\end{align}
which can be absorbed into translations of $(x_A,x_B)$.
Due to the $1/N$ factor in the solution, the moduli space is identified as the doubly extended torus: $\doublehat{T}^2$.

For the Yang-Mills fractional instanton with $Q_{\mathrm{top}}=\pm1/N$, the moduli space is four-dimensional.
There are two missing moduli compared to the one for the 4d Yang-Mills fractional instanton.
However, this discrepancy can be attributed to elimination of the 2d gauge field on $T^2_{AB}$ during the reduction to the 2d $\mathbb{C}P^{N-1}$ model.
Indeed, two moduli appear in the holonomies along $S_A^1$ and $S_B^1$ in the 't Hooft's solution, and thus it is consistent with our observation that they are not included within the 2d $\mathbb{C}P^{N-1}$ model.

\subsubsection{Minimal solution at \texorpdfstring{$Q_{\mathrm{top}}=-p/N$}{Qtop=-p/N} and \texorpdfstring{$L_AL_D/L_BL_C=1/(N-p)$}{LA LD/LB LC=1/(N-p)}}

Let us generalize the above discussion to the case with higher fractional charges.
If we introduce $p_c$ 't Hooft flux for $S_A^1 \times S_D^1$ and $p_s$ 't Hooft flux for $S_B^1 \times S_C^1$,
    \begin{align}
        \int_{S^1_{A} \times S^1_{D}} B = \frac{2\pi p_c}{N}, ~~\int_{S^1_{B} \times S^1_{C}} B = \frac{2\pi p_s}{N},
    \end{align}
then the topological charge is quantized as $Q_{\mathrm{top}}=-p/N + \mathbb{Z}$, where $p=p_s p_c$ mod $N$.

For simplicity, we only consider the cases with $p_c = 1,~p_s=p$.
Let us also temporarily assume $\operatorname{gcd}(N,p)=1$.
Then the boundary condition reads,
\begin{align}
[ \vec{\phi}(x_A+L_A,x_B)] &= [\vec{\phi}(x_A,x_B) - \vec{\mu}_1],\\
[\vec{\phi}(x_A,x_B+L_B)] &= [\vec{\phi}(x_A,x_B) - \tau p \vec{\mu}_1],
\end{align}
where $[\cdot]$ means that the identifications of the $2\pi$ periodicity and Weyl permutations are taken.

In Section \ref{sec:modulispace_quotient}, for $\operatorname{gcd}(N,p)=1$, we derived that the moduli space for the minimal fractional instanton is
\begin{align}
        \mathcal{M}^{N;(p,1)}_{0} \simeq \frac{\mathbb{C}P^{p-1} \times \doublehat{T}^2}{\mathbb{Z}_p^{c} \times \mathbb{Z}_p^{s}}\,.
\end{align}
Our aim here is to give another understanding of this moduli space via the reduction from the 4d Yang-Mills theory.

We use the following notations:
\begin{align}
       \vec{\textbf{0}}_{(\ell)} := \underbrace{(0,0,\cdots,0)}_\text{$\ell$ times} {}^T,~~~\vec{\textbf{1}}_{(\ell)} := \underbrace{(1,1,\cdots,1)}_\text{$\ell$ times}{}^T,~~~\vec{\rho}_{(\ell)}:= (0,1,2,\cdots,\ell-1)^T \,,
\end{align}
with $0< \ell <N$.
We also define the permutation of the latter $(N-p)$ components as,
\begin{align}
P_W^{(N-p)}&(\underbrace{\phi_1, \cdots,\phi_p}_\text{$p$}, \underbrace{\phi_{p+1}, \cdots, \phi_{N-2}, \phi_{N-1}}_\text{$N-p$} )^T \notag \\
&= (\underbrace{\phi_1, \cdots,\phi_p}_\text{$p$}, \underbrace{\phi_{p+2}, \cdots, \phi_{N-1},\phi_{p+1}}_\text{$N-p$} )^T.
\end{align}

Mimicking 't Hooft's construction of the constant-field-strength solutions, we can write the following configurations with $Q_{\mathrm{top}}=-p/N$ :
\begin{align}
    \vec{\phi}(x_A,x_B) = \frac{x_A}{N(N-p)L_A} 
    \begin{pmatrix}
(N-p)~ \vec{\textbf{1}}_{(p)} \\
- p~ \vec{\textbf{1}}_{(N-p)}
\end{pmatrix}
+
\frac{1}{N-p}
    \begin{pmatrix}
\vec{\textbf{0}}_{(p)} \\
\vec{\rho}_{(N-p)}
\end{pmatrix}
+ \tau \frac{x_B}{NL_B}
    \begin{pmatrix}
(N-p)~ \vec{\textbf{1}}_{(p)} \\
- p~ \vec{\textbf{1}}_{(N-p)}
\end{pmatrix} . 
\label{eq:solution_holonomy_p-twist}
\end{align}
Indeed, this configuration satisfies the boundary condition in the following way
\begin{align}
P_W^{(N-p)} \vec{\phi}(x_A+L_A,x_B) &= \vec{\phi}(x_A,x_B) - \vec{\mu}_1,\\
\vec{\phi}(x_A,x_B+L_B) &= \vec{\phi}(x_A,x_B) - \tau p \vec{\mu}_1, \label{eq:p-twist_solution_BC}
\end{align}
modulo $\mathbb{Z} -\tau \mathbb{Z}$.
The action of this configuration is expressed as,
\begin{align}
    S_{\mathrm{holonomy}} = \frac{4 \pi^2p}{N g^2} \left( \frac{L_B L_C}{ (N-p) L_A L_D}  + \frac{(N-p)L_A L_D}{ L_B L_C }  \right) \geq \frac{8 \pi^2 p}{N g^2},
\end{align}
and it saturates the BPS bound if the torus is tuned to $\frac{L_B L_C}{  L_A L_D} = (N-p)$.

Again, in a fixed topological sector, the solution is unique up to adding a constant:
\begin{align}
    \vec{\phi}(x_A,x_B) \rightarrow \vec{\phi}(x_A,x_B) + \vec{\phi}_0,
\end{align}
To keep the boundary condition (\ref{eq:p-twist_solution_BC}) intact, the constant must take the form of
\begin{align}
    \vec{\phi}_0 =     \begin{pmatrix}
c_1\\
\vdots\\
c_p \\
- \frac{(c_1 + \cdots+c_p)}{N-p}~ \vec{\textbf{1}}_{(N-p)}
\end{pmatrix},
\end{align}
where $c_1,\cdots,c_p$ are arbitrary complex numbers.

According to Ref.~\cite{Anber:2025yub}, 't Hooft's solutions are the only self-dual solutions at the tuned torus.
This suggests that our above configuration gives the general solution if the aspect ratio is tuned as $\frac{L_B L_C}{  L_A L_D} = (N-p)$. 
Then, the moduli space of the BPS solution with $Q_{\mathrm{top}}=-p/N$ is characterized by the arbitrary parameters $(c_1,\cdots,c_p)$.
Hence, we deduce that the moduli space is 
\begin{align}
        {\mathcal{M}'}^{N;(p,1)}_0 \simeq \frac{(T^2)^{p}}{S_p}.
\end{align}
The periodicity of $(c_1,\cdots,c_p) \in (T^2)^{p}$ is slightly complicated.
For the real part, we can identify 
\begin{align}
        (c_1,\cdots,c_p) \sim (c_1+n_1,\cdots,c_p+n_p), \label{eq:Re_c_periodicity}
\end{align}
where $(n_1,\cdots,n_p) \in \mathbb{Z}^p$.
By using the Weyl permutation $P_W^{(N-p)}$, one can eliminate the latter $(N-p)$ components produced from this shift.
However, this trick does not work for the imaginary part.
We identify
\begin{align}
        (c_1,\cdots,c_p) \sim (c_1-\tau m_1,\cdots,c_p - \tau m_p),\label{eq:Im_c_periodicity}
\end{align}
for $(m_1,\cdots,m_p) \in \mathbb{Z}^p$ satisfying $m_1 + \cdots+m_p \in (N-p)\mathbb{Z}$.

This periodicity would be clearer after decomposing the trace part $(c_1,\cdots,c_p) = (c,\cdots,c)$ and the traceless part $(c_1',\cdots,c_p')$.
We shall see later that the trace part $(c_1,\cdots,c_p) = (c,\cdots,c)$ becomes the translational moduli.
To this end, we reparametrize $(c_1,\cdots,c_p)\in (T^2)^{p}$ as $(c;c_1',\cdots,c_p')$ so that
\begin{align}
       (c_1,\cdots,c_p) = (c,\cdots,c) + (c_1',\cdots,c_p'),
\end{align}
and the latter part $(c_1',\cdots,c_p')$ is constrained as $c_1'+\cdots+c_p'=0$.
Correspondingly, we decompose $ (T^2)^{p}$ into
\begin{align}
        (T^2)^{p} = \frac{(T^2)_c\times S((T^2)^p) }{\sim}\ni (c;c_1',\cdots,c_p'),
\end{align}
where $(T^2)_c=\mathbb{C}/(\mathbb{Z} \times \tau (N-p) \mathbb{Z}) \ni c$, and $S((T^2)^p)$ consists of $(c_1',\cdots,c_p')$ with the constraint $c_1'+\cdots+c_p'=0$.
Under the constraint $c_1'+\cdots+c_p'=0$, the traceless part $(c_1',\cdots,c_p')$ are understood modulo $\mathbb{Z} - \tau  \mathbb{Z}$.

However, we need to introduce an identification $\sim$ for this parametrization $(c;c_1',\cdots,c_p')$.
An element in $(T^2)_c$ parameterizes the trace part:
\begin{align}
       (c_1,\cdots,c_p) = (c,\cdots,c).
\end{align}
Here, the periodicity of $c \in (T^2)_c$ is given by $c \sim c+1$ and $c \sim c -(N-p)\tau$ because of (\ref{eq:Re_c_periodicity}) and (\ref{eq:Im_c_periodicity}). Thus, we can write $(T^2)_c=\mathbb{C}/(\mathbb{Z} \times \tau (N-p) \mathbb{Z})$.
Note that, if $\operatorname{gcd}(N,p)\neq 1$, the periodicity along the $\tau$ direction is modified to $c \sim c -\frac{(N-p)}{\operatorname{gcd}(N,p)}\tau$.

There are several points which are also included in $S((T^2)^p)$: they are generated by
\begin{align}
       (c_1,\cdots,c_p) = (1/p,\cdots,1/p),~(-(N-p)\tau/p,\cdots,-(N-p)\tau/p),
\end{align}
where the sum vanishes in the torus, $c_1+\cdots+c_p \in \mathbb{Z}-\tau (N-p) \mathbb{Z}$.
Thus, the equivalence relation should be,
\begin{align}
    (c;c_1',\cdots,c_p') &\sim (c+1/p;c_1'-1/p+1,c_2'-1/p,\cdots,c_p'-1/p) \notag \\
    (c;c_1',\cdots,c_p') &\sim (c-\frac{(N-p)\tau}{p};c_1'+ \frac{(N-p)\tau(1-p)}{p},c_2'+ \frac{(N-p)\tau}{p},\cdots,c_p'+ \frac{(N-p)\tau}{p})\,, \label{eq:4d2d_toron_moduli_equi_relation}
\end{align}
where we have added $c_1'\mapsto c_1'+1$ or $c_1'\mapsto c_1'- (N-p)\tau $ to maintain the constraint $c_1'+\cdots+c_p'=0$.
We write this identification as $(\mathbb{Z}_p)_c$ for the former one and $(\mathbb{Z}_p)_s$ for the latter one.
Note that the latter one forms $\mathbb{Z}_p$ group as we assumed $\operatorname{gcd}(N,p)=1$.
The meaning of the subscripts of $(\mathbb{Z}_p)_c$ and $(\mathbb{Z}_p)_s$ will become evident shortly.

Hence, the moduli space can now be expressed as
\begin{align}
               {\mathcal{M}'}^{N;(p,1)}_0 \simeq \frac{(T^2)^{p}}{S_p}= \frac{(T^2)_c\times \left[S((T^2)^p)/ S_p\right] }{(\mathbb{Z}_p)_c \times (\mathbb{Z}_p)_s}  ,
\end{align}
Remember that $S((T^2)^p)/ S_p$ is isomorphic to $\mathbb{C}P^{p-1}$ (Appendix \ref{sec:YYmap_prodtosum}).
Let $[\vec{h}] $ denote an element of $\mathbb{C}P^{p-1}$.
The explicit map from $(c_1',\cdots,c_p')$ to $[\vec{h}] $ is given in Eq.~(\ref{eq:YY-map-K}).

Note also that the ``center'' transformations by $(\mathbb{Z}_p)_c$ and $(\mathbb{Z}_p)_s$ are represented by the clock and shift matrices in $\mathbb{C}P^{p-1}$.
Hence, we arrive at
\begin{align}
       {\mathcal{M}'}^{N;(p,1)}_0 \simeq  \frac{(T^2)_c\times \mathbb{C}P^{p-1}}{(\mathbb{Z}_p)_c \times (\mathbb{Z}_p)_s} \ni (c;[\vec{h}] ),
\end{align}
where the identifications by $(\mathbb{Z}_p)_c \times (\mathbb{Z}_p)_s$ are represented as:
\begin{align}
     (\mathbb{Z}_p)_c:& (c;[\vec{h}]) \mapsto (c + 1/p;[C^{-1}\vec{h}]) \notag \\
     (\mathbb{Z}_p)_s:& (c;[\vec{h}]) \mapsto (c - (N-p)\tau/p;[S^{-(N-p)}\vec{h}]). 
\end{align}

Lastly, the trace part $(T^2)_c$ is nothing but the position moduli.
Let $(x_A^*,x_B^*)$ denote the position moduli $(x_A \rightarrow x_A - x_A^*, x_B \rightarrow x_B -x_B^*)$.
The parameter $c$ is related to $(x_A^*, x_B^*)$ as,
\begin{align}
     c = - \frac{x_A^*}{NL_A} + \tau \frac{(N-p) x_B^*}{NL_B}.
\end{align}

Thus, in terms of the position moduli $(x_A^*, x_B^*)$, the torus $(T^2)_c$ becomes the doubly extended torus $\doublehat{T}^2$: $x_A^* \sim x_A^* + NL_A$ and $x_B^* \sim x_B^* + NL_B$.
The identification by $(\mathbb{Z}_p)_c \times (\mathbb{Z}_p)_s$ are rewritten as:
\begin{align}
     (\mathbb{Z}_p)_c:& (x_A^*, x_B^* ;[\vec{h}]) \mapsto (x_A^* - \frac{NL_A}{p}, x_B^* ;[C^{-1}\vec{h}]) \notag \\
     (\mathbb{Z}_p)_s:& (x_A^*, x_B^*;[\vec{h}]) \mapsto (x_A^* , x_B^* - \frac{NL_B}{p} ;[S^{-(N-p)}\vec{h}]). 
\end{align}
The moduli space is indeed,
\begin{align}
        {\mathcal{M}'}^{N;(p,1)}_0 \simeq \frac{\mathbb{C}P^{p-1} \times \doublehat{T}^2}{(\mathbb{Z}_p)_c \times (\mathbb{Z}_p)_s}\,,
\end{align}
identical to the result in Section \ref{sec:modulispace_quotient} on ${\mathcal{M}}^{N;(p,1)}_0 $.

This observation can be straightforwardly generalized to the case where $\operatorname{gcd}(N,p) \neq 1$.
In this case, the periodicity of $c$ is modified to become $(T^2)_c=\mathbb{C}/(\mathbb{Z} \times \tau \frac{N-p}{\operatorname{gcd}(N,p) }\mathbb{Z})$.
Thus, the periodicity of the doubly extended torus $\doublehat{T}^2$ is changed to $x_A^* \sim x_A^* + NL_A$ and $x_B^* \sim x_B^* + \frac{NL_B}{\operatorname{gcd}(N,p)}$, and the quotient by the equivalence relation (\ref{eq:4d2d_toron_moduli_equi_relation}) forms $(\mathbb{Z}_p)_c \times (\mathbb{Z}_{\frac{p}{\operatorname{gcd}(N,p)}})_s$ group.
Hence, the moduli space reads,
\begin{align}
       {\mathcal{M}'}^{N;(p,1)}_0 \simeq \frac{\mathbb{C}P^{p-1} \times \doublehat{T}^2}{(\mathbb{Z}_p)_c \times (\mathbb{Z}_{\frac{p}{\operatorname{gcd}(N,p)}})_s} \,.
       \label{eq:moduli_space_4d2d}
\end{align}
This result is indeed identical to the moduli space of the standard $\mathbb{C}P^{N-1}$ lumps under the conditions $p_c = 1, p_s = p,$ and $k=0$, which we derived in Section \ref{sec:modulispace_quotient}.
This agreement suggests that our qualitative findings, such as the topology of the moduli space, are robust.

\subsection{Comments on (fractional) instantons with higher topological charge}

By pursuing the line of analysis in the previous section, we expect to uncover a deeper correspondence between the moduli space of Yang-Mills instantons and that of the $\mathbb{C}P^{N-1}$ instantons.
Since such a deeper analysis is beyond the scope of this paper, we only give some comments on instantons on a periodic torus.

\begin{itemize}
    \item It is a well-established result that single instanton solutions are forbidden for $SU(N)$ Yang-Mills theory on a periodic $T^4$.
    This restriction can be understood from the Nahm duality \cite{Schenk:1986xe, Braam:1988qk} under which a $U(N)$ $k$-instanton would correspond to a $U(k)$ $N$-instanton.
    The $k=1$ instanton would be mapped to $U(1)$ instanton. 
    Then, there is no counterpart of $SU(N)$ instanton.
    
    \item Thus, the 2d holonomy model, which is the 2d $\mathbb{C}P^{N-1}$ model dimensionally reduced from 4d Yang-Mills theory, has no BPS solution with $Q_{\mathrm{top}}=1$ on periodic $T^2$.

    \item Correspondingly, for the standard $\mathbb{C}P^{N-1}$ model, our solution (in Sections \ref{sec:VS} and \ref{Sec:modulispace}) without any twist $(p_s=p_c=0)$ always suffers from a small-lump singularity at $k=1$ (see (\ref{eq:sol2})).
    Indeed, it is known that there is no 1-instanton solution on the periodic $T^2$ \cite{Richard:1981bv, Speight:1997zc}, because an elliptic function of order 1 is absent.

    \item Both the standard $\mathbb{C}P^{N-1}$ model and 2d holonomy model admit $k$-instanton configuration for $k \geq 2$.

    The 2d holonomy model on the periodic torus at the tuned aspect ratio $\frac{L_A L_D}{L_B L_C} = 1$ has $k=2$ solutions. For example, with a root vector of the form $\vec{\alpha} = \vec{e}_i-\vec{e}_j$, the configuration $\vec{\phi} = \left( \frac{x_A}{L_A}+\tau \frac{x_B}{L_B} \right) \vec{\alpha}$ is a BPS solution.

    However, unlike in the $k=0$ case, the number of moduli of the linear-type solutions is insufficient to account for the dimension predicted by the index theorem, and thus they cannot cover the entire moduli space.
    To cover the whole moduli space, it is necessary to exhaust all holomorphic solutions that satisfy the given boundary conditions, namely a certain class of elliptic functions.

    It would also be interesting to investigate the small-lump singularity in the 2d holonomy model; such an analysis requires the whole moduli space for higher topological charges.
    Since these tasks lie beyond the main scope of this paper and require considerable effort, we leave them for future work. 
\end{itemize}

Another possible direction for future research would be to revisit the Nahm transformation for the $\mathbb{C}P^{N-1}$ model from the perspective of dimensional reduction.
Ref.~\cite{Aguado:2001xg} indicates that the Nahm transformation does not exist for the standard $\mathbb{C}P^{N-1}$ model.
However, we can construct the $\mathbb{C}P^{N-1}$ model with a non-Fubini-Study metric, or the 2d holonomy model, by dimensionally reducing the 4d Yang-Mills theory.
Since the 4d Yang-Mills theory possesses a well-defined Nahm transformation, it is natural to expect that an analogue of Nahm-like duality survives the reduction process. 
This could suggest a novel relationship between $\mathbb{C}P^{N-1}$ models, originating from the Nahm transformation of the 4d Yang-Mills theory.

%%%%%%%%%%%%%%%%%%%%%%%%%%%%%%%%%%%%%%%%%%%

\section{Summary and Discussion}
\label{sec:SD}

In this paper, we have investigated BPS lump solutions in the $\mathbb{C}P^{N-1}$ nonlinear sigma model on a two-dimensional torus $T^2$ with a nontrivial ’t~Hooft twist. The study is motivated by the role of fractional topological excitations such as monopoles, vortices, and instantons in understanding the nonperturbative dynamics of compactified Yang--Mills theories, especially their connection to the $\mathbb{C}P^{N-1}$ model via dimensional reduction and vortex worldsheet dynamics.

Our investigation is performed in the `regularized' version of the $\mathbb{C}P^{N-1}$ model, i.e., an extended Abelian-Higgs model with $N$ complex scalar fields and a $U(1)$ gauge field. The original $\mathbb{C}P^{N-1}$ model is recovered in the strong-coupling limit $(e^2 \rightarrow \infty)$.
Our findings are valid for the extended model, but for the sake of clarity and conciseness, we summarize the main results for the $\mathbb{C}P^{N-1}$ limit here: (1) explicit forms of BPS solutions on a twisted torus, (2) determination of their moduli spaces, and (3) their connection to 4d Yang-Mills theory.

\medskip

{\flushleft \textbf{Setup}}:
    We consider the 2d $\mathbb{C}P^{N-1}$ model on a torus, $T^2$, with twisted boundary conditions. 
    This model possesses a $PSU(N) = SU(N)/\mathbb{Z}_N$ global symmetry, and we focus on a specific discrete subgroup, $(\mathbb{Z}_N)_s \times (\mathbb{Z}_N)_c \subset PSU(N)$. It is convenient to use the shift and clock matrices for the bases: $(S)^a{}_{b}\coloneqq \delta^a_{b-1}+\delta^a_{N}\delta^1_{b},~ (C)^a{}_{b}\coloneqq \omega^{a-1} \delta^a_{b}~{\rm with~} \omega=e^{\frac{2\pi i}{N}}. $
    The twisted boundary conditions are imposed as follows: $(\mathbb{Z}_N)_s$ twist along the $x_1$ direction ($x_1 \to x_1 + L_1$) and a $(\mathbb{Z}_N)_c$ twist along the $x_2$ direction ($x_2 \to x_2 + L_2$). 
    These twists are characterized by a pair of integer parameters $(p_s, p_c) \in \mathbb{Z}_N \times \mathbb{Z}_N$. 
    This twisted boundary condition forces the topological charge to be a fractional value: $k +\frac{p}{N}$ with $p=p_sp_c$, and the index theorem predicts the dimension of the moduli space ${\cal M}^{N;(p_s,p_c)}_{k}$ as:
\begin{align}
M\,:=\,{\rm dim}_{\mathbb C} {\cal M}^{N;(p_s,p_c)}_{k}= \frac{N}{2\pi} \int_{{\cal F}(T^2)}d^2x  F_{12}=Nk+p. \tag{\ref{eq:index_theorem_defM}}
\end{align}
    A more detailed description can be found in Section \ref{sec:model_and_BC}. 

\medskip

{\flushleft \textbf{(1) Explicit forms of BPS lumps on the twisted torus}}:
%    Section addresses the BPS equation, where the task of finding BPS solutions is reduced to the determination of $N$-component entire functions $\{ h_a (z) \}$ that satisfy the given boundary conditions: (\ref{eq:hzbcS}) and (\ref{eq:hzbcC}).
    To represent the solutions, we use fields $\{ h_a (z) \}$ that correspond to the $\mathbb{C}P^{N-1}$ field: $\vec{n} = \vec{H}/\sqrt{\xi}= \vec{h}/|\vec{h}|$ in the $\mathbb{C}P^{N-1}$ limit, and we write the complex coordinates as $z = \frac{x_1 + i x_2}{L_1}$ with the modulus of the torus $\tau = \frac{i L_2}{L_1}$.

    Sections \ref{sec:BPS_bound} and \ref{sec:VS} provide the explicit solutions of the BPS lumps on the twisted torus.
    Since the $S$-twist relates different components of the fields, it is sufficient to specify $h_{\bar{a}}(z)$ for $\bar{a}=1,\ldots, \gcd(N,p_s)$; the other flavor indices are automatically determined by the $S$-twist. 
    Up to some ambiguity of the representation (called the V-transformation), the general solution is given by
\begin{align}
    h^{\mathrm{rep}}_{\bar{a}}(z)=c_{\bar{a}}\prod_{n=1}^{\hat{M}} \theta_1 \left(\frac{\gcd(N,p_s)}{N}(z-z_{(\bar{a},n)})\Big| \frac{\gcd(N,p_s)}{N}\tau\right), 
    \tag{\ref{eq:BPSlump_generalps_abar}}
\end{align}
where $(c_{\bar{a}})_{\bar{a}=1,\ldots,\gcd(N,p_s)}\in \mathbb{C}^{\gcd(N,p_s)}\setminus\{\vec{0}\}$ and $z_{(\bar{a},n)}$ are moduli parameters, $\theta_1 \left(z| \tau\right)$ is the theta function (\ref{eq:thetadef}), and $\hat{M} = M/\gcd(N,p_s)$.
See Eq.~(\ref{eq:generalPsol}) for the full expression of all indices. A simpler expression for $p_s =1$ is given in Eq.~(\ref{eq:repr_ps1_extendedmoduli}).
%Some examples (Figure \ref{fig:taudependence}), the connection to the fractional instanton on $\mathbb{R}\times S^1$, and some remarks on small-lump singularity are included in Section \ref{sec:fractionalV}.

\medskip

{\flushleft \textbf{(2) Moduli spaces}}:
    %The above parametrization (by $c_{\bar{a}}$ and $z_{(\bar{a},n)}$) is ill-suited for describing the global structure of the moduli space.
    In Section \ref{Sec:modulispace}, we have introduced an alternative parametrization in terms of the basis (\ref{eq:basis}), which makes this global structure transparent.
    After this reparametrization, the moduli are the mean of the zeros $\langle z \rangle$ (of $h_1(z)$)\footnote{Recall that the sum of zeros of $h_{\bar{a}}(z)$ is constrained by the modified Abel's theorem, (\ref{eq:modified_abel_constraint_extended}): $\sum_{n=1}^{\hat M} z_{(\bar a,n)}= \hat M \langle z \rangle +\frac{p_c(\bar a-1)}{\gcd(N,p_s)}$, where $\langle z \rangle$ does not depend on $\bar{a}$.
    Since each zero $z_{(\bar a=1,n)}$ is subject to (at least) the following identification: $z_{(\bar a,n)} \sim  z_{(\bar a,n)} + \frac{N}{\operatorname{gcd}(N,p_s)} (n + \tau m)$ with $n,m \in \mathbb{Z}$, the mean $\langle z\rangle$ naturally takes a value in $\mathbb{C}/(\frac{N}{M}\mathbb{Z}+\frac{N}{M}\tau \mathbb{Z})$.
    } and coefficients $\{ \matC^n{}_{\bar a} \in \mathbb{C} ~|~ \substack{n=1,\cdots,\hat{M} \\ \bar{a}=1,\ldots,\gcd(N,p_s)}\}$ with the equivalence relations (\ref{eq:Zs}) and (\ref{eq:Zc}).
    This parametrization explains the fiber-bundle structure of the moduli space:
\begin{align}
    \mathbb{C}P^{M-1}\,\hookrightarrow\, \mathcal{M}^{N;(p_s,p_c)}_k \to \frac{\mathbb{C}}{\frac{N}{M}\mathbb{Z}+\frac{N}{M}\tau \mathbb{Z}}, 
    \tag{\ref{eq:modli_CPfiber}}
\end{align}
with $\{ \matC^n{}_{\bar a} \} \in \mathbb{C}P^{M-1}$ and $\langle z \rangle \in \frac{\mathbb{C}}{\frac{N}{M}\mathbb{Z}+\frac{N}{M}\tau \mathbb{Z}}$.
Equivalently, we can write
\begin{align}
{\cal M}^{N;(p_s,p_c)}_k\simeq \frac{\mathbb C P^{M-1}\times \doublehat{T}^2}{\mathbb Z_{M/\gcd(N,p_s)}^s\times \mathbb Z_{M/\gcd(N,p_c)}^c}, \tag{\ref{eq:moduli_space_ps_pc}}
\end{align}
where $    \doublehat{T}^2 := \frac{\mathbb C }{\frac{N}{\gcd(N,p_s)}\mathbb{Z}+\frac{N}{\gcd(N,p_c)}\tau \mathbb{Z}}$ is the doubly-extended torus, and the quotient represents the equivalence relations (\ref{eq:Zs}) and (\ref{eq:Zc}).
Note that this moduli space should be interpreted as the closure of the correct moduli space for the $\mathbb{C}P^{N-1}$ model, since the small-lump singularities are smoothed out in the Abelian-Higgs model and included in this expression.

Another advantage of this reparametrization is that the modular duality ($\tau \rightarrow -1/\tau$) can be explicitly shown as is done in Section \ref{sec:DFT}.

\medskip

{\flushleft \textbf{(3) On 2d (non-Fubini-Study) $\mathbb{C}P^{N-1}$ model from 4d Yang-Mills theory}:}
Section~\ref{sec:4d2d} discusses the relation between 2d $\mathbb{C}P^{N-1}$ model and 4d Yang-Mills theory based on Ref.~\cite{Yamazaki:2017ulc}. 
We have started with the $SU(N)$ Yang-Mills theory on $T^4 =     S^1_{A} \times S^1_{B} \times S^1_{C} \times S^1_{D} $ with two nontrivial ’t~Hooft fluxes on $S^1_{A} \times S^1_{D}$ and $S^1_{B} \times S^1_{C}$.
By performing a dimensional reduction along $S^1_{C} \times S^1_{D}$, one has obtained a 2d $\mathbb{C}P^{N-1}$ model on the torus $S^1_{A} \times S^1_{B}$ with the twisted boundary conditions. 
Apart from the fact that its metric is not the Fubini-Study one, the resulting 2d $\mathbb{C}P^{N-1}$ model is exactly what we have discussed.

When the aspect ratio of the torus is tuned, the 't Hooft-type fractional instantons are available in the 4d Yang-Mills theory.
Then, fractional instantons in this (non-Fubini-Study) $\mathbb{C}P^{N-1}$ model on the twisted torus can be explicitly constructed.
For these fractional instantons, we have explicitly determined their moduli spaces ${\mathcal{M}'}^{N;(p_s,p_c)}_k$ for simple cases ($p_s=p$, $p_c = 1$, and $k=0$):
\begin{align}
       {\mathcal{M}'}^{N;(p,1)}_0 \simeq \frac{\mathbb{C}P^{p-1} \times \doublehat{T}^2}{(\mathbb{Z}_p)_c \times (\mathbb{Z}_{\frac{p}{\operatorname{gcd}(N,p)}})_s}, \tag{\ref{eq:moduli_space_4d2d}}
\end{align}
which has the same structure as the moduli space ${\cal M}^{N;(p_s,p_c)}_k$ in the standard  $\mathbb{C}P^{N-1}$ model.
\medskip

We conclude by offering a further perspective on the moduli space, derived from the Yang-Mills-Higgs system. Recalling from the Introduction, the $U(N)$ gauge theory with $N$ fundamental Higgs scalar fields admits non-Abelian local vortex solutions, whose low-energy effective theory is the two-dimensional ${\mathbb C} P^{N-1}$ model. When this Yang-Mills-Higgs system is formulated on ${\mathbb R}^2 \times T^2$ incorporating the 't Hooft flux (and $S,~C$ flavor twist for the Higgs field), the resulting worldsheet theory for vortices wrapping $T^2$ corresponds exactly to the ${\mathbb C} P^{N-1}$ model on $T^2$ with the 't Hooft twist, as discussed throughout this work.

According to conventional wisdom \cite{Eto:2004rz,Nitta:2022ahj}, when a (confined) soliton is realized as a soliton within the effective theory of a higher-dimensional soliton, its moduli space is obtained by freezing certain directions within the original (unconfined) soliton's moduli space.
In our case, if we know the moduli space of the 4d fractional instanton $\mathcal{M}^{Q_{\mathrm{top}}=p/N}_{\mathrm{4d~YM}}$ (with $Q_{\mathrm{top}}=p/N$) for the ``unconfined soliton'', one expects that the moduli space of the 2d fractional instanton $\mathcal{M}^{Q_{\mathrm{top}}=p/N}_{\mathrm{2d~{\mathbb C} P^{N-1}}}$ (``confined soliton'' within the vortex) can be written as
\begin{align}
\mathcal{M}^{Q_{\mathrm{top}}=p/N}_{\mathrm{2d~{\mathbb C} P^{N-1}}} \overset{?}{=} \left. \mathcal{M}^{Q_{\mathrm{top}}=p/N}_{\mathrm{4d~YM}} \right|_{{\mathbb R}^2~\mathrm{fixed}}  .
\end{align}
If the 't Hooft toron configuration is employed as the 4d fractional instanton, the right-hand side would have already been determined in Section \ref{sec:torons_modulispace} albeit in a different context.
Hence, this conjecture holds for these simple cases, $Q_{\mathrm{top}}=p/N$.
This is still a preliminary observation, and we intend to leave a more in-depth physical investigation for future work.

\if0
Future problem

Extracting the small lump singularity from the moduli space.

Applications  Localization 
\fi

\begin{acknowledgments}
The authors appreciate the YITP-RIKEN iTHEMS conference ``Generalized symmetries in QFT 2024'' (YITP-W-24-15) and the YITP long-term workshop ``Hadrons and Hadron Interactions in QCD 2024'' (YITP-T-24-02) for providing the opportunities of useful discussions.  
This work was partially supported by Japan Society for the Promotion of Science (JSPS) Research Fellowship for Young Scientists Grant
No. 23KJ1161 (Y.H.), by JSPS KAKENHI Grant No. 23K22489 (Y.T.), 23K03425 (T.M.), 22H05118 (T.M.), 
JP22H01221 (M.N.), 
JP23K22492 (M.N.), 
by Center for Gravitational Physics and Quantum Information (CGPQI) at Yukawa Institute for Theoretical Physics, 
and by the WPI program ``Sustainability with Knotted Chiral Meta Matter (WPI-SKCM$^2$)'' at Hiroshima University (MN).

\end{acknowledgments}

%\newpage
%%%%%%%%%%%%%%%%%%%%%%%%%%%%%%%%%%%%%%%%%%%%%%%%%%%%%%
%%%%%%%%%%%%%%%%%%%%%%%%%%%%%%%%%%%%%%%%%%%%%%%%%%%%%%%%%%%
\appendix

\section{Action for $\psi$ and K\"ahler potential}\label{sec:existence}
\subsection{Existence and Uniqueness of $\psi$}
Here, let us comment on the existence and uniqueness of the solution to Eq.~\eqref{eq:master}.
For vortices in the Abelian-Higgs model with $N=1$ in the $\mathbb R^2$ spatial slice,
the existence of solutions has been proven in \cite{Taubes:1979tm}.
The following argument gives a brief and intuitive 'proof' for that of our model. 
While it  is not mathematically rigorous enough to be fully accepted by mathematicians, it provides an intuitive understanding that should be sufficient for physicists. It follows the discussion in Appendix A of \cite{Fujimori:2023wkd}, which deals with the case of  $\mathbb{R}^2$  or $ S^2$, and is adapted to fit the present discussion on the torus.

Let us introduce the following action for $\omega=2{\rm Re}\,\psi$ 
\begin{align}
    {\cal S}_{\rm vtx}[\omega]:=&\int_{{\cal F}(T^2)} d^2x \left\{\frac1{4e^2} (\p_i \omega)^2+\xi \left(\omega+e^{-\omega} |\vec h(z)|^2\right) \right\}\nonumber \\
    &{}-\frac1{e^2} {\rm Im}\left[\int_0^{L_2} d x_2\, \omega_s\, \p_2 v_s(z)\Big|_{x_1=0}-\int_0^{L_1}dx_1 \,\omega_c\,\p_1 v_c(z)\Big|_{x_2=0}\right]+{\cal S}_0, \label{eq:vtxaction}
\end{align}
where  $\omega_s,\omega_c$ are
\begin{align}
    \omega_s=\omega_s(x_2):= \frac{\omega(0,x_2)+\omega(L_1,x_2)}2,\quad \omega_c=\omega_c(x_1):=\frac{\omega(x_1,0)+\omega(x_1,L_2)}2,
\end{align}
such that this leads Eq.~\eqref{eq:master} as a saddle point equation
\begin{align}
    \frac{\delta {\cal S}_{\rm vtx}[\omega]}{\delta \omega} =-\frac1{2e^2} \p_i^2 \omega +\xi(1-e^{-\omega} |\vec h(z)|^2).
\end{align}
Here we require that $\omega$ satisfies the following boundary conditions,
\begin{align}
    \omega(x_1+L_1,x_2)=\omega(x_1,x_2)+2{\rm Re}\, v_s(z),\quad  \omega(x_1,x_2+L_2)=\omega(x_1,x_2)+2{\rm Re}\,v_c(z),
\end{align}
and the terms in the second line in Eq.~\eqref{eq:vtxaction} are introduced to cancel contributions from total derivative terms in the calculation of $\delta {\cal S}_{\rm vtx}[\omega]$.
${\cal S}_0$ is a certain term independent of $\omega$, but is introduced for the convenience of later explanation.

We can explicitly prepare an example of $\omega$ satisfying the above boundary conditions as,
\begin{align}
    \omega_0= \log |\vec h_0(z)|^2,%\quad {\rm with~} \exists \vec h_0(z):  \forall z, |\vec h_0(z)|^2\not = 0
\end{align}
where $\vec h_0(z)$ is some function satisfying the boundary conditions (\ref{eq:hzbcS}) and (\ref{eq:hzbcC}) and non-vanishing, i.e., $|\vec h_0(z)|^2\not = 0$ for all $z$.
Defining a linear function $\omega(s):=\omega_0+ s\, \delta \omega $ 
with an arbitrary doubly-periodic function $\delta \omega$,
${\cal S}_{\rm vtx}[\omega(s)]$ is always convex with respect to $s$ as
\begin{align}
    \frac{d^2{\cal S}_{\rm vtx}[\omega(s)]}{ds^2}= \int_{{\cal F}(T^2)} d^2x \left\{ \frac1{2e^2} (\p_i \delta \omega)^2+\xi e^{-\omega(s)}|\vec h(z)|^2\right\} > 0.
\end{align}
Furthermore we find 
\begin{align}
    \lim_{s\to \infty}\frac{d^2{\cal S}_{\rm vtx}[\omega(s)]}{ds^2} =0   \quad {\rm iff} \quad \delta \omega:~{\rm const.~~ and}\quad \delta \omega>0,
\end{align}
where the derivative in the limit of $s\to \infty$ keeps a finite value as
\begin{align}
    \lim_{s\to \infty}\frac{d{\cal S}_{\rm vtx}[\omega(s)]}{ds}&=\delta \omega \times \left(\xi \int_{T^2 }d^2x -
    \frac1{e^2} {\rm Im}\left[\int_0^{L_2} d x_2\, \p_2 v_s(z)\Big|_{x_1=0} -\int_0^{L_1}dx_1 \p_1 v_c(z)\Big|_{x_2=0}\right]\right)\\
    &=\delta \omega\times \left( \xi L_1L_2-\frac{2\pi}{e^2} \frac{M}N\right) >0,
\end{align}
as long as the Bradlow bound is satisfied, except when the bound is saturated.

Therefore, we find that the functional ${\cal S}_{\rm vtx}[\omega]$ is convex and coercive as
\begin{align}
   \forall \delta\omega:\quad 
  \frac{d^2{\cal S}_{\rm vtx}[\omega(s)]}{ds^2} >0,\quad \lim_{s\to \infty} {\cal S}_{\rm vtx}[\omega(s)]=+\infty. 
\end{align}
The above properties provide sufficient, if not mathematically rigorous, circumstantial evidence that there always exists only one solution $\omega=\omega_{\rm sol}$ to Eq.~\eqref{eq:master} with any given $\vec h(z)$.

\subsection{Manton approximation and K\"ahler potential}
The low-energy dynamics around the BPS states is described by the Manton approximation.
There, the moduli parameters $\{\phi^\alpha \}$ are lifted to functions $\{\phi^\alpha(t)\}$ with respect to time $t$ and their time derivative $\p_t$ are assumed to be sufficiently small and thus, $A_{\bar  z}, \vec H$ are given by
\begin{align}
    A_{\bar z}=i \p_{\bar z} \psi_{\rm sol} |_{\phi^\alpha \to \phi^\alpha(t)}+{\cal O}(\p_t^2),\quad 
     \vec H=\sqrt{\xi} e^{-\psi_{\rm sol}}\vec h(z)|_{\phi^\alpha \to \phi^\alpha(t)}+{\cal O}(\p_t^2),
\end{align} 
and the gauge potential $A_0$ is given by
\begin{align}
    A_0=i \bar \delta_t \psi_{\rm sol}-i\delta_t \bar \psi_{\rm sol}
    +{\cal O}(\p_t^3),\quad{\rm with~}  \quad \delta_t := \dot \phi^\alpha \frac{\p}{\p \phi_\alpha},\quad 
    \bar \delta_t := \dot {\bar \phi}^\alpha \frac{\p}{\p \bar \phi_\alpha}
\end{align}
as a solution to the equations of motion   around the BPS state. 
Then, the effective Lagrangian is given as the second order term of $\p_t$ in the original Lagrangian as follows
\begin{align}
    {\cal L}_{\rm eff}&= \int_{{\cal F}(T^2)} d^2x\left\{\frac1{2e^2} F_{0i}^2+|{\cal D}_0 \vec H|^2\right\} \Big|^{\rm sol}_{\phi\to \phi(t)}\\
    &= \int_{{\cal F}(T^2)} d^2x\left\{ \frac{2}{e^2L_1^2} |\p_{\bar z} \delta_t \omega|^2 +\xi e^{\omega}|\delta_t (e^{-\omega}\vec  h(z))|^2\right\}\Big|^{\rm sol}_{\phi\to \phi(t)}\\
&= \delta_t \bar \delta_t K_{\rm vtx}\,.    
\end{align}
This effective theory gives a non-linear sigma model of which the target manifold is the moduli space ${\cal M}^{N;(p_s,p_c)}_{k}$.
According to \cite{Eto:2006uw}, the K\"ahler potential $K_{\rm vtx}$ yielding the K\"ahler metric for the moduli space is given by substituting 
the solution $\omega=\omega_{\rm sol}$ into the functional ${\cal S}_{\rm vtx}[\omega]$ as
\begin{align}
    K_{\rm vtx}={\cal S}_{\rm vtx}[\omega_{\rm sol}],\quad {\rm with~} \frac{\delta {\cal S}_{\rm vtx}[\omega_{\rm sol}]}{\delta \omega}=0.
\end{align}
Strictly speaking, Ref.\cite{Eto:2006uw} assumes a spatial slice $\mathbb R^2$, so care is needed with total derivative terms when applying it to our model with $T^2$ as the spatial slice.
Here, ${\cal S}_0$ in Eq.~\eqref{eq:vtxaction} is the following term determined so that the above equation holds, 
\begin{align}
    {\cal S}_0=\frac{1}{2e^2} {\rm Im}\left[v_s(\tau)\bar v_c(0)-v_c(1)\bar v_s(0)\right],
\end{align}
where  $\bar \delta_t v_{s,c}=\delta_t \bar v_{s,c}=0$ are supposed.
With this term, we can confirm that ${\cal S}_{\rm vtx}$ is invariant under the V-transformation \eqref{eq:Vtrf},\eqref{eq:vsvcVtrf} up to the K\"ahler transformation as
\begin{align}
    {\cal S}_{\rm vtx}[\omega]\quad \mapsto\quad  {\cal S}_{\rm vtx}[\omega]+ (f[v,v_s,v_c]+{\rm c.c.}),
\end{align}
with a certain holomorphic function $f$.

%%%%%%%%%%%%%%%%%%%%%%%%%%%%%%%%%%%%%%%%%%%%%%%%%%%%%
%%%%%%%%%%%%%%%%%%%%%%%%%%%%%%%%%%%%%%%%%%%%%%%%%%%
\section{Proofs}
\label{sec:proofs_app}

For notational convenience, we write $\hat N:=N/ {\rm gcd}(N,p_s)$ and $\bar{N}=\gcd(N,p_s)$  throughout this appendix.

\subsection{The number of zeros}\label{sec:NumberZero}
Eq.~\eqref{eq:numberofzeros} is proved as follows.  

Since $h_a(z)$ has no poles, the number of zeros of $h_a(z)$ in the fundamental domain ${\cal F}(\hat T^2)$ of the extended torus $\hat T^2$ can be calculated by  
\begin{align}
|\hat \sigma_a|=\frac1{2\pi i}\oint_{\p {\cal F}(\hat T^2)} \frac{\p_z h_a(z)}{h_a(z)}dz
%= \Delta \alpha_a + \Delta\beta_a'- \Delta\alpha_a'- \Delta\beta_a
\end{align}
Here,
the contour integral along the boundary $\p {\cal F}(\hat T^2)$ is decomposed as follows:
\begin{align}
    \oint_{\p {\cal F}(\hat T^2)}=\int_{\alpha}+\int_{\beta'}-\int_{\alpha'}-\int_{\beta}
\end{align}
where paths $\alpha,\alpha',\beta,\beta'$ are defined as, with $s\in [0,1]\in \mathbb R$
\begin{align}
    &\alpha(s):= z_\epsilon+\hat N\,s,\quad &\alpha'(s):=z_\epsilon+\tau +\hat N\, s,\\ 
    &\beta(s):=z_\epsilon+\tau \,s,\quad &\beta'(s):=z_\epsilon+\hat N+\tau\,s.
\end{align}
Here, $z_\epsilon$ is introduced such that no zero appears on the paths.
The pseudo-periodic boundary conditions on $h_a(z)$ are explicitly given by
\begin{align}
h_a(z+1)=e^{v_s(z)} h_{a-p_s}(z),\quad
h_a(z+\tau)=e^{v_c(z)}h_a(z) e^{\frac{2\pi i}N (a-1)p_c},
\end{align}
with  $h_{a+N}(z)=h_a(z)$, and the first one gives
\begin{align}
  h_a(z+\hat N)=e^{\hat v_s(z)} h_{a}(z)\quad {\rm with~}\hat v_s(z):= \sum_{a=1}^{\hat N} v_s(z+a-1).  
\end{align} 
These conditions lead the following relations
\begin{align}
    \left[\int_{\alpha'}-\int_\alpha\right]   \frac{\p_z h_a(z)}{h_a(z)}dz = \int_\alpha \p_z v_c(z) dz=
   v_c(z_\epsilon+\hat N)-v_c(z_\epsilon), \\
     \left[\int_{\beta'}-\int_\beta\right]   \frac{\p_z h_a(z)}{h_a(z)}dz = \int_\beta 
     \p_z \hat v_s(z) dz=
   \hat v_s(z_\epsilon+\tau)-\hat v_s(z_\epsilon).
\end{align}
Note that $v_{s,c}$ themselves are multi-valued, but their differences are single-valued since they are supposed to be continuous functions.
Using these relations, we find 
\begin{align}
|\hat \sigma_a|
&=\frac{(\hat v_s(z_\epsilon+\tau)-\hat v_s(z_\epsilon))-(v_c(z_\epsilon+ \hat N)-v_c(z_\epsilon))}{2\pi i}\\
&=\sum_{a=1}^{\hat N}\frac{(v_s(z_\epsilon+a-1+\tau)- v_s(z_\epsilon+a-1))-(v_c(z_\epsilon+ a)-v_c(z_\epsilon +a-1))}{2\pi i}\\
&=\hat N \times \frac{M}N =\frac{M}{\bar N} =:\hat M,
\end{align}
where we also used the constraint  \eqref{eq:vCC} on $(v_s,v_c)$.

%%%%%%%%%%%%%%%%%%%%%%%%%%%%%%%%%%%%

\subsection{Modified Abel's theorem}\label{sec:Abel}
Imitating Abel's theorem on the sum of zeros and poles of an elliptic function,
we calculate the sum of the zeros of $h_a(z)$ as follows:
\begin{align}
    \sum_{w\in \hat \sigma_a} w = \frac1{2\pi i} \oint_{\p{\cal F}(\hat T^2)} z \,\frac{\p_z h_a(z)}{h_a(z)} dz .
\end{align}
The integration path can be divided into two parts and the integrations are performed as follows:
\begin{align}
    \left[\int_{\alpha'}-\int_\alpha\right] z \frac{\p_z h_a(z)}{h_a(z)}dz& =\int_\alpha \left(z  \p_z v_c(z)+\tau \left(\p_z v_c(z)+\frac{\p_z h_a(z)}{h_a(z)}\right)\right)dz\nonumber \\
    &=-\int_\alpha  v_c(z) dz +\left[(z+\tau) v_c(z)+\tau \log h_a(z)\right]^{z_\epsilon+\hat N}_{z_\epsilon},\\
   \left[\int_{\beta'}-\int_\beta\right] z \frac{\p_z h_a(z)}{h_a(z)}dz&  =-\int_\beta \hat  v_s(z) dz +\left[(z+\hat N) \hat v_s(z)+\hat N \log h_a(z)\right]^{z_\epsilon+\tau}_{z_\epsilon}.
\end{align}
Furthermore,  using the boundary conditions on $h_a(z)$ we find 
\begin{align}
  [\log h_a(z)]^{z_\epsilon+\hat N}_{z_\epsilon}&={\hat v_s(z_\epsilon)}-2\pi i\,l_a, \quad {\rm with~}\exists\, l_a\in \mathbb Z\\ 
[\log h_a(z)]^{z_\epsilon+\tau}_{z_\epsilon}&={v_c(z_\epsilon)}+2\pi i\,\frac{a-1}{N}p_c+2\pi i\,m_a  \quad {\rm with~} \exists\,m_a \in \mathbb Z
\end{align}
where undetermined integers $l_a,m_a$ are introduced to reflect the fact that $\log h_a(z)$ is multivalued.
By collecting those results, we finally obtain the following formula for the sum of the zeros as
\begin{align}
    \sum_{w\in \hat \sigma_a} w =\hat M \langle z\rangle + \frac{a-1}{\bar N} p_c+\hat N\,m_{a}+\tau\, l_a \quad {\rm with~} l_a,m_a \in \mathbb Z, 
\end{align}
where we also used the constraint  \eqref{eq:vCC} again and
the parameter $\langle z \rangle \in \mathbb C$ is defined as follows to summarize 
the $a$-independent terms:
\begin{align}
    \langle z\rangle := \frac1{2\pi i\hat M}\left\{ \int_\alpha v_c(z) dz-\int_\beta \hat v_s(z) dz  \right\}+z_\epsilon+\hat N+\tau.
\end{align}

%%%%%%%%%%%%%%%%%%%%%%%%%%%%%%%%
\subsection{Almost no-hair theorem}\label{sec:Uniquness}
We will discuss here that the solution of $\vec h(z)$ is determined almost exclusively by the information of those zeros.
%\begin{align}
%    h_a(z+\hat N)=e^{\hat v_s(z)}h_a(z), \quad \hat v_s(z):= \sum_{n=0}^{\hat N-1}v_s\left(z+n\right)
%\end{align}

Let us suppose that two different vectors $\vec h(z)$ and $\vec h'(z)$ of which entries satisfy 
\begin{align}
    h_a(z+1)=e^{v_s(z)}h_{a-p_s}(z),\quad h_a(z+\tau)=e^{v_c(z)}h_a(z)e^{\frac{2\pi i}N p_c(a-1)},\\
    h_a'(z+1)=e^{v_s'(z)}h'_{a-p_s}(z),\quad h'_a(z+\tau)=e^{v'_c(z)}h'_a(z)e^{\frac{2\pi i}N p_c(a-1)},
\end{align}
have the same set of zeros
such that a function $f_a(z):=h'_a(z)/h_a(z)$ has neither zeros nor poles.
Here note that $f_a(z)$ satisfies
\begin{align}
    f_a(z+1)=e^{v'_s(z)-v_s(z)}f_{a-p_s}(z),\quad f_a(z+\tau)=e^{v_c'(z)-v_c(z)}f_a(z),\label{eq:fbc}
\end{align}
and thus their ratios $f_a(z)/f_b(z)$ are an elliptic functions on the extended torus $\hat T^2$ as,
\begin{align}
    \frac{f_a(z+\hat N)}{f_b(z+\hat N)}=\frac{f_{a}(z)}{f_{b}(z)},\quad  \frac{f_a(z+\tau)}{f_b(z+\tau)}=\frac{f_a(z)}{f_b(z)}.
\end{align}
As a property of elliptic functions, their ratios must be constant. That is, $f_a(z)$ can be written as
\begin{align}
   \exists c_a, f(z):\quad f_a(z)= c_a\, f(z).
\end{align}
and thus $h'_a(z)$ and $h_a(z)$ are V-equivalent except for the difference in constants as
\begin{align}
 h'_a(z)=c_a f(z)\, h_a(z) \quad \stackrel{V}\sim \quad c_a h_a(z).   
\end{align}
Substituting this result into Eq.~\eqref{eq:fbc}, we find 
that ratios of the constants $c_a /c_{a-p_s}$ must be independent of the index $a$, 
which leads
\begin{align}
  c_a\,=\, c_{\bar r(a)}\, e^{\frac{2\pi i}{\hat N}\alpha (\hat s(a)-1)}\,\quad 
    {\rm with~}\quad \exists \, c_{\bar r(a)}\in \mathbb C\quad {\rm and}\quad \exists \,\alpha \in \mathbb Z/\hat N \mathbb Z.
\end{align}
\subsection{Equivalence relation on moduli}\label{sec:EqRel}
If there is a non-trivial equivalence relation on moduli parameters
 as $\{ \langle z\rangle, \matC\} \sim \{\langle z\rangle+\Delta,\, \matC^\Delta\}$,
there must be the following relation on the basis $\{\vartheta_{n,a}(z)| n=1,2,\cdots,\hat N\}$ 
\begin{align}
    \sum_{n=1}^{\hat M} \matC_{n, \bar r(a)} \vartheta_{n,a}(z-\langle z\rangle) \quad \stackrel{V}{\sim}\quad 
    \sum_{n=1}^{\hat M} \matC^\Delta_{n, \bar r(a)} \vartheta_{n,a}(z-\langle z\rangle -\Delta),
\end{align}
which means that there exists the following equation on $\vartheta_{n,a}(z)$ 
\begin{align}
    \vartheta_{n,a}(z+\Delta)=\sum_{m=1}^{\hat M} (g_{\bar r(a)}(z))_{n}{}^m\, \vartheta_{m,a}(z),
\end{align}
with a certain matrix valued function $g_{\bar r(a)}(z)$. This equation must be consistent with Eqs.~\eqref{eq:theta-ppdc1} and \eqref{eq:theta-ppdc2}.
By evaluating relations between $\vartheta_{n,a}(z+\Delta+1)$ ($\vartheta_{n,a}(z+\Delta+\tau)$) and $\vartheta_{m,a}(z)$ in two different ways,
we find the following equations
\begin{align}
g_{\bar r(a)}(z+1) =g_{\bar r(a)}(z),\quad g_{\bar r(a)}(z+\tau) = e^{-2\pi i \frac{M}N \Delta }g_{\bar r(a)}(z).
\end{align}
Form the first one, we can expand $g_{\bar r(a)}(z)$ as 
\begin{align}
    g_{\bar r(a)}(z)= \sum_{n=-\infty}^{\infty}  C_{n,\bar r(a)}e^{2\pi i n z }
\end{align}
and then  the second one requires
\begin{align}
   \exists m,l \in \mathbb Z: \quad \Delta =\frac{N}M (m+l\, \tau),\quad  g_{\bar r(a)}(z) \propto e^{-2\pi i\, l z}.
\end{align}
This result means any equivalence relation on $\{ \langle z\rangle, \matC\}$ is generated by only $\mathbb Z^s$ and $\mathbb Z^c$ in \eqref{eq:Zs} and \eqref{eq:Zc}.
%%%%%%%%%%%%%%%%%%%%%%%%%%%%%%%%%%%%%%%%%%%%%%%%%%%%%%%%%%%%%%%%%%%%%%%
\section{Product-to-sum formula and theta-function map}\label{sec:ThetaSpace}

We have parametrized the moduli space by locations of zeros in Section \ref{sec:VS}, and the product-to-sum formula (\ref{eq:product-to-sum-maintext}) is the key relation to bridge this description and our results in Section \ref{Sec:modulispace}.
In the presence of $K$ zeros, the locations of zeros form the space $T^{2K}/S_K$ with identifying permutations. 
Here, we decompose this parametrization $z$ into the average $\braket{z}$ and the relative coordinates $z - \braket{z}$.
The latter moduli form $S(T^{2K})/S_K \simeq T^{2(K-1)}/S_K$, which is the torus $(\phi_1,\cdots,\phi_K)$ modulo permutations with the constraint $\sum_{i=1}^K \phi_i = 0$.

Using theta functions, one can construct an explicit isomorphism from $T^{2(K-1)}/S_K$ to $\mathbb{C}P^{K-1}$, which is a common technique in algebraic geometry \cite{GriffithsHarris1994} (see also Ref.~\cite{Yamazaki:2017ulc} in the context of the reduction from 4d Yang-Mills theory to 2d $\mathbb{C}P^{N-1}$ model, which is reviewed in Appendix \ref{sec:4d2d}).
In this Appendix, we review this explicit construction and prove the product-to-sum formula (\ref{eq:product-to-sum-maintext}).

\subsection{An isomorphism from $T^{2(K-1)}/S_K$ to $\mathbb{C}P^{K-1}$ made of theta functions}
\label{sec:YYmap_prodtosum}

To prepare for the definition of this isomorphism, we introduce some notation as follows.

\begin{itemize}
    \item The element of $T^{2(K-1)}/S_K$ is represented by $\Vec{\phi} \in \mathbb{C}^K$ with constraint $\sum_{i=1}^K \phi_i = 0$.
    The torus is made from the periodicity $\phi_i \sim \phi_i +1$ and $\phi_i \sim \phi_i + \tau$, and $\sigma \in S_K$ is the permutation acting as $\phi_i \mapsto \phi_{\sigma(i)}$.
    \item Let $[H_1,\cdots,H_K]$ denote the homogeneous coordinates of $\mathbb{C}P^{K-1}$.
    \item Let $\mathbb{L}$ be the root lattice: $\mathbb{L} = \{ \Vec{\ell} = (\ell_1,\cdots,\ell_K) \in \mathbb{Z}^K;~\sum_{i=1}^K \ell_i = 0 \}$.
    \item  Let $(\Vec{e}_1,\Vec{e}_2,\cdots,\Vec{e}_K)$ denote the orthonormal basis of $\mathbb{R}^K$.
    \item We define $\Vec{\mu}_k$ as follows.
\begin{align}
    \Vec{\mu}_k &= \Vec{e}_1 + \cdots + \Vec{e}_k - \frac{k}{K} (\Vec{e}_1 + \cdots + \Vec{e}_K) \notag \\
    &= (\overset{1}{1},\cdots,\overset{k}{1},\overset{k+1}{0},\cdots,\overset{N}{0}) - \frac{k}{K} (1,1,\cdots,1).
\end{align}
    We also define $\Vec{\mu}_K = \Vec{0}$.
    In terms of the representation of $SU(K)$, these vectors are fundamental weights.

    \item We introduce the following theta functions:
\begin{align}
    \Theta_k (\Vec{\phi}) := \sum_{\Vec{\ell} \in \mathbb{L}} \rme^{\pi \im \tau (\Vec{\ell} + \Vec{\mu}_k)^2 + 2\pi \im (\Vec{\ell} + \Vec{\mu}_k) \cdot \Vec{\phi} }~~~(k=1,\cdots,K)
    \label{eq:def_large_Theta}
\end{align}
\end{itemize}

By using these notations, the explicit isomorphism from $T^{2(K-1)}/S_K$ to $\mathbb{C}P^{K-1}$ is given by,
\begin{align}
    T^{2(K-1)}/S_K &\longrightarrow \mathbb{C}P^{K-1} \notag \\
    \Vec{\phi}  & \mapsto [H_1 = \Theta_1(\Vec{\phi}), \cdots, H_K = \Theta_K(\Vec{\phi})]. \label{eq:YY-map-K}
\end{align}

We can check that this map is well-defined.
For root vectors $\vec{m},\vec{n} \in \mathbb{L}$, the theta functions transform as
\begin{align}
     \Theta_k(\Vec{\phi} + \tau \vec{m} + \vec{n}) = \rme^{-\pi \im \tau \vec{m}^2 - 2 \pi \im \vec{m} \cdot \vec{\phi}} \Theta_k(\Vec{\phi}),
\end{align}
in which the extra factor does not depend on $k$.
Thus, as a map to $\mathbb{C}P^{K-1}$, the definition (\ref{eq:YY-map-K}) is compatible with the periodicity of the torus.
One can also check the invariance under the permutation: $\Theta_k(\sigma \Vec{\phi}) = \Theta_k(\Vec{\phi})$, where $\sigma \in S_K$ and $\sigma \Vec{\phi} = (\phi_{\sigma(1)}, \cdots, \phi_{\sigma(K)})$.

With the help of the product-to-sum formula (\ref{eq:product-to-sum-App}) proved below, one can check that this map is isomorphic. See the main texts below (\ref{eq:product-to-sum-maintext}).
When the coefficients $[H_1,\cdots,H_K] \in \mathbb{C}P^{K-1}$ is given, the product-to-sum formula gives an algebraic equation to get $\vec{\phi } \in T^{2(K-1)}/S_K$ such that $[H_1 = \Theta_1(\Vec{\phi}), \cdots, H_K = \Theta_K(\Vec{\phi})]$.

\subsection{Proof of product-to-sum formula}
A product of the theta functions can be expanded as
\begin{align}
    \prod_{a=1}^K \theta_1(\Phi_a|\tau)=\sum_{\vec m \in \mathbb Z^K} q^{\left(\vec m+\frac12 \vec i\right)^2}e^{2\pi i \left(\vec m+\frac12 \vec i\right)\cdot\left(\vec \Phi -\frac12\vec i\right)}
\end{align}
with introducing $\vec \Phi=(\Phi_1,\Phi_2,\cdots,\Phi_K)\in \mathbb C$ and $\vec i:=(1,1,\cdots,1)$.

Here for $\vec m=(m_1,m_2,\cdots,m_K)$, its sum can always be written as  
\begin{align}
    \vec m \cdot \vec i =\sum_{a=1}^K m_a= K m+n-1, \quad {\rm with~}\, \exists \,m, n \in \mathbb Z,  \quad 1\le n \le K.
\end{align}
Using these integers $m,n$, we can define $\vec \ell\in \mathbb Z^K$  
so that $\vec \ell\cdot \vec i=0$  as
\begin{align}
    \vec \ell: =\vec m -\sum_{a=1}^{n-1}\vec e_a-m \,\vec i, \quad \Rightarrow \quad  \vec \ell \in \mathbb L \subset \mathbb Z^K, 
\end{align}
Furthermore, let us define $\vec \phi:= \vec \Phi-\langle \Phi \rangle \vec i $ with 
$\langle \Phi\rangle :=\frac1 K \vec \Phi\cdot \vec i$ so that $\vec \phi\cdot \vec i=0$.
Therefore, $\vec m$ and $\vec \Phi$ are decomposed as  
\begin{align}
    \vec m &=\vec \ell+ \vec \mu_{n-1}+\left(m+\frac{n-1}K\right) \vec i\\
    \vec \Phi&= \vec \phi+\langle \Phi\rangle \vec i.
\end{align}
And then, we find the following decompositions
\begin{align}
    \left(\vec m+\frac12 \vec i\right)^2&=\left(\vec \ell+\vec \mu_{n-1}\right)^2+K \left(m+\frac12 +\frac{n-1}K\right)^2\\
       \left(\vec m+\frac12 \vec i\right)\cdot\left(\vec \Phi-\frac12\vec i\right)&=\left(\vec \ell+\vec \mu_{n-1}\right)\cdot \vec \phi
       +K\left(m+\frac12 +\frac{n-1}K\right)\left(\langle \Phi\rangle -\frac12\right)
\end{align}
and therefore,
\begin{align}
 & \sum_{\vec m \in \mathbb Z^K} q^{\left(\vec m+\frac12 \vec i\right)^2}e^{2\pi i \left(\vec m+\frac12 \vec i\right)\cdot\left(\vec \Phi-\frac12\vec i\right)}\\
= & \sum_{n=1}^K \sum_{\vec \ell \in \mathbb L} q^{\left(\vec \ell+\vec \mu_{n-1}\right)^2}e^{2\pi i \left(\vec \ell+\vec \mu_{n-1}\right)\cdot \vec \phi}\sum_{m\in \mathbb Z} q^{K \left(m+\frac12 +\frac{n-1}K\right)^2}e^{2\pi iK\left(m+\frac12 +\frac{n-1}K\right)\left(\langle \Phi\rangle -\frac12\right)}\\
=& \sum_{n=1}^K \Theta_{n-1}(\vec \phi)\, \vartheta_n^K\left(\langle \Phi\rangle|\tau\right).
\end{align}

That is, we obtain the product-to-sum formula
\begin{align}
 \prod_{a=1}^K \theta_1(\Phi_a|\tau)=  \sum_{n=1}^K \Theta_{n-1}(\vec \phi)\, \vartheta_n^K\left(\langle \Phi\rangle|\tau\right).  \label{eq:product-to-sum-App}
\end{align}

%%%%%%%%%%%%%%%%%%%%%%%%%%%%%%%%%%%%%%%%%%%%%%%%%%%%%%%%%%%%%%%%%%%%%%%%%%%%%%%%%%%%%%%%%
\section{Calculations for explicit duality formulas}
\label{sec:DualityFormulas}

\subsection{Imaginary transformation on $\vartheta^{K}_n(z|\tau)$}
Here let us derive the extended version of Jacobi's imaginary transformation
\begin{align}
    \frac1{\sqrt{K}}\sum_{l=1}^Ke^{-2\pi i \frac{(n-1)(l-1)}K}\vartheta^{K}_l(z|\tau)
    =\sqrt{\frac{i}\tau} \left(ie^{-\pi i \frac{z^2}\tau}\right)^K \vartheta_n^K\left(\frac{z}\tau \Big|- \frac1\tau\right).
\end{align}
This identity can be confirmed as follows:
\begin{align}
    &\frac1{\sqrt{K}}\sum_{l=1}^Ke^{-2\pi i \frac{(n-1)(l-1)}K}\vartheta^{K}_l(z|\tau)\\
=&\frac1{\sqrt{K}}\sum_{l=1}^K\sum_{m\in \mathbb Z}q^{K\left(m+\frac12 +\frac{l-1}K\right)^2}e^{2\pi i K\left(m+\frac12 +\frac{l-1}K\right)\left(z-\frac12\right)-2\pi i \frac{(n-1)(l-1)}K}\\
\stackrel{(1)}{=}&\frac1{\sqrt{K}}\sum_{l\in \mathbb Z}q^{\frac1K\left(l-1+\frac K2 \right)^2}e^{2\pi i \left(l-1+\frac K2 \right)\left(z-\frac12\right)-2\pi i \frac{(n-1)(l-1)}K}\\
=&\frac1{\sqrt{K}}\sum_{l\in \mathbb Z}e^{\frac{\pi i \,\tau}K\left(l-1+\frac K2 \right)^2}e^{2\pi i \left(l-1+\frac K2 \right)
\left(z-\frac12-\frac{n-1}K\right)-\pi i (n-1)}\\
\stackrel{(2)}{=}&\sqrt{\frac{i}\tau}\sum_{m\in\mathbb Z} e^{-\frac{\pi i \,K}{\tau}\left(z-\frac12 -\frac{n-1}K-m\right)^2+\pi i m K-\pi i (n-1)}\\
=&\sqrt{\frac{i}\tau} \sum_{m\in\mathbb Z} e^{-\frac{\pi i \,K}{\tau}\left(m+\frac12 +\frac{n-1}K\right)^2+2\pi i K\left(m+\frac12 +\frac{n-1}K\right)\frac{z}\tau -\pi i K \frac{z^2}\tau-\pi i ( Km +n-1+\frac{K}2)+\pi i \frac{K}2}\\
=&\sqrt{\frac{i}\tau} \left(ie^{-\pi i \frac{z^2}\tau}\right)^K \vartheta_n^K\left(\frac{z}\tau \Big|- \frac1\tau\right),
\end{align}
where we used the following tricks:
\begin{enumerate}
    \item[(1)]  rearrangement of summation
    \begin{align}
        \sum_{l=1}^K \sum_{m\in \mathbb Z} f\left(Km+l-1\right)=\sum_{l\in\mathbb Z} f(l-1),
    \end{align}
    \item[(2)] Poisson summation formula 
    \begin{align}
        \sum_{l\in\mathbb Z} e^{\pi i\,a (l+c)^2+2\pi i\, b(l+c)} 
        =\sqrt{\frac{i}a}\sum_{m\in\mathbb Z}e^{-\frac{\pi i}a (b-m)^2 +2\pi i\, m \,c}\quad{\rm with~}{\rm Im}\, a>0.
    \end{align}
%    \begin{align}
%        &\sum_{l\in\mathbb Z} e^{\pi i\,a (l+c)^2+2\pi i\, b(l+c)} \quad{\rm with~} {\rm Im}\, a>1\\
%        =&\sum_{m\in \mathbb Z}\int_\mathbb R e^{\pi i\,a(x+c)^2+2\pi i\,b(x+c)-2\pi i m x} dx\\
%        =&\sum_{m\in \mathbb Z}\int_\mathbb R e^{\pi i\, a\left(x+c+\frac{b-m}a\right)^2-\frac{\pi i}a (b-m)^2 +2\pi i m c} dx \\
%        =&\sqrt{\frac{i}a}\sum_{m\in\mathbb Z}e^{-\frac{\pi i}a (b-m)^2 +2\pi i\, m \,c}
%    \end{align}
\end{enumerate}
%%%%%%%%%%%%%%%%%%%%%%%%%%%%%
\subsection{Duality formula between $(p_s,p_c)=(1,p)$ and $(p_s,p_c)=(p,1)$}
Here let us derive the duality relation between the two basis, 
$\{\vartheta^{k,N;(1,p)}_{n,a}\}$ and $\{\vartheta^{k,N;(p,1)}_{n,a}\}$.
Those two basis are explicitly given as
%\begin{align}
%\vartheta^{k,N;(p_s,p_c)}_{n,a}(z|\tau)
%:= e^{\pi i \frac{u(a)}N}\vartheta_n^{\hat M}\left(\frac{1}{\hat N}\left(z-\frac{u(a)}{M}\right)\Big| \frac{\tau}{\hat N} \right),
%\end{align}
\begin{align}
     & \vartheta^{k,N;(1,p)}_{n,a}(z|\tau)
= e^{\pi i \frac{M}N(a-1)}\vartheta_n^{ M}\left(\frac{z-a+1}{N}\Big| \frac{\tau}{ N} \right)\\
&=\sum_{m\in\mathbb Z}e^{\pi i\,\tau \frac{M}N\left(m+\frac12 +\frac{n-1}{M}\right)^2+2\pi i\frac{M}N\left(m+\frac12 +\frac{n-1}{M}\right)\left(z'-\frac12\right)-2\pi i\frac{M}N\left(m+\frac{n-1}{M}\right)(a-1) },  
\end{align}
and
\begin{align}
    &\vartheta^{k,N;(p,1)}_{n,a}\left(z_D\Big|-\frac1{\tau}\right)
= e^{\pi i \frac{M}Nq_b(a-1)}\vartheta_n^{ M}\left(\frac{z_D-q_b(a-1)}{N}\Big| -\frac{1}{ N\tau} \right)\\
&=\sum_{m\in\mathbb Z}e^{-\frac{\pi i}{\tau}\,\frac{M}N\left(m+\frac12 +\frac{n-1}{M}\right)^2
+2\pi i\frac{M}N\left(m+\frac12 +\frac{n-1}{M}\right)\left(\frac{z'}{\tau}+\frac12\right)-2\pi i\frac{M}N\left(m+\frac{n-1}{M}\right)q_b(a-1) }\\
&=\sum_{m\in\mathbb Z}e^{-\frac{\pi i}{\tau}\,\frac{M}N\left(m+\frac12 +\frac{n-1}{M}\right)^2
+2\pi i\frac{M}N\left(m+\frac12 +\frac{n-1}{M}\right)\left(\frac{z'}{\tau}+\frac12\right)-\frac{2\pi i}N\left(m+q_b(n-1)\right)(a-1) },
\end{align}
with $z':=z-(N-1)/2=\tau(z_D-(N+1)/2)$.
The duality relation between them can be derived as follows:
\begin{align}
    &\frac1{\sqrt{MN}}\sum_{n=1}^M\sum_{a=1}^N e^{2\pi i\frac{\alpha (n-1)}M}e^{-2\pi i\frac{(b-1)(a-1)}N}\vartheta^{k,N;(1,p)}_{n,a}(z|\tau)\\
    \stackrel{(1)}=&
   \frac1{\sqrt{MN}}\sum_{a=1}^N \sum_{m\in\mathbb Z}e^{\frac{\pi i\,\tau}{MN} \left(m+\frac  M 2\right)^2+\frac{2\pi i}N\left(m+\frac M 2 \right)\left(z'-\frac12\right)-\frac{2\pi i}N \left(m+b-1\right)(a-1)+2\pi i\frac{\alpha m }M }\\
   \stackrel{(1')}=&\sqrt{\frac{N}M} \sum_{m\in\mathbb Z}e^{\pi i\,\tau\frac{N}{M} \left(m+\frac{M}{2N}-\frac{b-1}N\right)^2+2\pi i\left(m+\frac{M}{2N}-\frac{b-1}N\right)\left(z'-\frac12\right)+2\pi i\frac{\alpha (Nm-(b-1) )}M }\\
   =&\sqrt{\frac{N}M} \sum_{m\in\mathbb Z}e^{\pi i\,\tau\frac{N}{M} \left(m+\frac{M}{2N}-\frac{b-1}N\right)^2+2\pi i\left(m+\frac{M}{2N}-\frac{b-1}N\right)\left(z'-\frac12+\alpha \frac{N}M\right)-\pi i \alpha }\\
   \stackrel{(2)}=& \sqrt{\frac{i}\tau}\sum_{m\in\mathbb Z}e^{-\frac{\pi i}{\tau}\frac{M}N\left(m-\alpha \frac{N}M+\frac12 -z'\right)^2+2\pi i m \left(\frac{M}{2N}-\frac{b-1}N\right)-\pi i \alpha }\\
   =& \sqrt{\frac{i}\tau} e^{-\frac{\pi i}{\tau}\frac{M}N(z')^2-\frac{\pi i}2\frac{M}N}\sum_{m\in\mathbb Z}e^{-\frac{\pi i}{\tau}\frac{M}N\left(m-\alpha \frac{N}M+\frac12 \right)^2+2\pi i \frac{M}N\left(m-\alpha \frac{N}M+\frac12 \right)\left(\frac{z'}\tau+\frac12\right)-2\pi i\frac{m(b-1)}N}\\
   \stackrel{(2')}=&
   \sqrt{\frac{i}\tau} e^{-\frac{\pi i}{\tau}\frac{M}N(z')^2-\frac{\pi i}2\frac{M}N}\nn \\ & \quad \times\sum_{m'\in\mathbb Z}
   e^{-\frac{\pi i}{\tau}\frac{M}N\left(m'+\frac12+\frac{n'-1}M\right)^2+2\pi i \frac{M}N\left(m'+\frac12+\frac{n'-1}M \right)\left(\frac{z'}\tau+\frac12\right)-2\pi i\frac{m'+q_b(n'-1)}N(b-1)}\\
   =&\sqrt{\frac{i}\tau} e^{-\frac{\pi i}{\tau}\frac{M}N(z')^2-\frac{\pi i}2\frac{M}N } 
   \vartheta^{k,N;(p,1)}_{n',b}\left(z_D\Big|-\frac1{\tau}\right)
\end{align}
where the following  tricks were used in addition to $(1)$ and $(2)$ in the previous subsection:
\begin{enumerate}
    \item[($1'$)] discrete Fourier transformation
    \begin{align}
       \sum_{a=1}^N \sum_{m\in\mathbb Z}f(m)e^{2\pi i\frac{(m+b-1)(a-1)}N} =N\sum_{m\in \mathbb Z}f(N m-b+1),
    \end{align}
    \item[($2'$)] rearrangement of the summation index $m':=m-q_b(n'-1)$ by setting $\alpha =\nu (n'-1)$ with $\nu\in\mathbb Z$ satisfying $q_b M-\nu  N=1$
    \begin{align}
        m-\alpha\frac{N}M=m-(n'-1)\frac{\nu N}M=(m-q_b(n'-1))+\frac{n'-1}M=m'+\frac{n'-1}M.
    \end{align}
\end{enumerate}
%%%%%%%%%%%%%%%%%%%%

%%%%%%%%%%%%%%%%%%%%%%%%%%%%%%%%%%%%%%%%%%%%
%%%%%%%%%%%%%%%%%%%%%%%%%%%%%%%%%%%%%%%%%%%

%%%%%%%%%%%%%%%%%%%%%%%%%%%%%%%%%%%%%%%%%%%

%%%%%%%%%%%%%%%%%%%%%%%%%%%
%\bibliographystyle{apsrev4-1}
%\bibliographystyle{jhep}
%\bibliographystyle{ieeetr}
%\bibliography{references}

\bibliographystyle{utphys}
\bibliography{./QFT,./refs,./references}

\end{document}